\documentclass[twocolumn]{aastex6}
\usepackage{subcaption}
\usepackage{amsmath,amssymb,amstext}
\usepackage{natbib}
\usepackage{xspace}

\newcommand{\Fermi}{\emph{Fermi}\xspace}

\newcommand{\be}{\begin{linenomath*}\begin{equation}}
\newcommand{\ee}{\end{equation}\end{linenomath*}}
\newcommand{\ba}{\begin{linenomath*}\begin{eqnarray}}
\newcommand{\ea}{\end{eqnarray}\end{linenomath*}}
\newcommand{\fracb}[2]{\left(\frac{#1}{#2}\right)}
\newcommand{\Granotpaper}{2008ApJ...677...92G}
\newcommand{\Hascoetpaper}{2012MNRAS.421..525H}
\newcommand{\eqt}[1]{eq.~(#1)}
\newcommand{\eqp}[1]{(eq.~#1)}

\newcommand*{\vcenteredhbox}[1]{\begingroup
\setbox0=\hbox{#1}\parbox{\wd0}{\box0}\endgroup}

\graphicspath{{figures/}}

\begin{document}

\title{The Bright and the Slow -- GRBs 100724B \& 160509A with high-energy cutoffs at $\lesssim100\;$MeV
}

\author{G. Vianello\altaffilmark{1}}
\author{R. Gill\altaffilmark{2,3}}
\author{J. Granot\altaffilmark{2,4}}
\author{N. Omodei\altaffilmark{1}}
\author{J. Cohen-Tanugi\altaffilmark{5}}
\author{F. Longo\altaffilmark{6,7}}
\altaffiltext{1}{Hansen Experimental Physics Lab, Stanford University, 452 Lomita Mall, Stanford, CA 94305-4085, United States}
\altaffiltext{2}{Department of Natural Sciences, The Open University of Israel, 
1 University Road, PO Box 808, Raanana 4353701, Israel}
\altaffiltext{3}{Physics Department, Ben-Gurion University, P.O.B. 653, 
Beer-Sheva 84105, Israel}
\altaffiltext{4}{Department of Physics, The George Washington University, Washington, DC 20052, USA}
\altaffiltext{5}{Laboratoire Univers et Particules de Montpellier, Universit\'{e} de 
Montpellier, CNRS/IN2P3, Montpellier, France}
\altaffiltext{6}{Istituto Nazionale di Fisica Nucleare, Sezione di Trieste, I-34127 Trieste, Italy}
\altaffiltext{7}{Dipartimento di Fisica, Universit`a di Trieste, I-34127 Trieste, Italy}

\begin{abstract}
We analyze the prompt emission of GRB~100724B and GRB~160509A, two of the brightest Gamma-Ray Bursts (GRBs) observed by \Fermi at $\lesssim{\rm MeV}$ energies but surprisingly faint at $\gtrsim100\;$MeV energies. Time-resolved spectroscopy reveals a sharp high-energy cutoff at energies $E_c\sim20-60\;$MeV for GRB~100724B and  $E_c\sim80-150\;$MeV for GRB~160509A. We first characterize phenomenologically the cutoff and its time evolution. We then fit the data to two models where the high-energy cutoff arises from intrinsic opacity to pair production within the source ($\tau_{\gamma\gamma}$): (i) a Band spectrum with $\tau_{\gamma\gamma}$ from the internal-shocks motivated model of \citet{\Granotpaper}, and (ii) the photospheric model of \citet{GT14}. Alternative explanations for the cutoff, such as an intrinsic cutoff in the emitting electron energy distribution, appear to be less natural. Both models provide a good fit to the data with very reasonable physical parameters, providing \textbf{an estimate} of bulk Lorentz factors in the range $\Gamma\sim 100-400$, on the lower end of what is generally observed in \Fermi GRBs. Surprisingly, their lower cutoff energies $E_c$ compared to other \Fermi/LAT GRBs  arise not predominantly from the lower  Lorentz factors, but also 
at a comparable level from differences in variability time, luminosity, and high-energy photon index. Finally, particularly low $E_c$ values  may prevent detection by \Fermi/LAT, thus introducing a bias in the \Fermi/LAT GRB sample against  GRBs with low Lorentz factors or variability times.
\end{abstract}

\maketitle

\section{Introduction}

The $\gamma$-ray emission from Gamma-Ray Bursts (GRBs) is believed to originate within an ultra-relativistic jet, which is launched
during the collapse of a massive star \citep[for long duration GRBs that last $\gtrsim2\,$s, ][]{1999ApJ...524..262M} and likely also during the merger of two compact objects \citep[for short duration GRBs that last $\lesssim2\,$s, ][]{2011ApJ...732L...6R}. However, the mechanisms that produce the prompt emission of GRBs are still debated \citep[see e.g. the recent review by ][]{2015PhR...561....1K}. 
An important question is the composition of the jet, which remains unresolved, and for which two scenarios have been proposed: a baryonic jet where particles are accelerated converting thermal energy into bulk motion (fireballs) \citep{1994ApJ...430L..93R}, or a Poynting-flux-dominated jet \citep{2001MNRAS.321..177L}. The composition of the jet in turn determines the dominant dissipation mechanism that converts the energy content of the jet into heat and accelerated particles that radiate the observed prompt emission. For example, in baryonic jets energy dissipation can be attributed to internal shocks \citep[e.g.][]{1994ApJ...430L..93R, 0004-637X-723-1-267, 0004-637X-767-1-19},  and/or collisional heating due to inelastic collisions between neutrons and protons \citep{2010MNRAS.407.1033B}. On the other hand, in a Poynting-flux-dominated jet, where most of the energy is stored in the magnetic field, magnetic reconnection occurring 
in an outflow with a striped magnetic field structure or due to magnetohydrodynamic turbulence can dissipate magnetic energy and power the prompt emission \citep[e.g.,][]{1994MNRAS.270..480T, 2003astro.ph.12347L,2011ApJ...726...90Z}.

In the context of fireball models, the dominant emission mechanism was thought to be synchrotron radiation, possibly also accompanied by synchrotron self-Compton. In particular, the highly-variable prompt emission has been attributed to synchrotron emission from particles accelerated in multiple internal shocks, i.e., shocks that occur when a faster shell ejected by the central engine collides with a slower shell within the outflow. Such a scenario has been used to explain the non-thermal spectrum that characterizes GRBs. The efficiency that internal shocks can achieve in converting energy into radiation appears to be insufficient to explain the luminosity of some GRBs \citep{1999MNRAS.309L..13L, 1997ApJ...490...92K}, unless the spread in Lorentz factor between the colliding shells is large \citep{2001ApJ...551..934K}. Also, a non-negligible fraction of GRBs show spectra that are difficult to explain with pure synchrotron emission \citep{2002ApJ...581.1248P, 2015A&A...583A.129Y,2015MNRAS.451.1511B, 2015MNRAS.447.3150A}. For this reason, some GRBs have been modeled with phenomenological models adding a thermal component to the non-thermal one \citep{2005ApJ...625L..95R, 2011ApJ...727L..33G,2012ApJ...757L..31A, 2013ApJ...770...32G, 2014ApJ...784...17B, 2015A&A...573A..81Y,2015ApJ...807..148G, 2017A&A...598A..23N}.

Because of these issues with the so-called ``standard'' fireball paradigm, another class of fireball models has emerged, which we call for simplicity photospheric models \citep[for example][]{2004ApJ...614..827R, 2005ApJ...635..476P, 2010MNRAS.407.1033B, 2011ApJ...738...77V, 2013ApJ...765..103L}. In this class of models the spectrum of a GRB is explained as reprocessed quasi-thermal radiation coming from the photosphere, i.e. the surface where radiation and matter decouple, typically after the acceleration of the fireball has ended for thermal acceleration, or possibly during the acceleration phase for magnetic acceleration (which is slower than thermal acceleration). A thermal or quasi-thermal initial spectrum is reprocessed within the jet to produce the non-thermal spectrum commonly observed in GRBs. The differences between the various photospheric models lie in the mechanisms responsible for the reprocessing of the thermal spectrum, which in turn requires different ingredients: strongly-magnetized or non-magnetized jets, baryon-dominated or baryon-poor, or other factors.

Here we present the analysis of the prompt emission of GRB~100724B and GRB~160509A, both detected by the \textit{Fermi Gamma-ray Space Telescope} instruments. These two GRBs are very bright at low energy, but they do not show any emission above 1 GeV during the prompt phase, which sets them apart from bursts of comparable low-energy fluence such as GRB~080916C, GRB~090902B and GRB~090926 \citep{LATGRBcat}. Moreover, the high-energy emission above 1 GeV, widely thought to originate from a different mechanism than the prompt emission (for example, external shock), picks up after the prompt phase is finished. This gives us the rare possibility of studying the prompt emission without any contamination from the high-energy component. Both GRBs show a very evident spectral cutoff in the $10-200$ MeV energy range with respect to the extrapolation of the low-energy component. \textbf{We interpret it} as pair production opacity, which allows for a measurement of the bulk Lorentz factor of the jet. While other cases of sub-GeV cutoffs have been reported \citep{LATGRBcat, 2015ApJ...806..194T}, GRB~100724B and GRB~160509A are by far the two brightest ones, and allow for an in-depth analysis impossible in the other cases. We also perform a detailed time-resolved analysis and measure the time evolution of the bulk Lorentz factor in both GRBs. Our detailed analysis allows us also to verify the viability of specific physical models. We choose to consider one model related to the ``standard'' fireball picture and one photospheric model. In particular, among many possibilities, we choose the semi-phenomenological internal-shock model of \citet{2008ApJ...677...92G} featuring a detailed modeling of the pair production opacity, and the photospheric model of \citet{GT14}. These models provide a natural explanation for the spectral cutoff, and we have readily available numerical codes which provide the spectra foreseen by the two scenarios as a function of physical parameters (see section \ref{sec:physical_modeling} for more details).

In \S~\ref{sec:Fermi} we present the \Fermi observatory. We then present the main features of GRB~100724B (\S~\ref{sec:100724B}) and GRB~160509A (\S~\ref{sec:160509A}). In particular, we establish phenomenologically that the high-energy data cannot be modeled extrapolating the low-energy spectrum, requiring instead a high-energy cutoff in the $10-200\;$MeV energy range. Next, in \S~\ref{sec:physical_modeling} we interpret such a feature in the context of physical models. We finally discuss our results (\S~\ref{sec:dis}) and provide our conclusions (\S~\ref{sec:conc}).
Throughout this paper we will use the ``Planck 2015'' flat cosmology \citep{2016A&A...594A..13P}, with $H_{0} = 67.8$ km s$^{-1}$ Mpc$^{-1}$ and $\Omega_{m} = 0.308$. 

\section{The \Fermi observatory}
\label{sec:Fermi}
\Fermi orbits the Earth at an altitude of $\sim 565$ km. Its pointing is continuously changing in a pattern that allows its instruments to survey the entire sky approximately every 3 hours. 

The Large Area Telescope (LAT) \citep{2009ApJ...697.1071A} is a pair-conversion telescope operating in the energy range from around 20 MeV up to over 300 GeV. For this study we use the P8\_TRANSIENT020E class of LAT data, and the corresponding instrument response function, and the \textit{LAT Low-Energy} data (LLE), available on the \Fermi Science Support Center (FSSC) website\footnote{http://heasarc.gsfc.nasa.gov/W3Browse/fermi/fermille.html}. When compared to P8\_TRANSIENT020E data, LLE data feature a higher acceptance especially below 100 MeV, at the expense of a higher background contamination and a very limited spatial resolution. It is designed for the spectral analysis of short-duration transients such as GRBs and solar flares.

On board \Fermi is also the Gamma-Ray Burst Monitor (GBM). It is comprised of 12 sodium iodide (NaI) detectors sensitive in the 8 keV $-$ $\sim$1 MeV energy range, and 2 bismuth germanate (BGO) detectors sensitive in the 200 keV $-$ $\sim$40 MeV energy range. The detectors are arranged to allow GBM to probe continuously all the sky not occulted by the Earth, with the exception of the time interval when the spacecraft is going through the South Atlantic Anomaly and data taking is suspended. \textbf{In this work we use the GBM data and tools publicly available on the FSSC website.}

\section{GRB 100724B}
\label{sec:100724B}

\subsection{Observations}

\begin{figure*}
\centering
\vcenteredhbox{\includegraphics[width=0.55\textwidth]{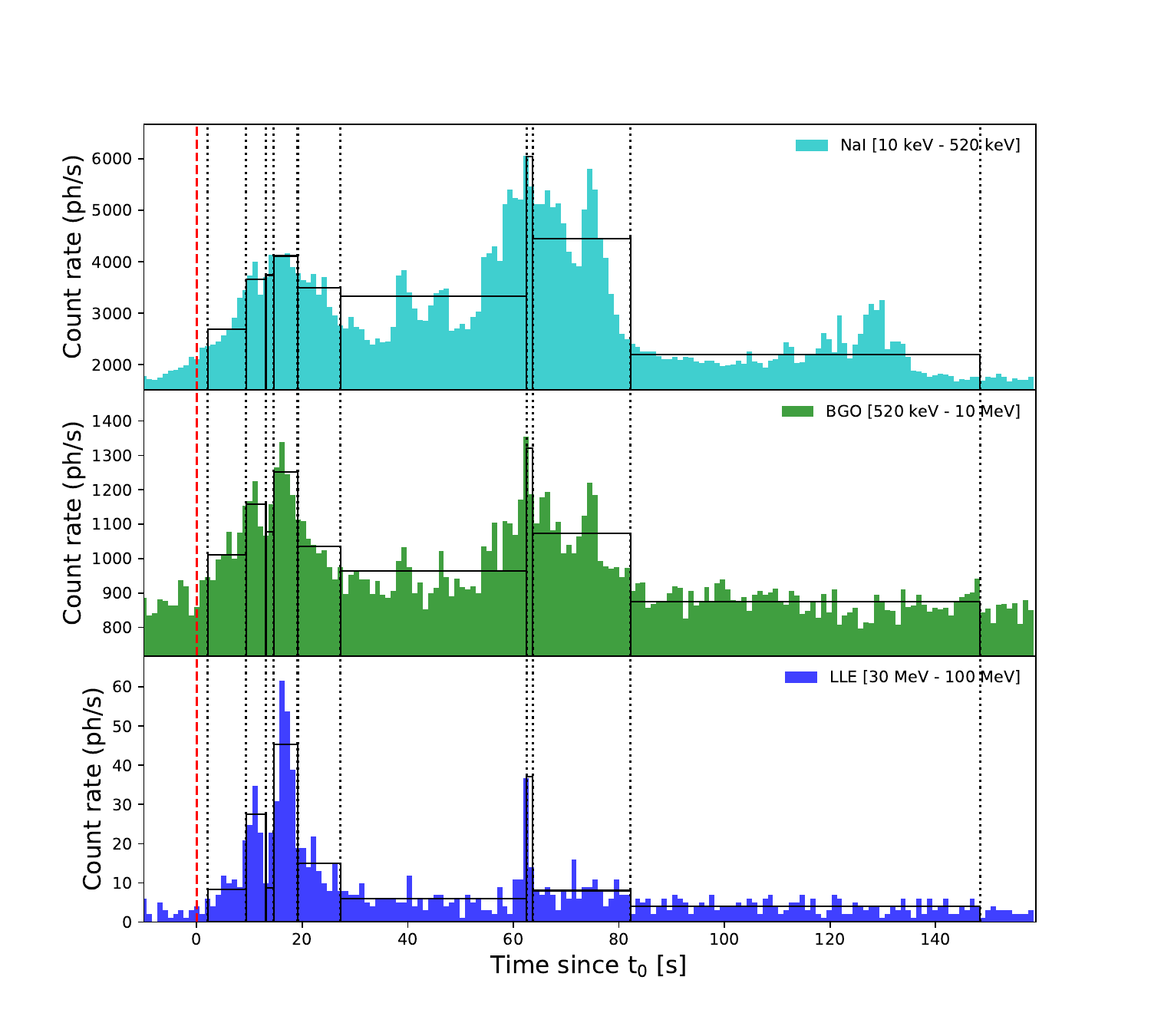}}
\vcenteredhbox{\includegraphics[width=0.40\textwidth]{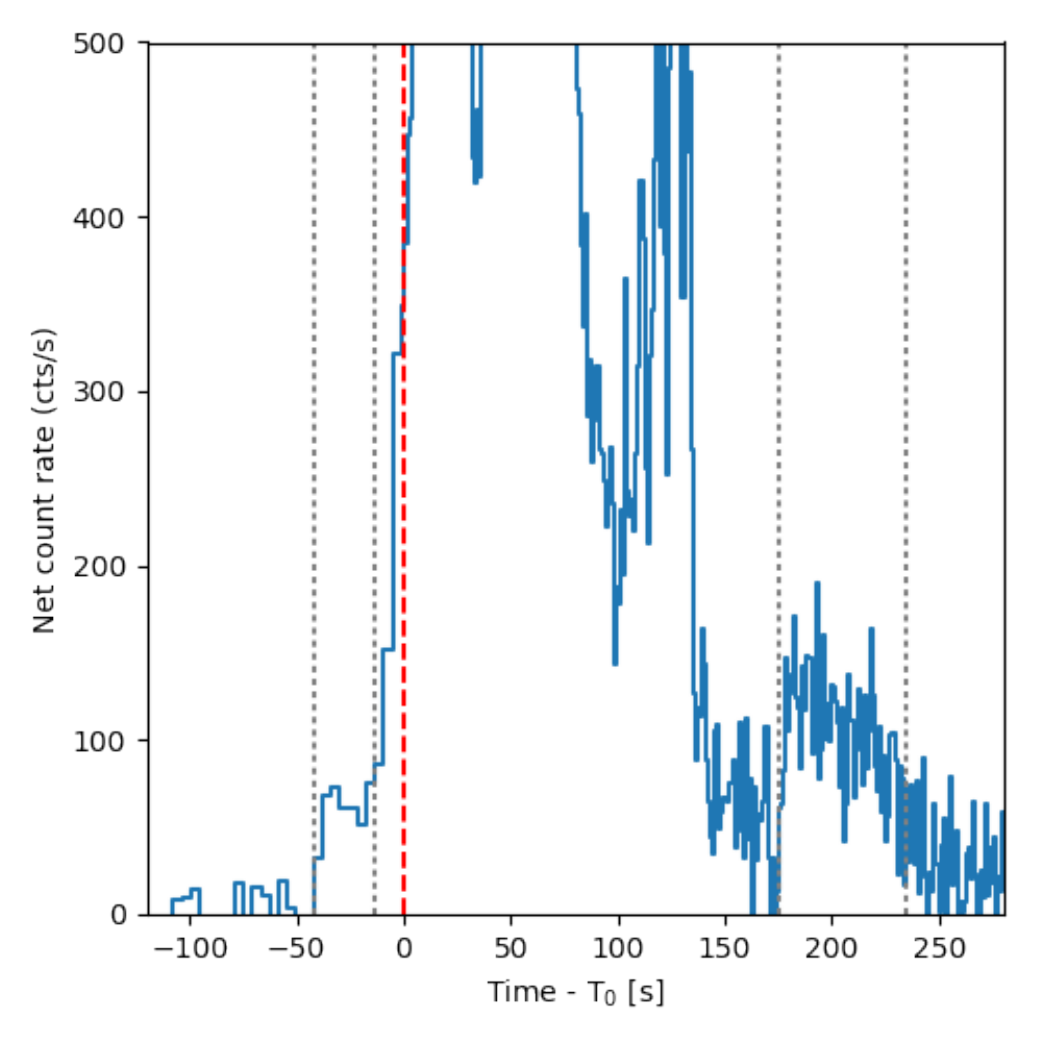}}
\caption{Left: Composite light curve of GRB~100724B showing NaI, BGO and LAT/LLE data. There is no photon spatially and temporally associated with the GRB with energy above 100 MeV, thus we do not show LAT standard data. The dashed red vertical lines represent the trigger time, while the other vertical lines correspond to the intervals obtained with the Bayesian Blocks algorithm. The black bars show the light curve obtained using the blocks as bins.  Right: zoomed background-subtracted light curve for the NaI low-energy detectors. We used here a bin size of 4 s to highlight the precursor and the late time soft emission (dashed vertical lines).}
\label{fig:bblc}
\end{figure*}

The bright GRB 100724B triggered \Fermi/GBM \citep{2009ApJ...702..791M} at 00:42:05.99 on 2010-07-24 \citep{2010GCN..10977...1B} ($t_{0}$ in the following). It was also detected by \Fermi/LAT and a preliminary localization was reported \citep{2010GCN..10978...1T}. GRB~100724B was also detected by {\it Konus-Wind} \citep{2010GCN..10981...1G}, {\it AGILE} \citep{2010GCN..10994...1M,2010GCN..10996...1G,2011A&A...535A.120D} and {\it Suzaku} \citep{2010GCN..10995...1U}. This burst has the third greatest fluence to date at low energy ($<10\;$MeV) among all the LAT-detected GRBs, exceeded only by GRB 090902B and the record-breaking GRB~130427A \citep{LATGRBcat, 2014Sci...343...42A}.  The initial localization has been improved in \citet{LATGRBcat}. We use in this paper an even more refined localization, $R.A. = 123.47\,^{\circ}$ and $Dec. = 75.88\,^{\circ}$ (J2000), obtained as described in Appendix~\ref{sec:localization}.

Another burst, GRB~100724A, was detected few seconds later by Swift \citep{2010GCN..10968...1M} in a position occulted by the Earth for \Fermi. Therefore, even if \Fermi/GBM is a non-imaging full-sky monitor, this second GRB was not observed by any GBM detector and does not therefore affect the analysis presented in this work. However, follow up efforts focused on this second GRB and therefore no multi-wavelength data are available for GRB~100724B.

The light curve of GRB~100724B is shown in Figure~\ref{fig:bblc}. During the main emission episode the signal in the LAT was exceptionally intense in the 30$\;$MeV--100$\;$MeV energy range, but nothing was detected above 100$\,$MeV. There is a faint ``precursor'' peak before $t_{0}$, the main emission episode going from $t_{0}$ to $\sim t_{0} + 150\,$s, and then a late soft peak starting at $\sim t_{0} + 180$ s.

\begin{table*}
\centering
\begin{tabular}{l|l|l}
\hline
Model & Description & Ref. \\
\hline
$f_{Band}$ & Band function & \eqt{\ref{eq:band}} \\
$f_{BB}$ & Band function plus blackbody & \eqt{\ref{eq:bb}} \\
$f_{BHec}$ & Band function with high-energy exponential cutoff & \eqt{\ref{eq:bhec}} \\
$f_{Bbkpo}$ & Band function with broken power-law spectrum above the peak & \eqt{\ref{eq:band_bknpo}} \\
$f_{Bgr}$ & Band function with a gradual break in power-law spectrum above the peak & \eqt{\ref{eq:Bgr}} \\
$f_{BG}$ & Band function with high-energy spectral break due to $\gamma\gamma$ pair opacity & \S5.1; \citep{2008ApJ...677...92G} \\
$f_{GT}$ & Spectrum from delayed pair breakdown model in a strongly magnetized jet & \S5.2; \citep{GT14}\\
$f_{th}$ & Quasi-thermal spectrum described by a power-law plus a Wein peak & \eqt{\ref{eq:mg14_qt}}; \citep{GT14} \\
\hline
\end{tabular}
\caption{Summary of various spectral models used in this work.}
\label{tab:models}
\end{table*}

\subsection{Spectral analysis of the prompt emission}
\label{sec:spectral_analysis}

We consider GBM detectors NaI 0 and 1, because they are the only two low-energy detectors seeing the GRB at an off-axis angle of less than $40\,^{\circ}$. Furthermore, we select the BGO detector closest to the GRB direction (BGO 0). We use \textit{Time-Tagged-Events} data provided by the GBM team and publicly available on the FSSC website. We generate custom response matrices (\texttt{rsp2} files, with one new response every time the spacecraft slew by 0.5 deg) using the public tool \textit{gbmrspgen}\footnote{http://fermi.gsfc.nasa.gov/ssc/data/analysis/scitools/gbmrspgen.html} and using our best localization for the source. We use NaI data in the energy range 8 keV $-$ $\sim$1 MeV but excluding the energy range $30-40$ keV, which contains the K-edge feature. We use BGO data from channel 2 to channel 125 (\textbf{corresponding to the energy range $\sim 217$ keV - $38$ MeV}). We also use LAT LLE data above 30 MeV.

We estimate the background for all GBM detectors and for LLE data by fitting off-pulse intervals with one polynomial function for each channel, and then interpolating such fit to the on-pulse interval \citep[for details see][]{LATGRBcat}. \textbf{This way the time-varying background --and in particular the Earth Limb contribution -- is naturally taken into account.}

We first perform a time-integrated analysis using the same time interval used in \citet{LATGRBcat}, i.e., the GBM $t_{90}$ time interval, from $t_{0} + 8.195\;$s to $t_{0} + 122.882\;$s. We use the Multi-Mission Maximum Likelihood framework (3ML) for all spectral analysis performed in this paper \citep{3ML}. We find very similar results: the spectrum can be successfully modeled by a Band function, a phenomenological model traditionally used in describing GRB spectra \citep{1993ApJ...413..281B}, multiplied by an exponential cutoff. The formulas for the Band function and the Band with exponential cutoff function are reported in Appendix~\ref{sec:spmodels}, eqs.~(\ref{eq:band}) and (\ref{eq:bhec}). The best fit parameters for the time-integrated analysis are $\alpha = -0.69 \pm 0.02$, $\beta =  -2.01^{-0.02}_{+0.03}$, $E_{p} = 330 \pm 10\;$keV and $E_{c} = 48 \pm 6\;$MeV. The fluence in the 1$\;$keV$\,$--$\,10\;$GeV energy range is $(4.7 \pm 0.3) \times 10^{-4}\;{\rm erg\;cm^{-2}}$. \textbf{Given the very high signal-to-noise ratio of this time-integrated analysis we can rebin the spectra in order to have at least $30$ counts in each bin, and then we can use a standard $\chi^2$ test. We obtain $\chi^2 = 376.2$ for $391$ d.o.f., corresponding to a p-value of $\sim 0.7$.}

In order to further study this feature, we then perform a time-resolved spectral analysis. The choice of the time intervals requires a trade-off. Choosing many time bins gives good time resolution but low sensitivity for detecting features, due to the decreased statistics in each spectrum. On the contrary, choosing few bins gives good sensitivity at the risk of smearing the time evolution of the parameters. In this paper we are mainly interested in the study of the cutoff, thus we choose to focus on LLE data, which cover the energy range where the cutoff is measured, and we decide our time bins based on the variability seen in the LLE light curve. In particular, we apply the Bayesian Blocks algorithm  \citep{2013ApJ...764..167S} that finds the most probable segmentation of the observation into time intervals during which the photon arrival rate has no statistically significant variations, i.e., it is perceptibly constant. We have used the implementation provided in the tool \texttt{gtburst} of the \Fermi Science Tools, using a probability of false positives of $p_{0}=0.01$. Applying it to LLE data we find the 9 intervals between 0 and $\sim 150\;$s shown in Figure~\ref{fig:bblc}.

We note that these intervals do not cover the faint and soft ``precursor'' peak that can be seen between $t_{0} - 42\;$s and $t_{0} - 13.3\;$s, nor the faint and soft late peak between $t_0+175\;$s and $t_0+235\;$s, since they do not show any LLE emission. For the precursor, we find that it is well described by the power-law with exponential cutoff model $f_{PHec}$ (eq.~\ref{eq:PHec}), with parameters $\alpha = -0.4^{-0.3}_{+0.4}$ and $E_{p} = 130^{-50}_{+90}$ keV\textbf{, and a p-value computed as above of $p = 0.62$}. The faint and soft late peak is described again by $f_{PHec}$, with parameters $\alpha = -1.35^{-0.12}_{+0.13}$ and $E_{p} = 142^{-35}_{+60}$ keV (\textbf{$p=0.45$).}

We now focus on the main emission episode. \textbf{We extract the spectra and compute the response matrices for each detector and each interval.} Initially we consider a pool of commonly used phenomenological spectral models (summarized in Table \ref{tab:models}) in order to characterize the spectra without having to assume a specific theoretical framework. We will consider two specific physical models later on (see section \ref{sec:physical_modeling}). Our phenomenological models are based on the Band model: a) the Band model itself $f_{Band}$ \eqp{\ref{eq:band}}; b) a Band plus black-body model $f_{BB}$ \eqp{\ref{eq:bb}}, which was used for the modeling of this GRB in \citet{2011ApJ...727L..33G}; c) the Band model multiplied by an exponential cutoff $f_{BHec}$ \eqp{\ref{eq:bhec}}; d) a Band model where the high-energy power law changes photon index abruptly at a cutoff energy ($f_{Bbkpo}$, \eqt{\ref{eq:band_bknpo}}); and e) a Band model with a smooth spectral break, suggested on theoretical grounds in \citet{2008ApJ...677...92G} ($f_{Bgr}$, \eqt{\ref{eq:Bgr}}). We also apply a group of alternative models, namely the log-parabolic spectral shape \citep{2010ApJ...714L.299M}, a broken power law,  and the smoothly broken power law of \citet{1998tx19.confE..83R}. However, they yield large residuals and in all time intervals considered here they describe the data significantly worse than the models based on the Band function. Therefore, we disregard them from now on. We also use a procedure to mitigate effects due to inter-calibration issues between the instruments. We take one instrument as reference (NaI 0), and then we introduce a multiplicative constant for every other detector. Such constant is left free to vary in the fit between 0.7 and 1.3, corresponding to an inter-calibration uncertainty of up to 30\%. This ``effective area correction'' reduces the biases due to systematic errors in the total effective area of the instruments with respect to the reference one.

For each time interval we measure separately the significance of the black body in model $f_{BB}$ and of the exponential cutoff in model $f_{BHec}$ with respect to the Band model alone $f_{Band}$. We rely on the Likelihood Ratio Test, which uses as Test Statistic ($TS$) twice the difference in log-likelihood between the null hypothesis (the Band model in our case) and the alternative hypothesis (either $f_{BHec}$ or $f_{BB}$ in our case). The details of this procedure can be found in Appendix \ref{sec:LRT}. \textbf{We also measure the p-value for a goodness-of-fit test using a procedure equivalent to the classic $\chi^2$ test but more appropriate for Poisson data. In particular, we follow the method proposed by \citet{cousins2013generalization} based on Monte Carlo simulations. It is well known that the goodness-of-fit p-value $p$ can be misleading when the data points have very different uncertainties, because the points with smaller errors will dominate. This is the case in our situation, where GBM data provide a much larger statistic than LLE data. Therefore, we measure separately the null-hypothesis probability for the entire dataset ($p$) and for the LLE data alone ($p_{\rm LLE}$) in order to investigate whether a model is able to describe the data both at low and at high energy.}

\begin{table*}
\centering
\begin{tabular}{|r|c|c|c|c|cc|c|c|cc|c|c|}
\hline
 \# &   Time interval & $S$ of $f_{Band} $ & p & p$_{\rm LLE}$ & \multicolumn{2}{c|}{TS of $f_{BB}$} & p & p$_{\rm LLE}$ & \multicolumn{2}{c|}{TS of $f_{BHec}$} & p & p$_{\rm LLE}$\\
 \hline
 1 &     2.11 - 9.39 &   1374.3 &  $< 10^{-3}$ & $< 10^{-3}$ & 16.8 & ($3.5~\sigma$) &  0.3 &       0.002 &  27.0 & ($5.2~\sigma$)  & 0.38 & 0.23 \\
 2 &    9.39 - 13.11 &   1079.6 &  $< 10^{-3}$ & $< 10^{-3}$ & 74.6 & ($8.3~\sigma$) & 0.05 & $< 10^{-3}$ &  74.2 & ($8.6~\sigma$)  & 0.26 & 0.52 \\
 3 &   13.11 - 14.61 &    604.9 &         0.05 &       0.002 & 12.8 & ($2.9~\sigma$) & 0.13 &       0.004 &  16.4 & ($4~\sigma$)    & 0.13 & 0.05 \\
 4 &   14.61 - 19.14 &   1227.2 &  $< 10^{-3}$ & $< 10^{-3}$ & 97.6 & ($9.5~\sigma$) & 0.26 & $< 10^{-3}$ & 127.2 & ($11.2~\sigma$) & 0.93 & 0.61 \\
 5 &   19.14 - 27.23 &   1448.3 &  $< 10^{-3}$ & $< 10^{-3}$ & 12.0 & ($2.8~\sigma$) & 0.12 & $< 10^{-3}$ &  57.3 & ($7.6~\sigma$)  & 0.88 & 0.17 \\
 6 &   27.23 - 62.52 &   2174.9 &          0.1 &       0.001 & 50.8 & ($6.7~\sigma$) & 0.25 & $< 10^{-3}$ &  27.1 & ($5.2~\sigma$)  & 0.21 & 0.09 \\
 7 &   62.52 - 63.65 &    544.7 &         0.09 &        0.03 &  3.8 & ($3.1~\sigma$) &  0.4 &       0.001 &   7.8 & ($2.8~\sigma$)  & 0.59 & 0.82 \\
 8 &   63.65 - 82.23 &   1888.2 &  $< 10^{-3}$ & $< 10^{-3}$ & 51.8 & ($6.8~\sigma$) & 0.26 & $< 10^{-3}$ &  40.6 & ($6.4~\sigma$)  & 0.21 & 0.36 \\
 9 &  82.23 - 148.41 &   2393.0 &         0.07 &        0.12 &  7.2 & ($1.9~\sigma$) & 0.58 &        0.15 &   3.2 & ($1.8~\sigma$)  & 0.78 & 0.32 \\
\hline
\end{tabular}
\caption{Value of the -log(likelihood) $S$ for the Band model $f_{Band}$, and the $TS$ obtained respectively with a Band + Black body ($f_{BB}$) and a Band with exponential cutoff ($f_{BHec}$) as alternative hypotheses, for GRB~100724B. In parenthesis we report the significance of the improvement. In the $p_{\rm null}$ columns we also report the null-hypothesis probability for the models for LLE data. Given the limited number of simulations used to measure p$_{\rm null}$, we cannot reliably measure probabilities smaller than $10^{-3}$. Therefore, we report p$_{\rm null} < 10^{-3}$ in these cases.}
\label{tab:model-selection}
\end{table*}

\begin{figure*}
\centering
\includegraphics[width=0.4\textwidth]{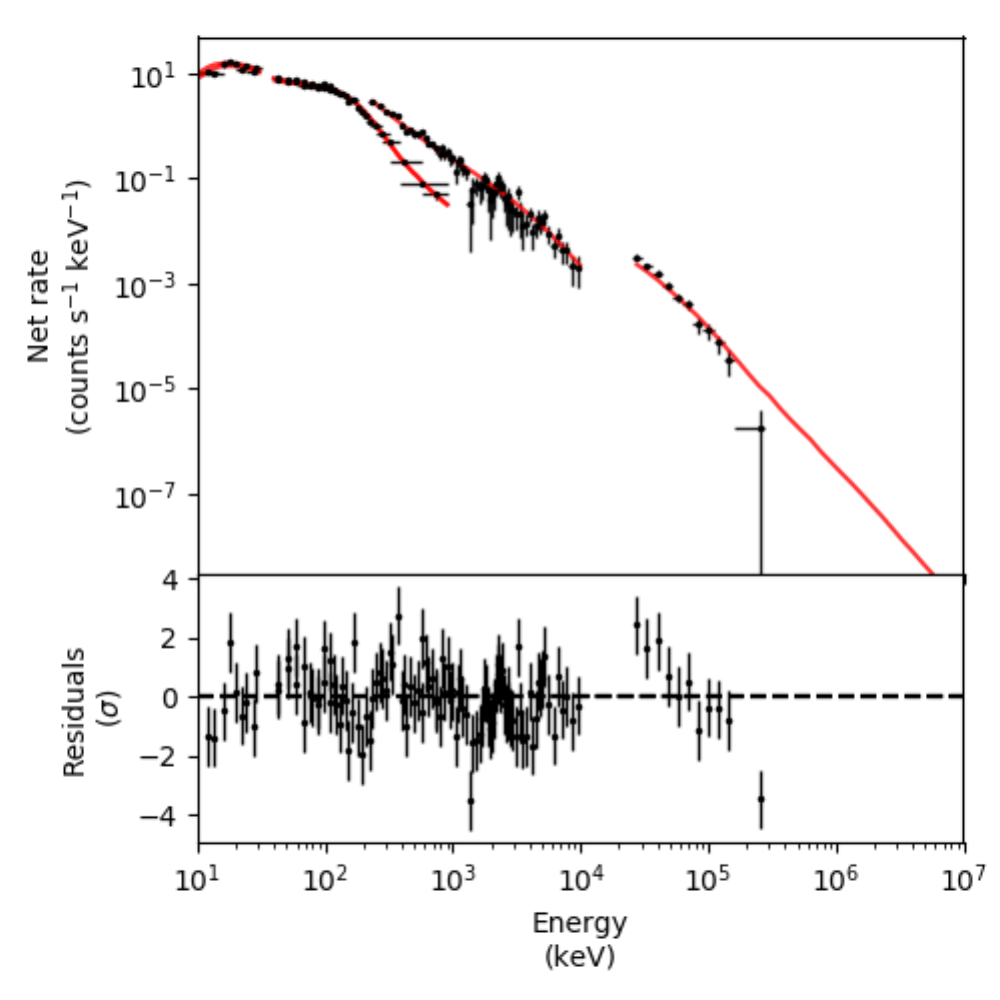}
\includegraphics[width=0.57\textwidth]{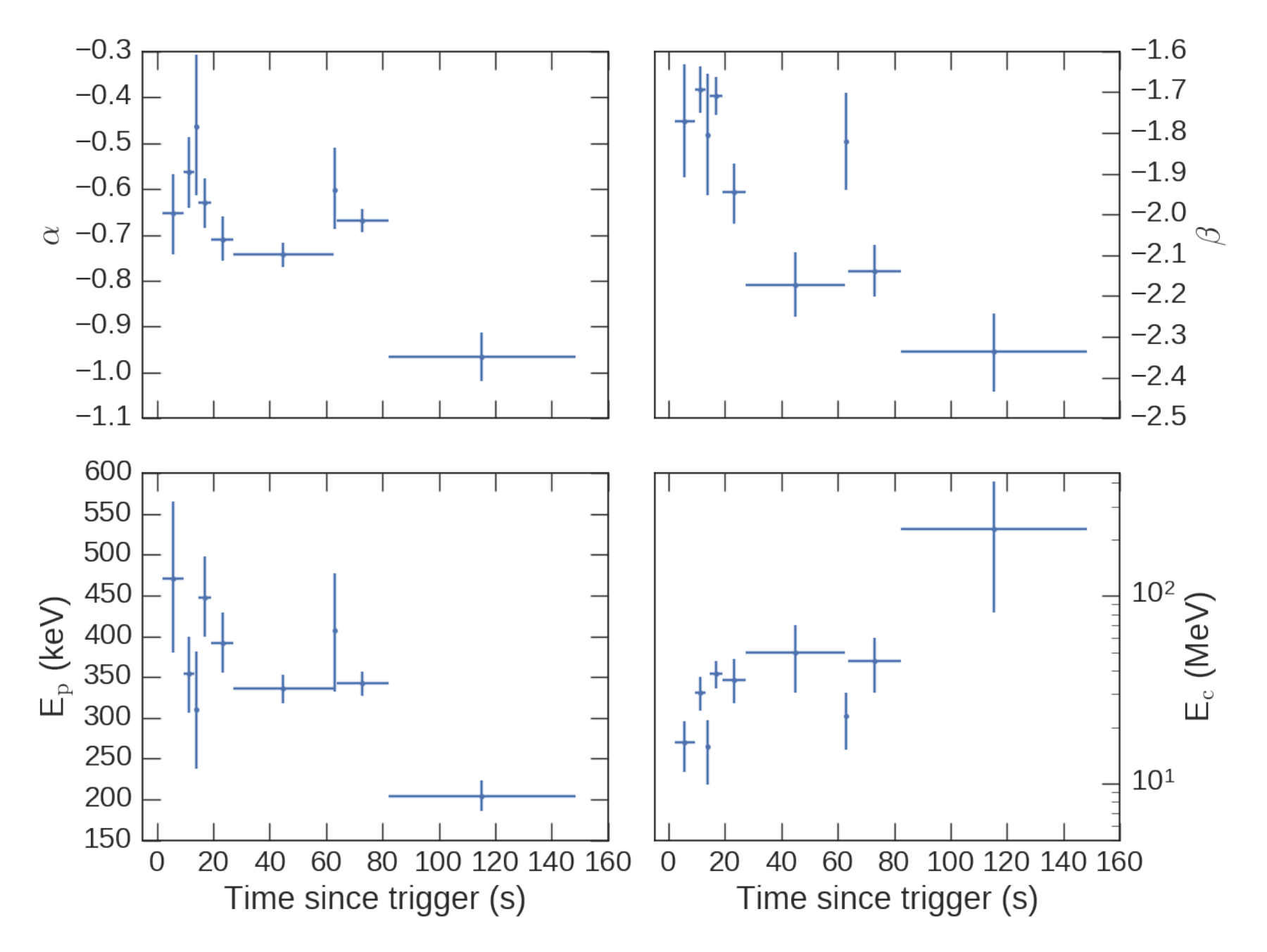}
\caption{Left: \textbf{Fit of the $f_{BB}$ model (red line) to interval 4. While the model provides a good description of the low-energy data, it is not a good description of the LLE data, as shown by the structured residuals.} Right: Temporal evolution of the parameters of the $f_{BHec}$ model for GRB~100724B. These parameters are defined in \eqt{\ref{eq:bhec}}.}
\label{fig:spEvolution}
\end{figure*}

We report the results in Table~\ref{tab:model-selection}. The TS of $f_{BHec}$ with respect to $f_{Band}$ and the corresponding significance of the improvement (\textbf{9}th column) is large ($> 4\sigma$) for all intervals except for the two where the GRB is faint. \textbf{The quality of the fit is good both overall and for LLE data in particular, as shown by the p-values $p$ and $p_{\rm LLE}$ (last two columns). On the other hand, the improvement obtained with $f_{BB}$ with respect to $f_{Band}$ is large ($> 4\sigma$) for 4 intervals (6th column). The values for the overall null-hypothesis probability $p$ for $f_{BB}$ seem to indicate a good fit (7th column), however while the model describes well the low-energy data it does not describe well LLE data, as shown by $p_{\rm LLE}$ (8th column, see also fig.~\ref{fig:spEvolution}). Hence, we can conclude that while $f_{BB}$ models well the low-energy data -- as already concluded in \citet{2011ApJ...727L..33G} -- it fails to describe well LLE data.}

Summarizing, $f_{BHec}$ is a more parsimonious model than $f_{BB}$ (it has one less free parameter) and also provides a better description of LLE data in all intervals. It is therefore our model of choice. This result appears to be at odds with what is reported in \citet{2011ApJ...727L..33G}. We note however that these authors did not use LLE data, \textbf{which is where the advantage of $f_{BHec}$ over $f_{BB}$ becomes evident,} and used different time intervals for the time-resolved spectral analysis. They also used a different localization for the GRB, provided by the \Fermi/GBM with a large localization error, which has an impact on the response matrices used for GBM data and therefore on the modeling of the spectrum \citep{2015ApJS..216...32C, 2018MNRAS.476.1427B}. This GRB has also been studied by \citet{2011A&A...535A.120D} using {\it AGILE} data. The spectrum they measured is much harder than what \Fermi measured, and with a much larger flux. If the characteristics measured by {\it AGILE} were true, we would have detected with the LAT a large number of photons above 100 MeV which we do not see. We discuss in Appendix~\ref{appendix:agile} a plausible motivation for this discrepancy. 

The procedure described here considers only statistical uncertainties. A study of the effects of systematic uncertainties on the significance of the cutoff that are not neutralized by the use of the ``effective area correction'' is reported in Appendix~\ref{sec:sys_effects}, and demonstrates that the improvement given by the cutoff is unlikely to be due to systematic uncertainties \textbf{in the response of the instruments}. 

The significance of the cutoff with respect to the simple extrapolation of the low-energy spectrum being established, we compare $f_{BHec}$ with the two models with power-law shape after the cutoff ($f_{Bbkpo}$ and $f_{Bgr}$) to assess whether the spectrum is curved (exponential cutoff) or not (power law break) after the break. We find that $f_{Bbkpo}$ and $f_{Bgr}$ never provide a better fit as measured by $p_{null}$ with respect to the exponential shape despite having more parameters, which favors a curved spectrum above the cutoff. 

In Figure~\ref{fig:spEvolution} we show the best fit parameters for $f_{BHec}$ for the intervals of the main emission episodes. The parameters $\alpha$ and $\beta$ decrease during the first peak, increase in the second peak, and then decrease again. $E_{\rm p}$ shows a similar evolution. This tracking behavior is common in GRBs \citep{1995ApJ...439..307F,2006ApJS..166..298K}. The cutoff energy $E_{\rm c}$ increases slightly with time.

\section{GRB 160509A}\label{sec:160509A}

\begin{figure*}
\centering
\vcenteredhbox{\includegraphics[width=0.65\textwidth]{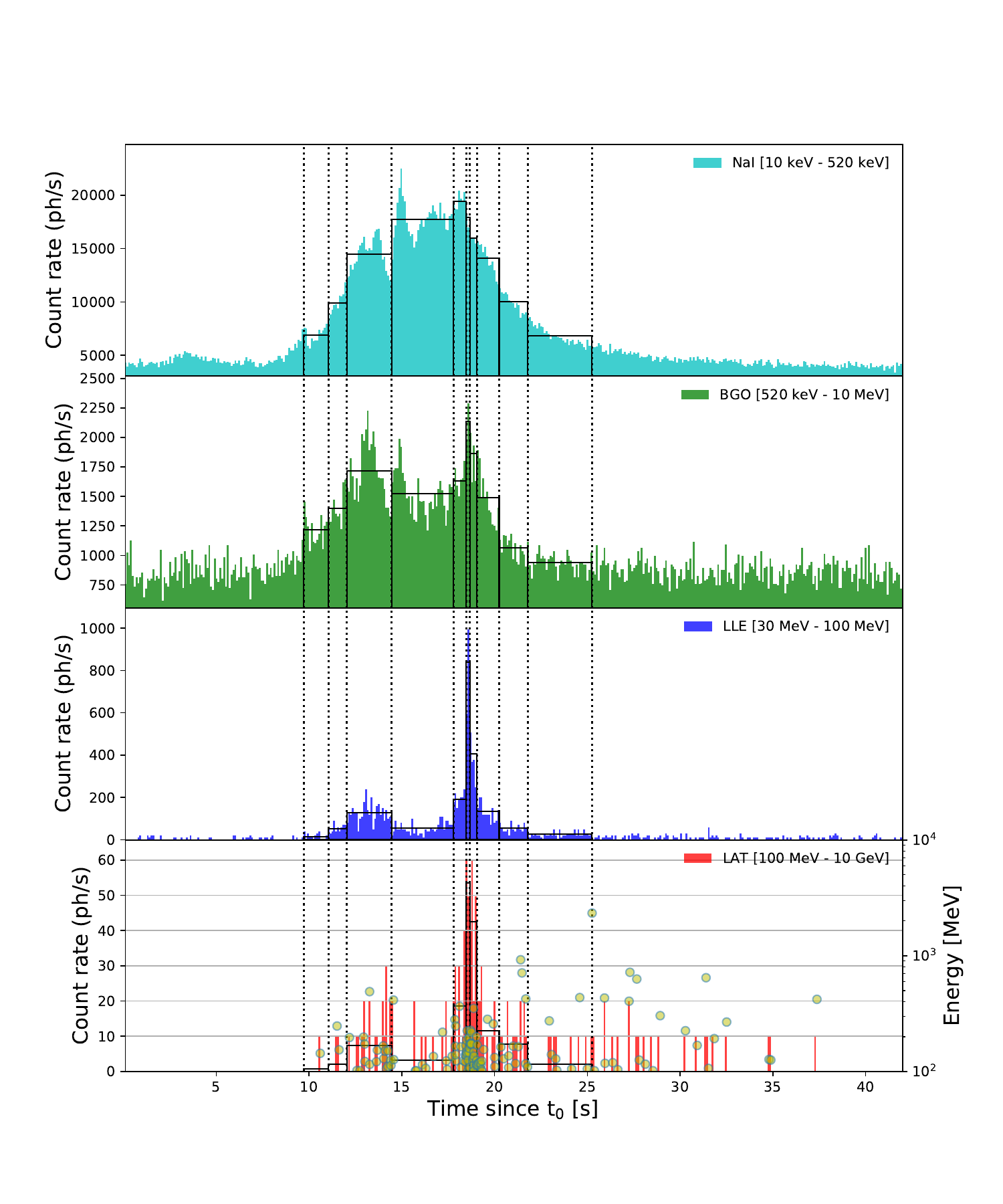}}
\vcenteredhbox{\includegraphics[width=0.33\textwidth]{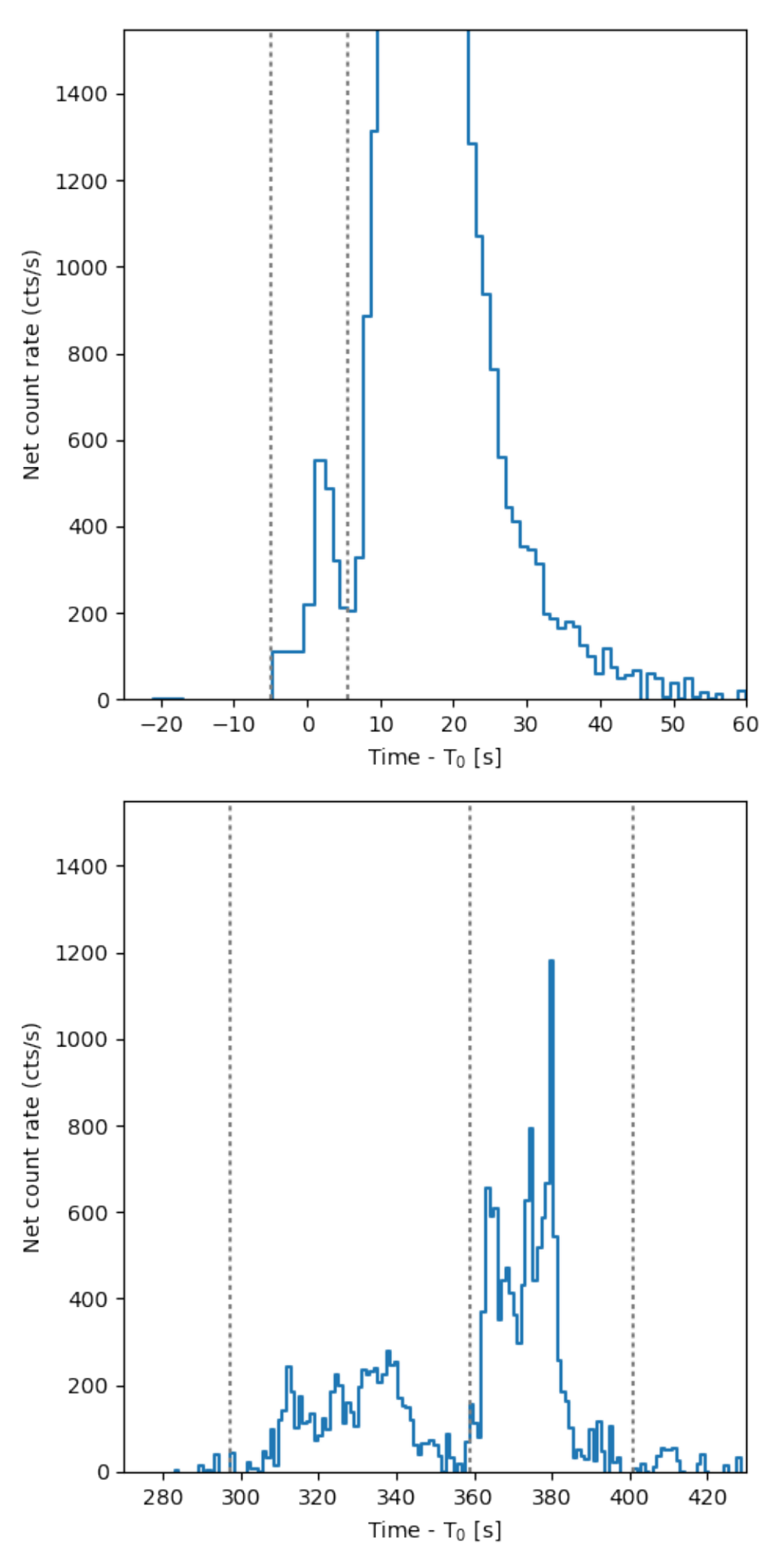}}
\caption{Left: Composite light curve of GRB~160509A showing NaI, BGO, LAT/LLE and LAT standard data. The green dots in the bottom panel represent single photons detected by the LAT and associated with the GRB, and their energy is provided by the right y-axis. The vertical lines correspond to the intervals obtained with the Bayesian Blocks algorithm and the bars indicate the rebinning of the data accordingly. Right: early-time (upper panel) and late- time (lower panel) light curve in the low-energy NaI detectors. The dashed lines indicate the intervals used to analyze the precursor and the soft emission.}
\label{fig:bb_160509}
\end{figure*}

\subsection{Observations}
GRB~160509A triggered \Fermi/GBM on 2016-05-09 at $t_{0} =$ 08:58:45.22 UTC. It was also localized on-board by \Fermi/LAT \citep{2016GCN..19403...1L}, one of only 5 cases over 8 years of mission. This allowed for a quick follow up and localization by {\it Swift} \citep{2016GCN..19408...1K}, which in turn allowed for a redshift measurement by Gemini/North of $z=1.17$ \citep{2016GCN..19419...1T} when the afterglow was still bright. We adopt the position of the afterglow measured by Gemini North (R.A. $= 311.7538\,^{\circ}$, Dec. $= 76.1081\,^{\circ}$). The prompt emission (Figure~\ref{fig:bb_160509}) consists of a soft ``precursor'' peak between $t_{0} - 5.0$ and $\sim t_{0} + 5.5$ s, followed by a much brighter main episode which lasts until $t_{0} + 40$ s. After a quiescent time, there is another very soft emission episode, visible only in the low-energy detectors, from $\sim t_{0} + 300 $\,s until $\sim t_{0} + 400$\,s. \textbf{This excess localizes in roughly the same direction as the main episode, although with large statistical uncertainty (GBM team, private communication), therefore it is likely to be associated with GRB~160509A.} Similarly to GRB~100724B, during the main emission episode the LAT detected many photons associated with the GRB in the 30$\;$--100$\;$MeV energy range, but surprisingly few above 100$\;$MeV (see last two panels in Figure~\ref{fig:bb_160509}).

\subsection{Spectral analysis of the prompt emission}

We use here the same technique and energy selections discussed in section~\ref{sec:spectral_analysis}.
For the first and second episode we use data from GBM detectors NaI 0, NaI 3 and BGO 0, which are the detectors in the most favorable position to observe the GRB. Since the pointing of the \Fermi satellite changed between the first two emission episodes and the third one, for the latter we used NaI 0, NaI 6, NaI 9 and BGO 1, which were the detectors closest to the direction of the GRB at that time. Contrary to the case of GRB~100724B, \textbf{the Earth Limb was far from the direction of  GRB~160509A during the prompt emission}. For all intervals we hence use LAT LLE data from 30 MeV up to 100 MeV, and LAT standard data above 100 MeV.

\begin{table*}
\centering
\begin{tabular}{|r|c|c|c|c|cc|c|c|}
\hline
 \# &   Time interval & $S$ of $f_{Band}$ & p & p$_{\rm LAT}$ & \multicolumn{2}{c|}{TS of $f_{BHec}$} & p & p$_{\rm LAT}$ \\
 \hline
1 &   9.712-11.045 &   970.5 &        0.23 &        0.12 &  7.1 & ($2.4~\sigma$) & 0.14 & 0.15 \\
2 &  11.045-12.042 &  894.07 &        0.36 &        0.05 &  7.6 & ($2.5~\sigma$) & 0.59 & 0.31 \\
3 &  12.042-14.449 & 1516.65 & $< 10^{-3}$ & $< 10^{-3}$ & 93.8 & ($9.6~\sigma$) & 0.23 & 0.33 \\
4 &  14.449-17.783 & 1620.88 & $< 10^{-3}$ & $< 10^{-3}$ & 42.1 & ($6.4~\sigma$) & 0.62 & 0.48 \\
5 &  17.783-18.480 &  847.16 &        0.08 &       0.003 & 21.2 & ($4.5~\sigma$) & 0.44 & 0.23 \\
6 &  18.480-18.667 &  233.18 & $< 10^{-3}$ & $< 10^{-3}$ & 67.5 & ($8.1~\sigma$) & 0.59 & 0.77 \\
7 &  18.667-19.044 &  527.27 & $< 10^{-3}$ & $< 10^{-3}$ & 43.9 & ($6.5~\sigma$) & 0.09 & 0.24 \\
8 &  19.044-20.249 & 1105.23 & $< 10^{-3}$ & $< 10^{-3}$ & 63.1 & ($7.8~\sigma$) & 0.87 & 0.89 \\
9 &  20.249-21.787 & 1124.36 &        0.56 &        0.09 &  5.1 & ($2.0~\sigma$) & 0.07 & 0.35 \\
10 & 21.787-25.254 & 1460.61 &        0.15 &        0.23 &  6.3 & ($2.2~\sigma$) & 0.19 & 0.22 \\
\hline
\end{tabular}
\caption{Value of the -log(likelihood) $S$ for the Band model $f_{Band}$, and the $TS$ obtained with a Band with exponential cutoff ($f_{BHec}$) as alternative hypothesis, for GRB~160509B. In parenthesis we report the significance of the improvement. We also report the null-hypothesis probability p$_{\rm null}$ for each model, for LAT and LLE data.}
\label{tab:model-selection2}
\end{table*}

The spectrum accumulated over the entire duration of the GRB, from $t_{0}$ to $t_{0} + 400$\,s, can be well described with the $f_{BHec}$ model. \textbf{Thanks to the high signal-to-noise ratio we have many counts in each bin in the spectrum; thus we can assume Gaussian statistics and apply a normal $\chi^2$ test. We obtain $\chi^2 = 392.4$ with 381 d.o.f, corresponding to a p-value of $p=0.33$.}. The cutoff is required with high significance ($> 10\sigma$) with respect to the Band model alone. We measure a fluence of $(3.2 \pm 0.5) \times 10^{-4}$ erg cm$^{-2}$ in the 1 keV  -- 10 GeV energy range, corresponding to an isotropic emitted energy of $E_{\rm iso} = (2.6 \pm 0.4) \times 10^{54}$\,erg. The contribution to this quantity by the precursor and the late emission episode is negligible.

The spectrum of the ``precursor'' peak is well described by a power law with exponential cutoff (eq.~\ref{eq:PHec}), with $\alpha=-1.03^{-0.06}_{+0.07}$, $E_{c}=410^{-80}_{+100}\;$keV, and $K=1.42^{-0.3}_{+0.4}\;$ph.~cm$^{-2}$~s$^{-1}$. This is very similar to the ``precursor'' peak in GRB 100724B.

The third, late episode is faint and soft as well. We divide it in two intervals, $297.45-358.9\;$s and $358.9-400.88\;$s from $t_{0}$. Their spectra are both well described by a Band model. The best fit parameters are respectively $\alpha = -1.21^{-0.07}_{+0.06}$, $\beta = -1.98^{-0.24}_{+0.10}$, $E_{p} = 270^{-50}_{+90}\;$keV, and $\alpha = -1.27^{-0.04}_{+0.05}$, $\beta = -2.20^{-0.13}_{+0.09}$, $E_{p} = 138^{-13}_{+15}\;$keV. Adding an exponential cutoff, or any other component like a thermal component, does not significantly improve the fit. This can of course either be intrinsic, or just due to the lack of sufficient statistics, especially at high energies.

We focus then on the main episode, much brighter than the other two. The Band model $f_{Band}$ overestimates the amount of LLE signal by a large amount and the improvement obtained by adding an exponential cutoff to the Band model is very large. We obtain $TS = 278$ for $f_{BHec}$, corresponding to a significance of $16.6~\sigma$. The addition of a black body, instead, returns a lower $TS = 120$. Moreover, the $f_{BB}$ model does not describe well LAT data, yielding very large residuals. \textbf{This is reflected by the p-values returned by the $\chi^2$ test -- which again we can apply in virtue of the very high statistics -- that are respectively $0.25$ for the $f_{BHec}$ model and $\sim 10^{-6}$ for the $f_{BB}$ model.}. We therefore do not consider $f_{BB}$ as a viable model for \textbf{the time integrated analysis for} this GRB and our current energy and interval selection.

As for GRB~100724B, we run the Bayesian Blocks algorithm on LLE data with the same setup to determine the time intervals for the time-resolved spectral analysis of the main episode. We show these intervals as the black lines in Figure~\ref{fig:bb_160509}.

\begin{figure*}[t!]
 \centering 
 \leavevmode 
 \includegraphics[width=0.55\textwidth]{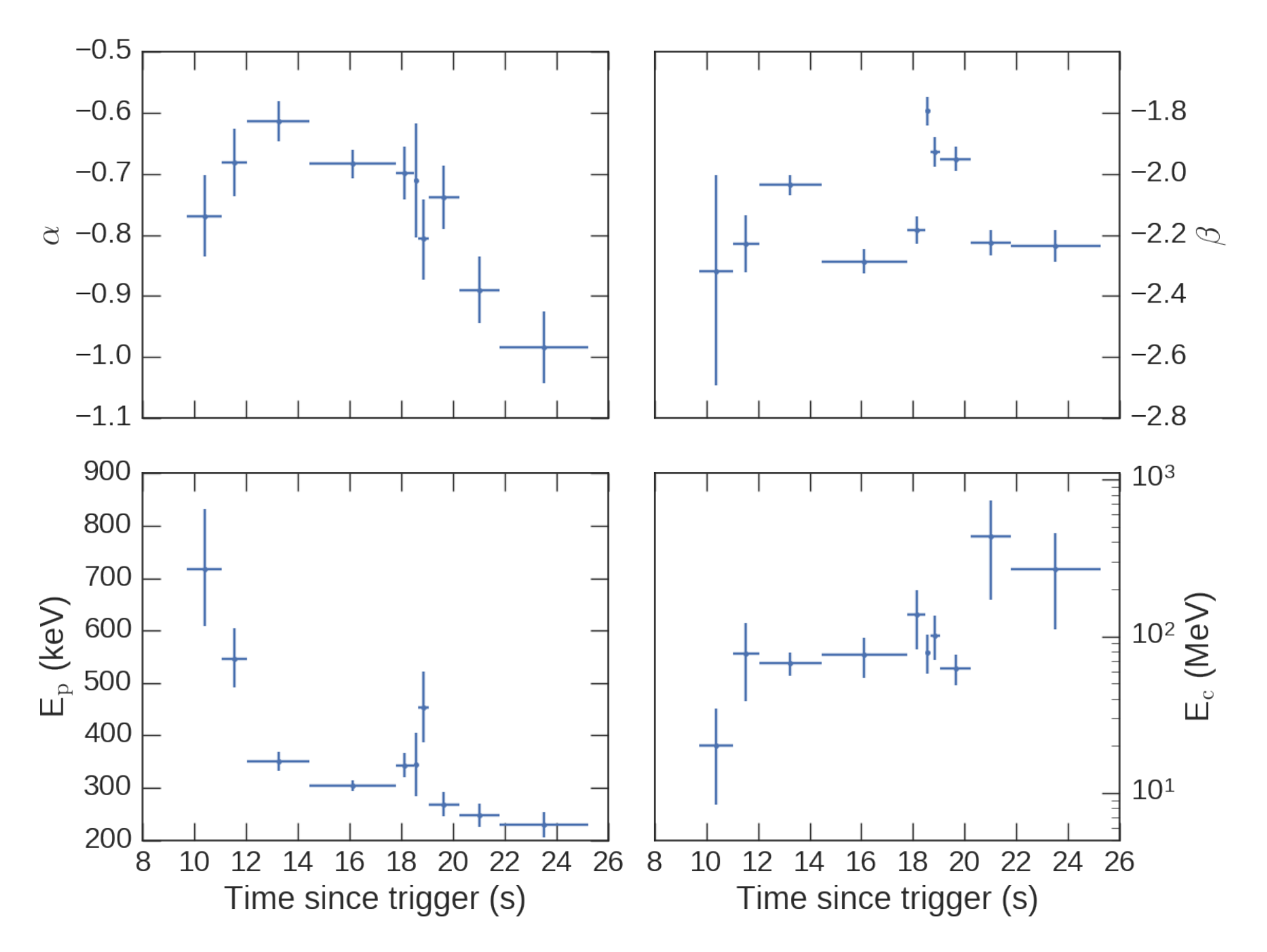}%
 \hfil
 \includegraphics[width=0.4\textwidth]{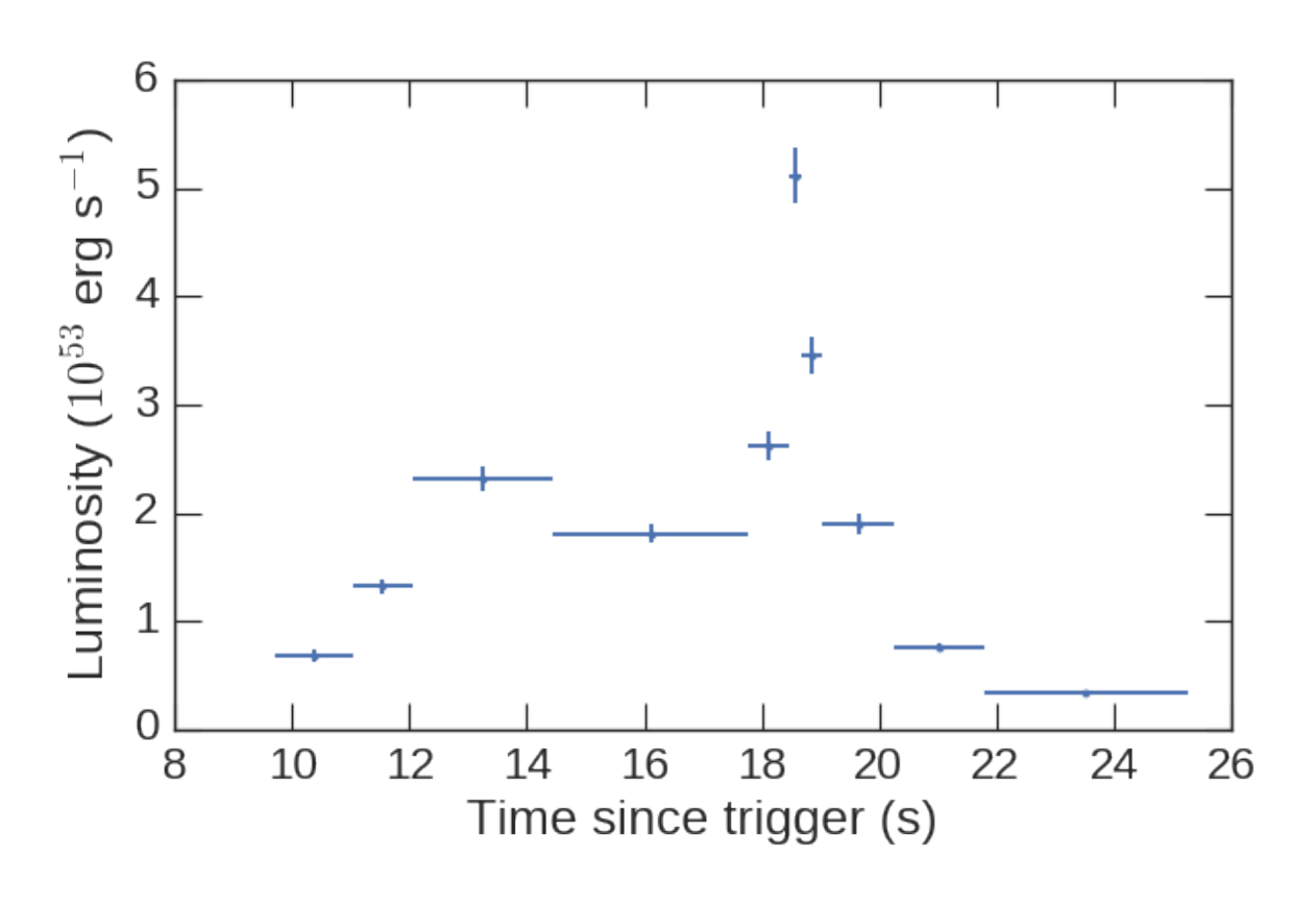}
\caption{Temporal evolution of the parameters of the $f_{BHec}$ model and isotropic-equivalent luminosity as a function of time for GRB 160509A. The values shown here for the luminosity are the averages for each time interval.}
\label{fig:spEvolution2}
\end{figure*}

In the 6 intervals where the GRB is bright we find again that the addition of an exponential cutoff to the Band spectrum improves the fit significantly ($>4\sigma$), as shown by the 5th column in table~\ref{tab:model-selection2}. \textbf{There we also report the p-value for the goodness of fit test for the entire dataset ($p$) and for LAT and LLE data ($p_{\rm LAT}$), computed as described in section~\ref{sec:spectral_analysis}. It shows that the $f_{BHec}$ model provides a good description both overall and for LLE and LAT data in particular. In the case of this GRB, contrary to GRB~100724B, the addition of a black body to the Band spectrum (i.e., the $f_{BB}$ model) does not yield a significant improvement in most intervals. Moreover, residuals in LLE and LAT data are very significant. Therefore, we avoid computing the p-values for the goodness-of-fit test for this model, which is very computationally intensive, and do not consider it as a good model for this GRB given our selections.}

As for GRB 100724B we tested whether the two models with a power-law shape after the cutoff ($f_{Bbkpo}$ and $f_{Bgr}$) provide a better fit than $f_{BHec}$. The $p_{null}$ obtained with these models is worse for all intervals despite their increased complexity with respect to $f_{BHec}$. Hence, as for GRB 100724B, the Band with exponential cutoff model $f_{BHec}$ provides a better fit with less parameters and is therefore statistically preferred, and the shape of the spectrum after the cutoff appears to be curved. 

The best fit parameters as a function of time are shown in the left panel in Figure~\ref{fig:spEvolution2}. The time evolution of $\alpha$ and $E_{\rm p}$ is similar to the case of GRB~100724B: there is a decreasing trend with some variability tracking the light curve in correspondence of the second peak, while the cutoff energy appears to increase slightly with time. Also $\beta$ is tracking the flux, becoming harder when the flux increases, but there is no decreasing trend. The luminosity as a function of time is shown in the right panel of Figure~\ref{fig:spEvolution2}, computed in the energy range 1 keV -- 10 GeV: the values of a few $10^{53}$ erg s$^{-1}$ are quite typical for long-duration GRBs \citep{2004ApJ...609..935Y}.

\section{Interpretation and physical modeling}
\label{sec:physical_modeling}

In the previous sections we have established phenomenologically that an high-energy exponential cutoff in GRB~100724B and GRB~160509A must be added to the extrapolation of the low-energy Band spectrum in order to successfully model LAT data. In this section we provide some possible interpretations.

\textbf{Among many possibilities \citep[for example][]{2004ApJ...614..827R, 2005ApJ...635..476P, 2010MNRAS.407.1033B, 2011ApJ...738...77V, 2013ApJ...765..103L}}, we consider two scenarios: i) the cutoff is due to pair-production opacity that attenuates a non-thermal spectrum (produced for example by synchrotron emission during internal shocks); or ii) a photospheric model where the cutoff arises due to the development of an electron-positron pair cascade in a highly magnetized, dissipative and baryon-poor outflow. In the first scenario we adopt \textbf{a hybrid approach by considering} the phenomenological Band model, traditionally used in modeling the non-thermal spectrum of GRBs, and we multiply it by a $\gamma$-$\gamma$ attenuation factor \textbf{computed from first principles} in \citet{2008ApJ...677...92G}. It features a self-consistent semi-analytic calculation of the impulsive emission from a thin spherical ultra-relativistic shell (model $f_{BG}$, see eq.~\ref{eq:band_granot}). The calculation accounts for the fact that, in impulsive relativistic sources, the timescale for significant variations in the properties of the radiation field within the source is comparable to the total duration of the emission episode, and therefore, the dependence of the opacity to pair production on space and time cannot be ignored. In the second scenario we instead adopt the photospheric model of \citet{GT14}, which we will call $f_{GT}$ in the following, as this model produces spectra which are strikingly similar to the phenomenological $f_{BHec}$ model that is a good description of the data. 
We describe both models in some detail next.

\subsection{Pair Opacity Break in Impulsive Relativistic Outflows - the $f_{BG}$ model}

The model of \citet{2008ApJ...677...92G} features an expanding ultra-relativistic spherical thin shell. The emission, which 
may arise from internal shock heated electrons, is assumed to be isotropic in the shell's comoving frame. Its comoving 
luminosity scales as a power law with dimensionless energy $\varepsilon'=E'_{\rm ph}/m_ec^2$, where $m_e$ is the electron 
mass and $c$ is the speed of light, and with radius $R$, $L'_{\varepsilon'}\propto(\varepsilon')^{1+\beta}R^b$, 
where $\beta$ is the (high-energy) photon index. The shell's Lorentz factor (LF) is also assumed to vary as a power law with radius, 
$\Gamma\propto R^{-m/2}$. The emission episode lasts between radii $R_0$ and $R_0+\Delta R$, where the fractional radial 
width $\Delta R/R_0$ determines how impulsive the emission is.


The optical depth $\tau_{\gamma\gamma}$ to pair production ($\gamma\gamma\to e^+e^-$) is calculated 
along the trajectory of each test photon that reaches the observer. Its contribution to the observed flux is attenuated by a factor of $\exp(-\tau_{\gamma\gamma})$, leading to a quasi-exponential (after adding contributions 
from different emission radii and angles) cutoff in the instantaneous spectrum. Depending on the value of $\Delta R/R_0$, the time 
integrated spectrum either features a smoothly broken power law cutoff ($\Delta R/R_0\sim 1$) or a quasi-exponential 
cutoff that asymptotes into a power law ($\Delta R/R_0\gtrsim 1$).

In order to enable a (semi-) analytic calculation, the effects of the pairs that are produced in this process are neglected. This is a reasonable approximation as long as their Thomson opacity is $\tau_{{\rm T},\pm}\ll 1$. Below we will examine the validity of this approximation.

In practice, we compute the attenuation factor\footnote{We note that here that $-\beta$ plays the role of the parameter called $\alpha$ in \citet{2008ApJ...677...92G}.} $\Lambda(\beta, \Delta R / R_0, \psi, m, b)$ due to the $\gamma$-$\gamma$ opacity through a numerical code which implements the computation described in \citet{2008ApJ...677...92G}. We then define $f_{BG}$ as:
\be
f_{BG}(E) = f_{Band}(\alpha, \beta, E_{p}, K)~ \Lambda(\beta, \frac{\Delta R}{R_0}, \psi, m, b),
\label{eq:band_granot}
\ee
where $f_{Band}$ is the Band model \eqp{\ref{eq:band}}, and $\psi = E / \varepsilon_z$ where $\varepsilon_z=(1+z)\varepsilon$ is the dimensionless photon energy in the source's cosmological frame at redshift $z$, while $\varepsilon$ is the value measured at Earth. In order to reduce the number of free parameters to a manageable number, we fix $m=0$, which corresponds to a shell in coasting phase as expected from an internal shock scenario, and $b=0$, which corresponds to assuming a comoving spectral emissivity independent of radius. Therefore, we have 6 free parameters: $\alpha, \beta, E_{p}, K, \Delta R/R_0, \psi$.

\begin{figure}
\centering
\includegraphics[width=0.48\textwidth]{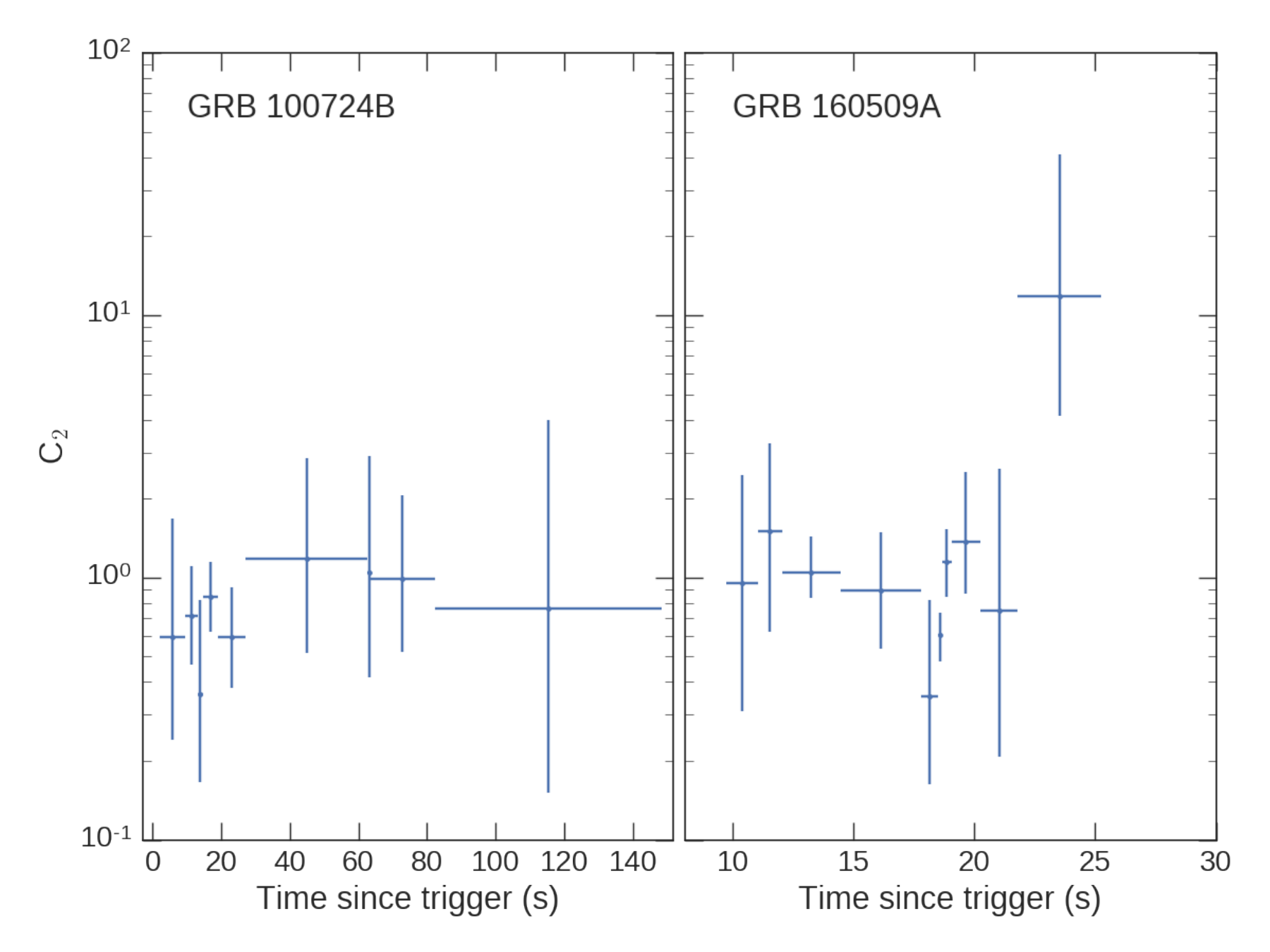}
\caption{The parameter $C_2$ obtained from fitting the $f_{BG}$ model to both 
GRBs, shown here for different time bins. $C_2$ appears in \eqt{126} of 
\citet{\Granotpaper} and is used to determine $\Gamma$ from observed quantities.}
\label{fig:c2}
\end{figure}

The LF $\Gamma_0 = 100\Gamma_2$ can be estimated by using this relation:
\footnote{\label{foot:correction}The numerical coefficient in the expression here for $\Gamma_2 = \Gamma_0/100$ is larger by a factor of $\pi^{1/(1-\beta)}$ compared to Eq.~(126) of \citet{\Granotpaper}, correcting an error in the latter equation.}
\be
\Gamma_2 =
\left[3.969\,\frac{10^{4(2 + \beta)}L_{0,52}}{(
1+z)^{\beta}
f^{\beta+1}}\left(\frac{- \beta}{2}\right)^{-\frac{5}{3}}\frac{33.4\,{\rm ms}}{t_{\rm v}}\right]^{1/(2-2 \beta)}.
\label{eq:lorentz_factor_bg}
\ee
Here $L_{0,52} = 4\pi d_{L}^2(1+z)^{-\beta-2} F_{0} / (10^{52}\;{\rm erg~s^{-1}})$,    where $d_{L}$ is the luminosity distance of the burst, $F_{0}$ is the (unabsorbed) energy flux ($\nu F_\nu$) foreseen by the high-energy power law of the Band model at 511 keV. The 
parameter $f = \psi / m_{e}c^2$ relates to the parameter 
$C_2 = 10^{-(6+4\beta)}\left(\frac{f}{(E_c/5.11~{\rm GeV})}\right)^{\beta+1}$ that 
appears in \eqt{126} of \citet{\Granotpaper}. Its exact value is not known a priori and 
can only be determined numerically. 
To that end, we obtain the value of $C_2$ from model fits to both GRBs considered 
in this work and, as expected in \citet{\Granotpaper}, find that its value is of 
order unity (see Figure \ref{fig:c2}).
We extract from the data $t_{\rm mv}$, which is the minimum variability time scale detected in the light curve, defined as the rise time of the shortest significant structures. Therefore, $t_{\rm mv} \approx (1+z)\Delta R/2c\Gamma_0^2 = t_0\Delta R/R_0$ where $t_0=(1+z)R_0/2c\Gamma_0^2$ is the arrival time of the first photons to the observer. Since we chose to define the variability timescale as $t_v \equiv 2t_0$ when deriving \eqt{\ref{eq:lorentz_factor_bg}}, we obtain that it can be expressed in terms of $t_{\rm mv}$ as follows: $t_v = 2~t_{\rm mv} / (\Delta R/R_{0})$.

We determine $t_{\rm mv}$ using a wavelet analysis. Similar techniques have been already used by many authors to study the variability of GRBs \citep{2000ApJ...537..264W, 2013MNRAS.432..857M, 2015ApJ...811...93G}. In contrast to these authors, we adopt the Continuous Wavelet Transform (CWT) in place of the Discrete Wavelet Transform (DWT), as the CWT allows for a much better resolution in the spectrum \citep{torrence1998practical}. 
\begin{figure*}
\centering
\includegraphics[width=0.47\textwidth]{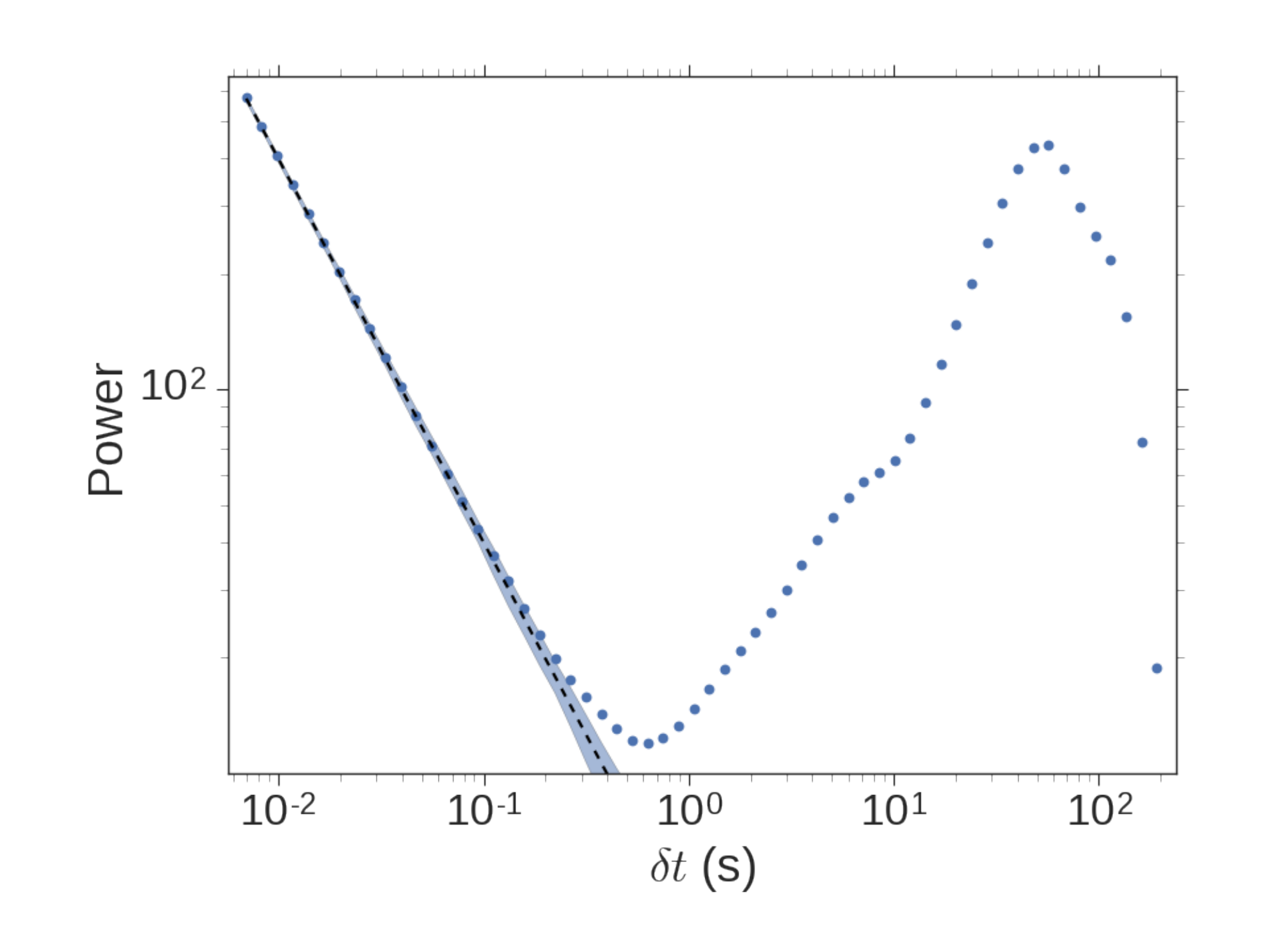}
\includegraphics[width=0.47\textwidth]{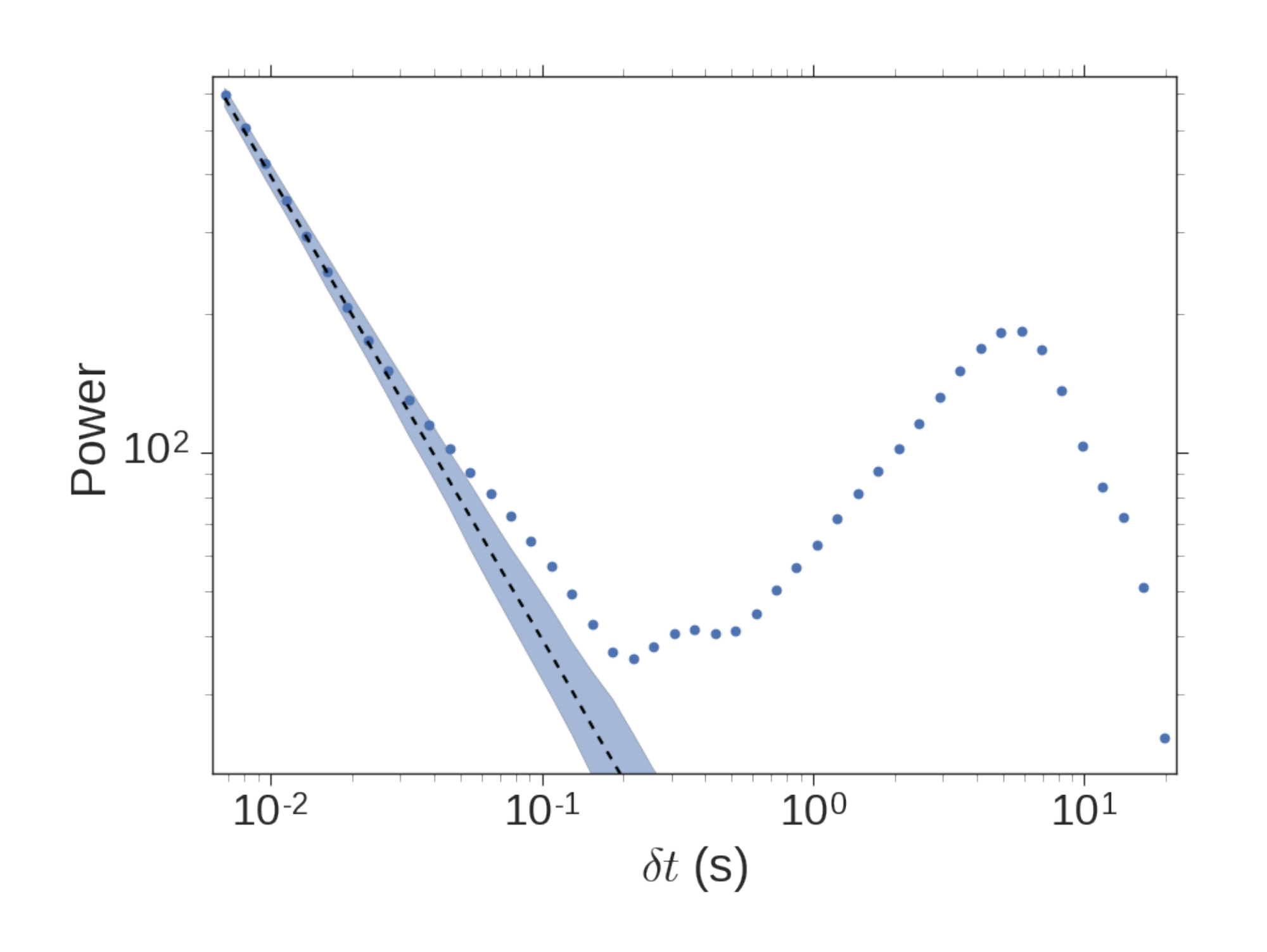}
\caption{Wavelet power spectrum for GRB~100724B (left) and GRB~160509A (right). The points represent the measured spectrum, the black dashed line represent the median background spectrum obtained from Monte Carlo simulations, while the blue shaded region represents the uncertainty on the background spectrum (see text for details).}
\label{fig:mvts}
\end{figure*}
We start by obtaining a light curve with a bin size of $10^{-4}$ s of the entire time interval with a bright LLE emission (respectively $t_0 + 8.195$ -- $t_0 + 122.882\;$s for GRB~100724B and $t_0 + 10$ -- $t_0 + 25\;$s for GRB~160509A). Then we compute the wavelet power spectrum $W$ as a function of the time scale $\delta t$, as described in \citet{torrence1998practical}, with the correction suggested by \citet{liu2007rectification}. The result is shown in Figure~\ref{fig:mvts} (dots). In order to measure the variance of the power spectrum due to the Poisson fluctuations of the background, we generate 10 thousand simulated background light curves with the same duration and binning as the original light curve, and a background rate estimated in an off-pulse interval, measuring the wavelet spectrum for each realization. We then plot the 99\% containment interval for each time scale $\delta t$ (blue shaded region) centered on the median (dotted line). In the wavelet power spectrum, Poisson noise follows a power law $W \propto \delta t^{-1}$. This is evident for very short time scales, where the data are dominated by noise. The first time scale that deviates from the noise power law outside the 99\% c.l. region represents our estimate of $t_{\rm mv}$. We obtain $t_{\rm mv} \sim 0.3$ s for GRB~100724B and $t_{\rm mv} \sim 0.05$ s for GRB~160509A. We also run the Bayesian Blocks algorithm on the GBM+LLE dataset and confirm that we find the shortest significant structures with a duration of respectively $\sim 0.6$ s and $\sim 0.1$ s, corresponding to $\sim 2~t_{\rm mv}$ as expected.

\subsection{Delayed Pair-Breakdown in a High-$\sigma$ Relativistic Jet - the $f_{GT}$ model}
\label{sec:GT_model}

The model presented by \cite{GT14} considers the breakout of a strongly magnetized, baryon-poor 
jet from the confining envelope of a Wolf-Rayet (WR) star at a breakout radius 
\be
R_{\rm br} \sim 2\Gamma_{\rm br}^2c\mathcal{R}t_{\rm eng} = 
5.4\times10^{12}\Gamma_{\rm br,3}^2\mathcal{R}_1t_{\rm eng,0}~{\rm cm}~.
\ee
The outflow bulk-LF at breakout $\Gamma_{\rm br} = 3\Gamma_{\rm br,3}$ is modest and ranges 
from $\sim3 - 10$, and $t_{\rm eng} = 1~t_{\rm eng,0}~{\rm s}$ represents the typical time 
scale over which the central engine (a black hole in this case) remains active. 
$\mathcal{R} = 10\mathcal{R}_1$ is a factor that governs the geometry of the ouflow at 
deconfinement ($\mathcal{R} < 1$ for `jet' geometry and $\mathcal{R} > 1$ for `pancake' geometry). 
It was shown by \citet{2014ApJ...791...46T} that the advected quasi-thermal radiation field at breakout has a 
relatively flat spectrum ($E'f'_{\rm th}(E')\propto E'^{1+\alpha}$, with $\alpha\sim -1$) below 
the Wien peak at energy $E'_{\rm pk,br}\simeq 0.1m_ec^2$ in the fluid-frame. The enthalpy 
density of the jet at breakout is dominated by the magnetic field, with compactness 
$\ell_{B,\rm br}'\gtrsim \ell_{\rm th,br}'$, where the advected quasi-thermal radiation 
field has compactness
\ba
\ell_{\rm th,br}' &&\equiv \sigma_T\frac{U'_{\rm th}}{m_ec^2}
\frac{R_{\rm br}}{\Gamma_{\rm br}} 
= \frac{3\sigma_TE_{\gamma,{\rm iso}}}{32\pi\mathcal{R}\Gamma_{\rm br}^5m_ec^4t_{\rm eng}^2} \\
&&= 10^{10}E_{\gamma,54}\Gamma_{\rm br,3}^{-5}\mathcal{R}_1^{-1}t_{\rm eng,0}^{-2}~,
\nonumber 
\ea
and where $\sigma_T$ is the Thomson cross-section and 
$E_{\gamma,{\rm iso}} = 10^{54}E_{\gamma,54}$ erg is the total isotropic equivalent 
energy of the radiation field.

Post jet-breakout, the outflow is accelerated to high bulk-LF $\Gamma \sim 10^2 - 10^3$, 
and the radiation field compactness $\ell'_{\rm th}(R)\propto R^{-4}$ and optical depth 
of the flow $\tau_{T,\pm}(R)\propto R^{-3}$ drop with radius. The thin baryonic layer, that was lifted 
from the WR envelope during breakout, suffers a corrugation instability (akin to a Rayleigh-Taylor instability) 
as it feels an effective gravity in its rest frame $g'_{\rm eff} = -c^2d\Gamma/dR$ due to the 
acceleration of the outflow. This breaks the baryonic layer into multiple plumes, which lose radiation 
pressure support at a critical radiative compactness, $\ell'_{\gamma,{\rm crit}} \sim m_p/Y_em_e \simeq 4\times10^3$, 
where $m_p$ is the proton mass and $Y_e\approx 0.5Y_{e,1/2}$ is the electron fraction in a long-GRB, 
and begin to lag behind as the magneto-fluid continues to accelerate. This differential motion between 
the two components leads to strong inhomogeneities in the magnetofluid and dissipation of the 
magnetic energy in the form of a turbulent cascade. The dissipation zone is radially localized at
\ba
R_{\rm diss} &&= R_{\rm br}\fracb{\ell'_{\rm th,br}}{\ell'_{\rm th}}^{1/4}
\gtrsim \left(\frac{Y_em_e}{m_p}\ell'_{\rm th,br}\right)^{1/4}R_{\rm br}\quad\quad \\ 
&&= 2.2\times10^{14}Y_{e,1/2}^{1/4}E_{\gamma,54}^{1/4}
\Gamma_{\rm br,3}^{3/4}\mathcal{R}_1^{3/4}t_{\rm eng,0}^{1/2}~{\rm cm} \nonumber
\ea
and the corresponding bulk-LF of the outflow is
\ba
\Gamma_{\rm diss} &&= \Gamma_{\rm br}\fracb{\ell'_{\rm th,br}}{\ell'_{\rm th}}^{1/4} \\
&&\gtrsim 125Y_{e,1/2}^{1/4}E_{\gamma,54}^{1/4}\Gamma_{\rm br,3}^{-1/4}
\mathcal{R}_1^{-1/4}t_{\rm eng,0}^{-1/2} \nonumber
\ea
The Thomson depth of the pairs at the dissipation radius is 
$\tau_{T,\pm,{\rm diss}} \lesssim 10^{-4}$ and the dissipated magnetic 
energy with compactness $\ell'_{\rm heat}$ goes into heating the pairs. The initially 
relativistically hot pairs inverse-Compton (IC) scatter the peak thermal photons to high 
energies. As the average energy of the pairs drops, due to pair production, the IC scattered 
peak gradually moves to lower energies, and finally merges with the thermal peak.

The total radiative compactness of the flow after dissipation can be written as
\be
\ell_{\rm tot}' = \ell'_{\rm th} + \ell'_{\rm heat} = (1+\xi_{\rm th})\ell'_{\rm heat} \lesssim \ell'_{\gamma,{\rm crit}}
\ee
where $\xi_{\rm th} = \ell'_{\rm th}/\ell'_{\rm heat}$ sets the heating compactness 
$\ell'_{\rm heat}$ relative to the thermal compactness $\ell'_{\rm th}$. 
The quasi-thermal soft seed photon spectrum is described as
\be
f_{th}'(E')\propto\left\{\begin{array}{ll}
{E'}_{\rm pk}^{2-\alpha}\exp({-E'_{\rm pk}/k_BT'_{\rm th}}){E'}^\alpha & E' < E'_{\rm pk} \\
{E'}^2\exp(-E'/k_BT'_{\rm th}) & E' > E'_{\rm pk}\end{array}\right.
\label{eq:mg14_qt}
\ee
where $\alpha$ sets the spectral power-law index below 
$E'_{\rm pk} = 3k_BT'_{\rm th}$, $T'_{\rm th}$ is the temperature of the radiation field 
and $k_B$ is the Boltzmann constant. The free parameters of this model are: 
$\alpha$, $T'_{\rm th}$, $\xi_{\rm th}$, and 
$\ell'_{\rm tot}$. The comoving radiation spectrum is then formed using a one-zone 
time-dependent kinetic code that involves integro-differential equations for both 
the radiation and particle distributions in the frame of the outflow 
(see \citealp{GT14} for further details of the numerical scheme).

To reduce the number of independent model parameters, so as to make the 
fitting procedure computationally tractable, we set the low-energy power-law index $\alpha$ to that obtained from fitting the $f_{BHec}$ profile. In addition, an estimate of the average comoving radiation field compactness can be obtained from the burst luminosity, such that
\be
R = \frac{3\sigma_T}{16\pi m_ec^3}\frac{L_{\rm iso}}{\Gamma^3\ell'_{\rm tot}}~.
\ee
Demanding that $R>R_{\rm diss}$, the above equation can be iterated to determine the correct $\ell'_{\rm tot}$ that satisfies the radius constraint. For example, taking the value for the luminosity $L_{\gamma,\rm iso} = 2.32\times10^{53}~{\rm erg~s}^{-1}$ in the third time-interval of GRB 160509A from Figure \ref{fig:spEvolution2}, we find an emission radius $R = 6.7\times10^{14}~{\rm cm}$ for $\Gamma = 200$ and $\ell'_{\rm tot} = 70$. Ideally, $\ell'_{\rm tot}$ should remain as an independent 
parameter since it depends strongly on radius. Redshift information is not 
available for GRB 100724B, which makes it less trivial to ascertain the correct 
dissipation radius and $\ell'_{\rm tot}$. Therefore, for simplicity, we use the same 
radiative compactness $\ell'_{\rm tot}=70$ for this burst as well, with the underlying 
assumption being that GRB 100724B had a similar intrinsic brightness as GRB 160509A. This appears to be a reasonable assumption, given the similarity between the two bursts.
\subsection{Model Fitting and Results} 
\label{sec:model_fitting_and-results}
Next, we fit the two physical models described in the previous subsections to the data.
In order to compute the attenuation factor for $f_{BG}$ we have implemented the semi-analytical computation described in \citet{2008ApJ...677...92G} in a code that, in the spirit of reproducible research \citep{donoho2009reproducible}, we make publicly available\footnote{https://github.com/giacomov/pyggop}. 

To fit the $f_{GT}$ model to the data we use templates that are produced by a numerical code which is very computing-intensive, requiring a medium-sized computer farm. We therefore release, in place of the code, the templates that can be used to reproduce our results. These templates are interpolated during the fit procedure to give the final results. As explained in section \ref{sec:GT_model} the model has 3 parameters, including the low-energy photon index $\alpha$. The computer code returns differential photon flux as a function of dimensionless rest frame energy. Therefore, in order to fit the data in the observer frame, we need to multiply the dimensionless rest frame energy by $m_{e}c^2$ and by a scale factor $\eta$ so that $\Gamma = \eta (1+z) / (m_{e}c^2)$, where $\Gamma$ is the bulk Lorentz factor. We also of course need a normalization for the model, for a total of 5 free parameters. 
In order to reduce the number of templates we need to generate, we fix $\alpha$ to the index measured with the $f_{BHec}$ model. Indeed, the fit would converge there anyway since $\alpha$ is the only parameter affecting the spectrum at low energies. This of course does not reduce the number of free parameters of the model, since $\alpha$ is still measured on the data, but it allows us to reduce the complexity of the problem.

\begin{figure*}
\gridline{\fig{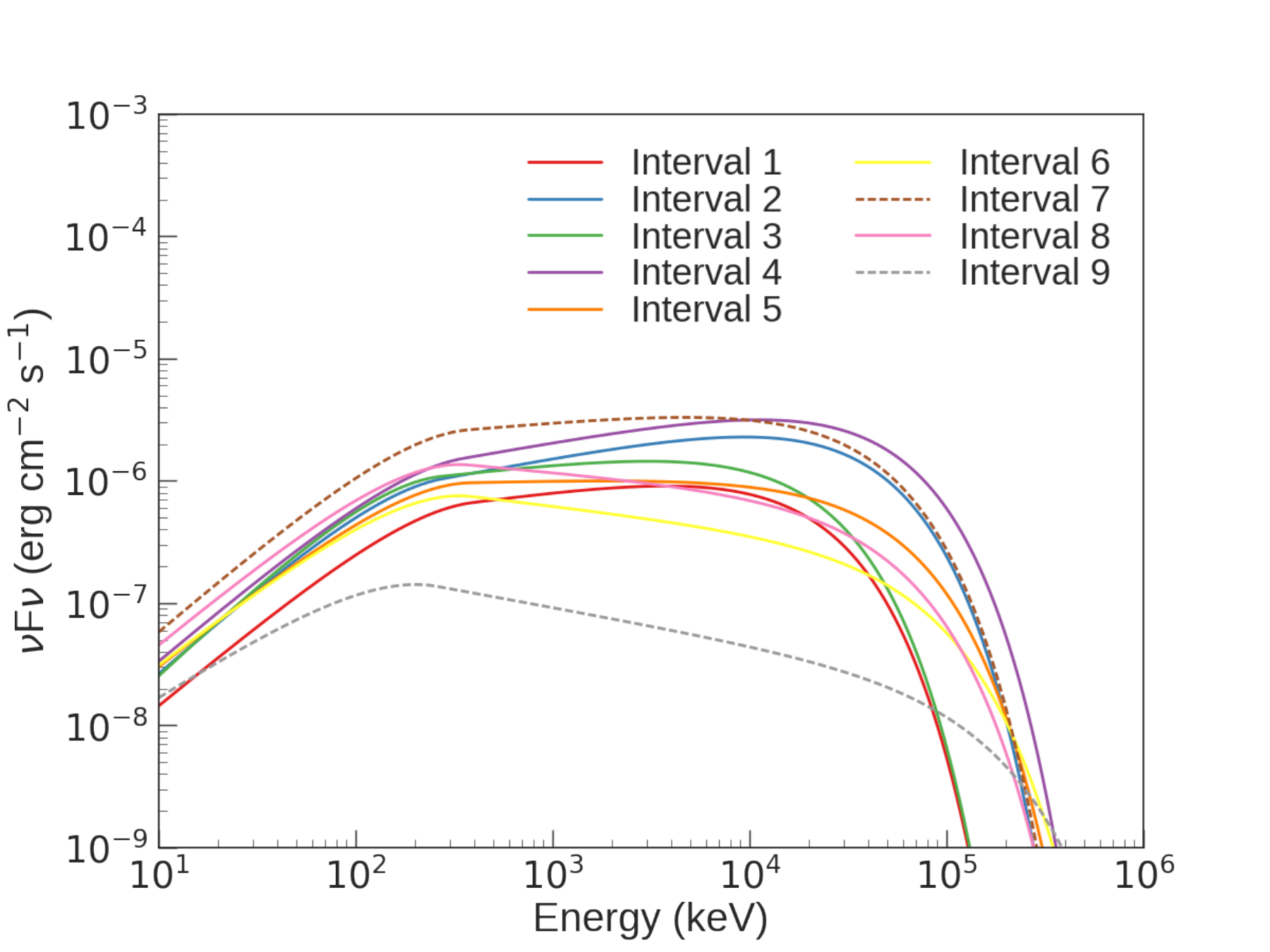}{0.5\textwidth}{(GRB~100724B, $f_{BHec}$ model)}
          \fig{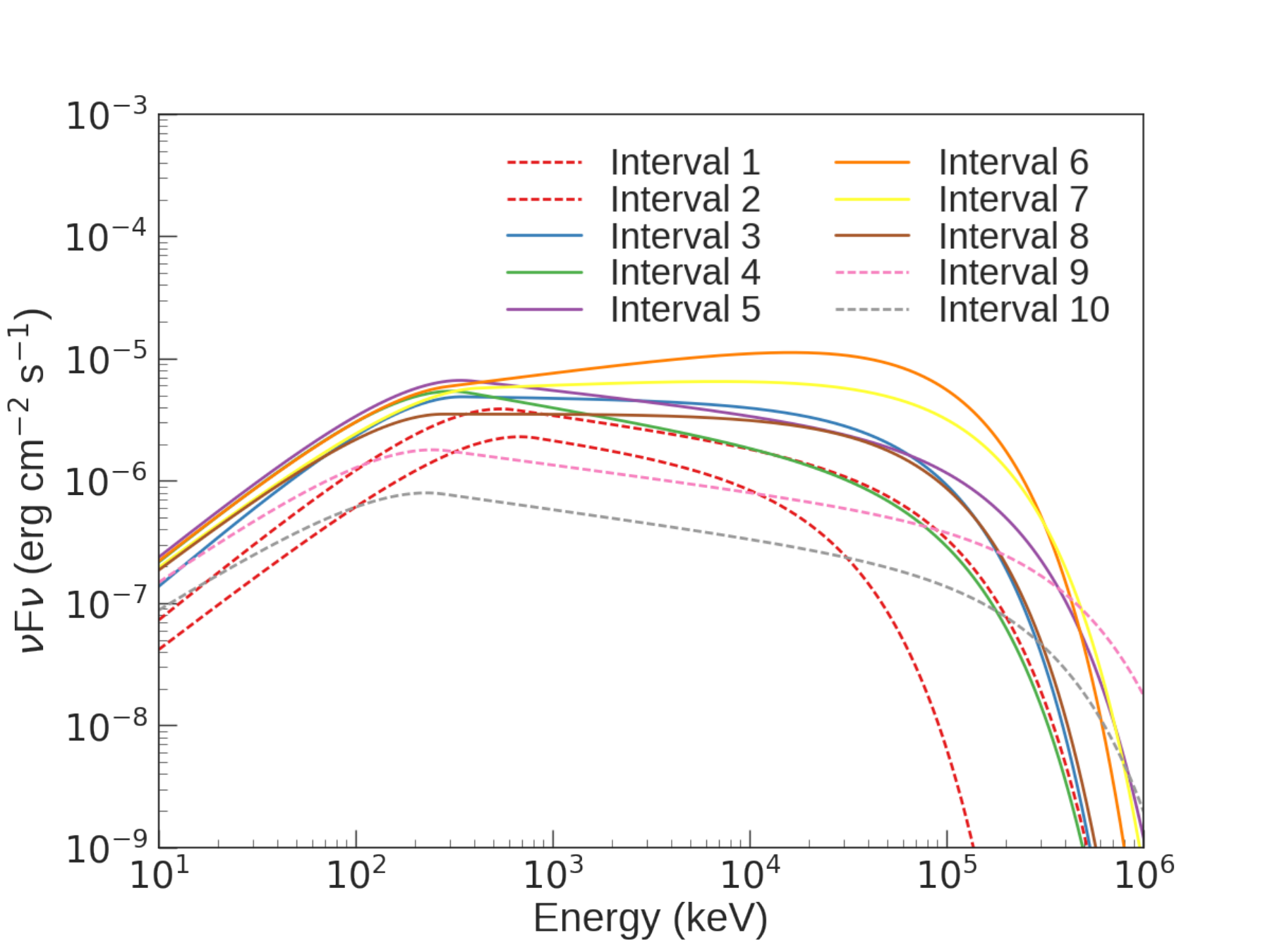}{0.5\textwidth}{(GRB~160509A, $f_{BHec}$ model)}}
          \vspace{-0.7cm}
\gridline{\fig{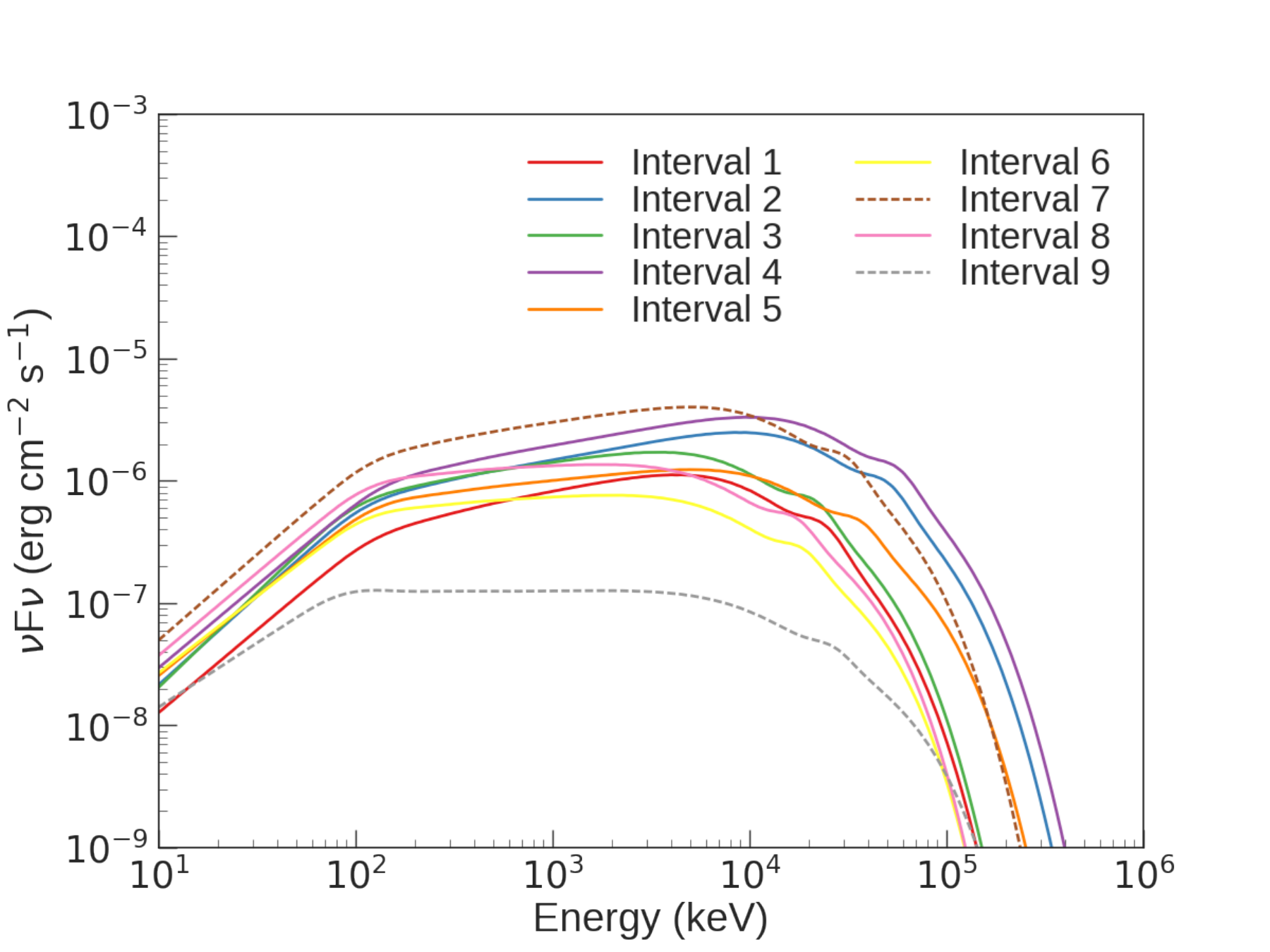}{0.5\textwidth}{(GRB~100724B, $f_{GT}$ model)}
          \fig{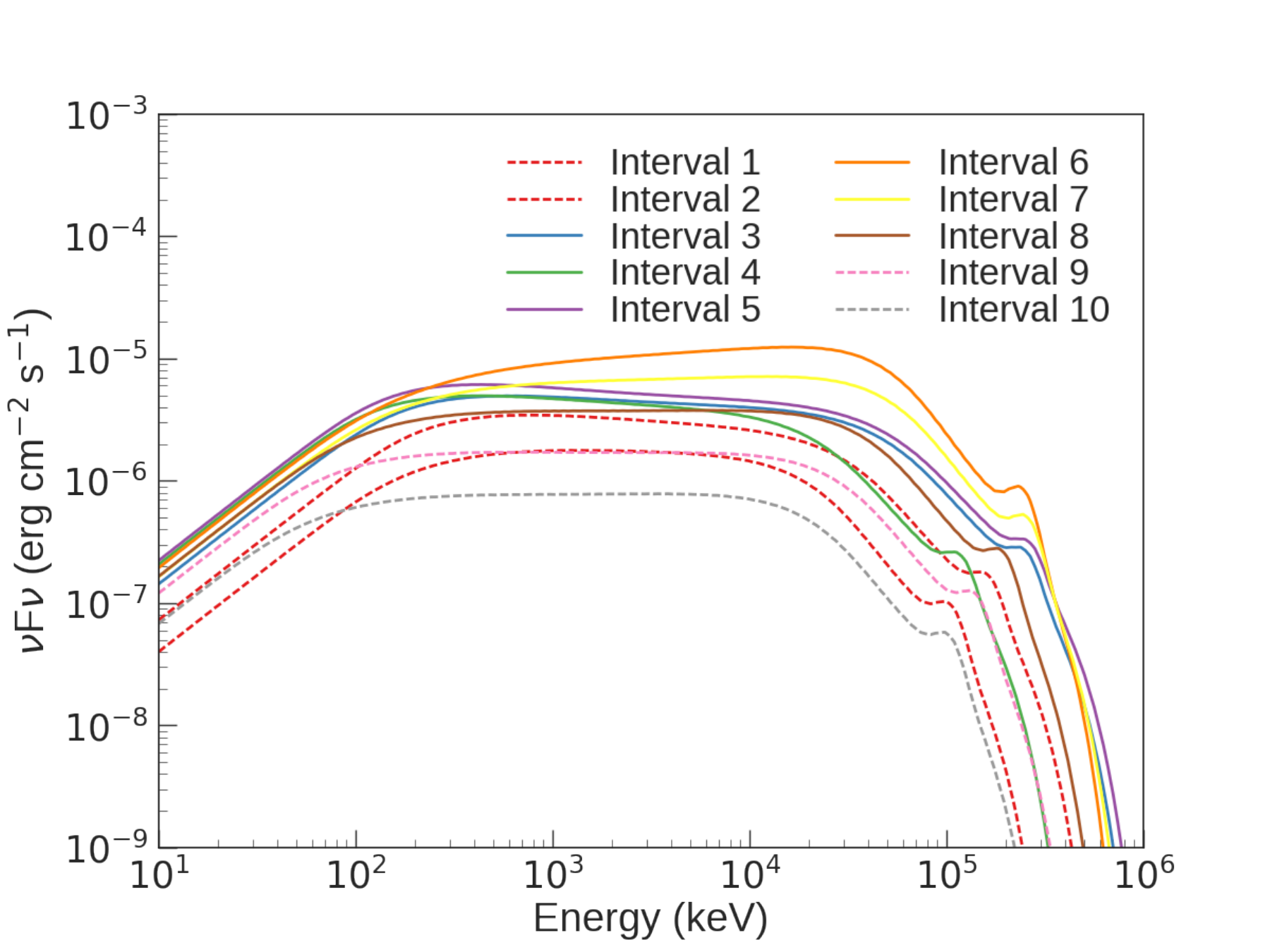}{0.5\textwidth}{(GRB~160509A, $f_{GT}$ model)}}
          \vspace{-0.7cm}
\gridline{\fig{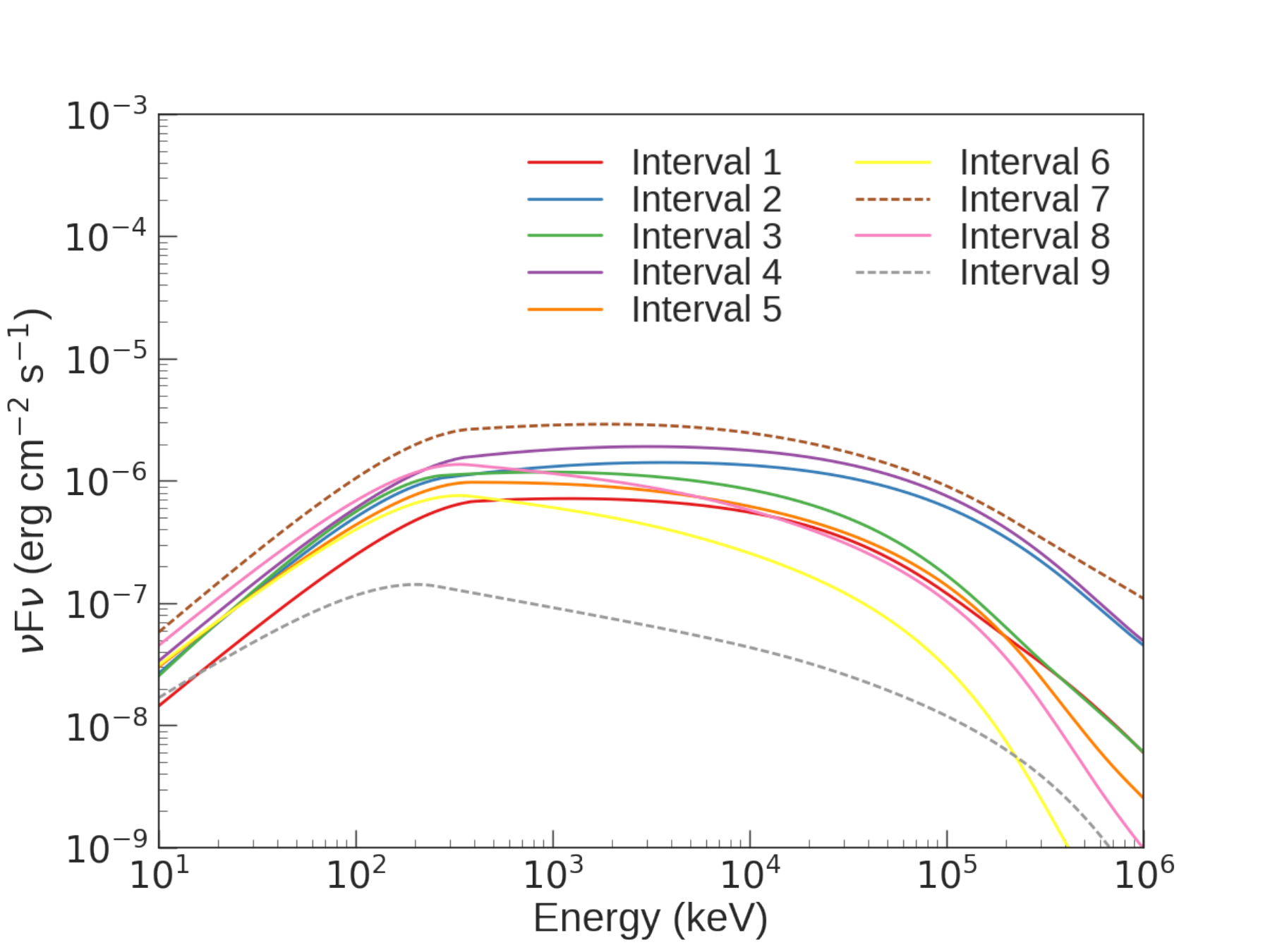}{0.5\textwidth}{(GRB~100724B, $f_{BG}$ model)}
          \fig{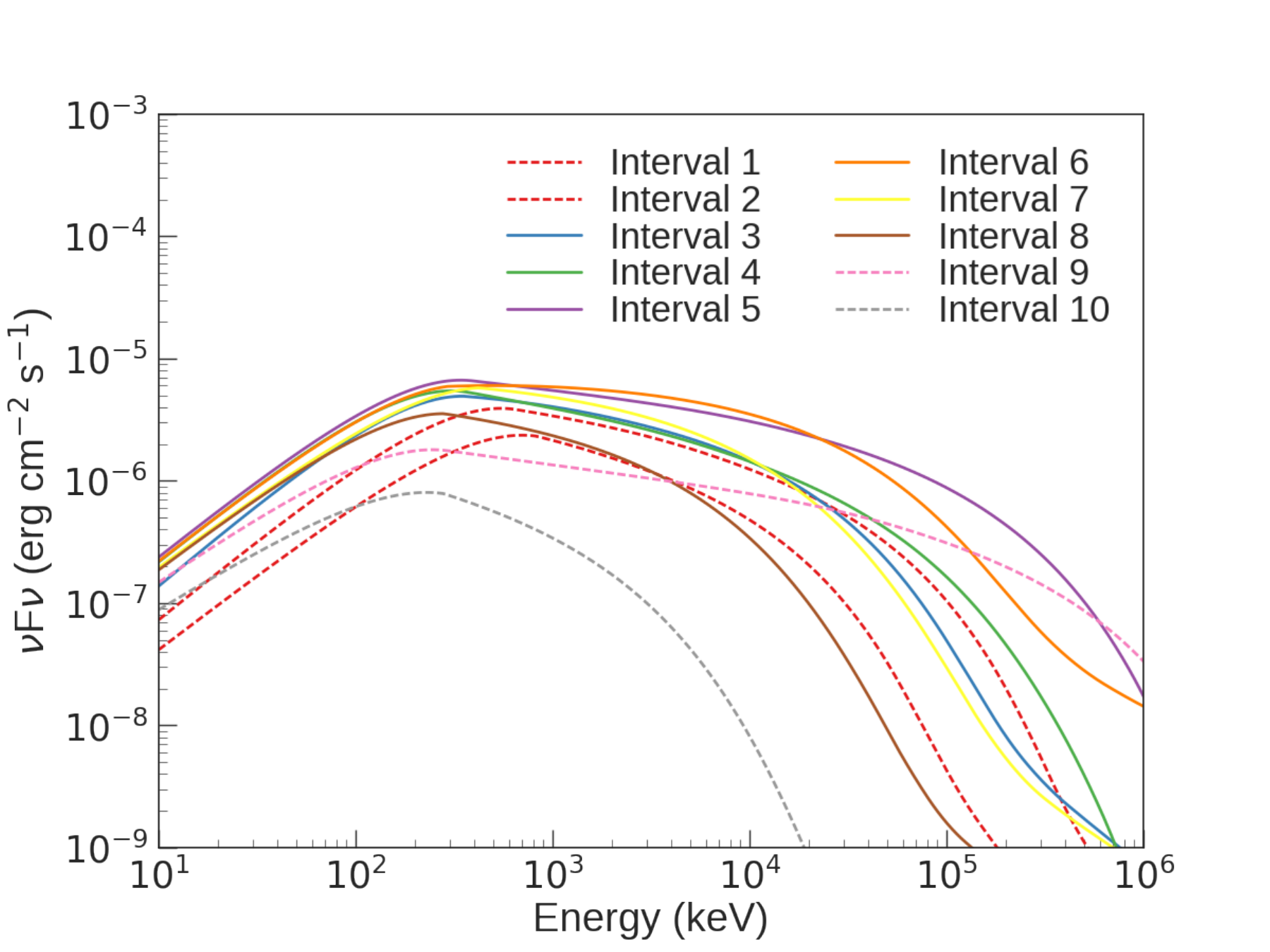}{0.5\textwidth}{(GRB~160509A, $f_{BG}$ model)}}
\caption{Best fit $\nu F_{\nu}$ spectra for GRB~100724B (left) and for GRB~160509A (right), for the $f_{BHec}$, $f_{GT}$ and $f_{BG}$ models. The dashed lines mark intervals where the improvement given by the addition of the cutoff is lower than $3~\sigma$.}
\label{fig:nuFnu_phys_models}
\end{figure*}

\begin{figure*}
\gridline{
\fig{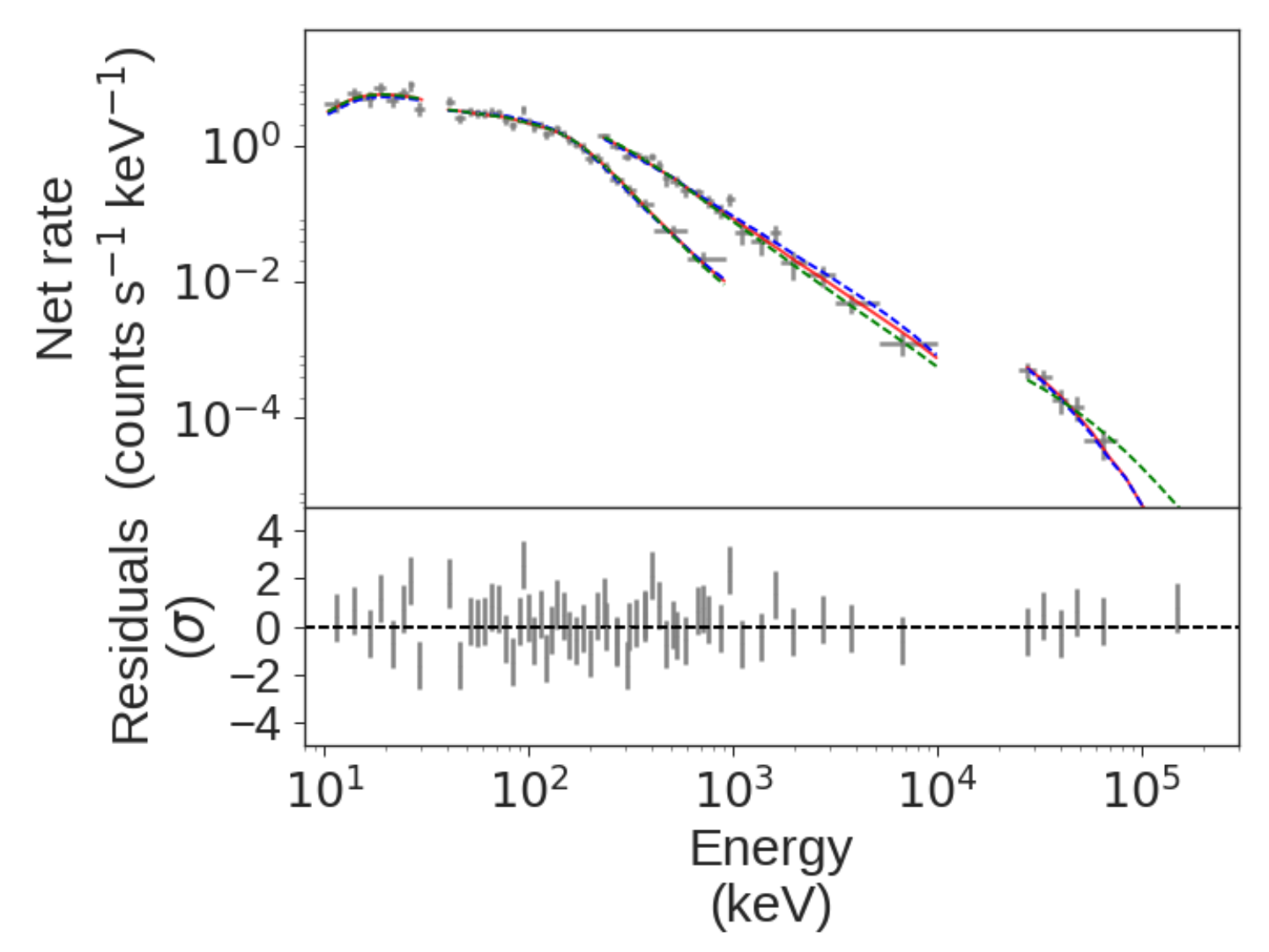}{0.333\textwidth}{(Interval 1)}
\fig{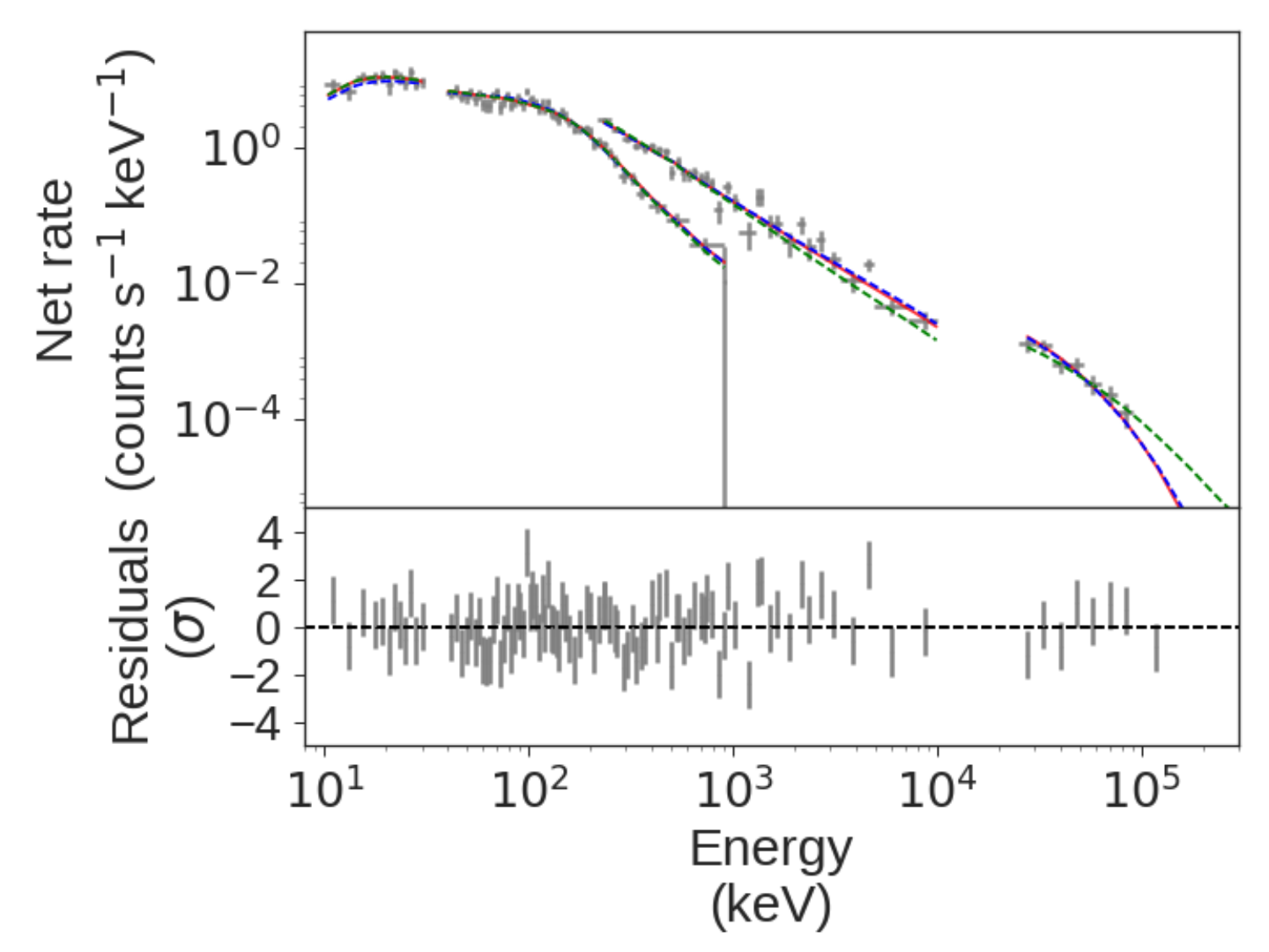}{0.333\textwidth}{(Interval 2)}
\fig{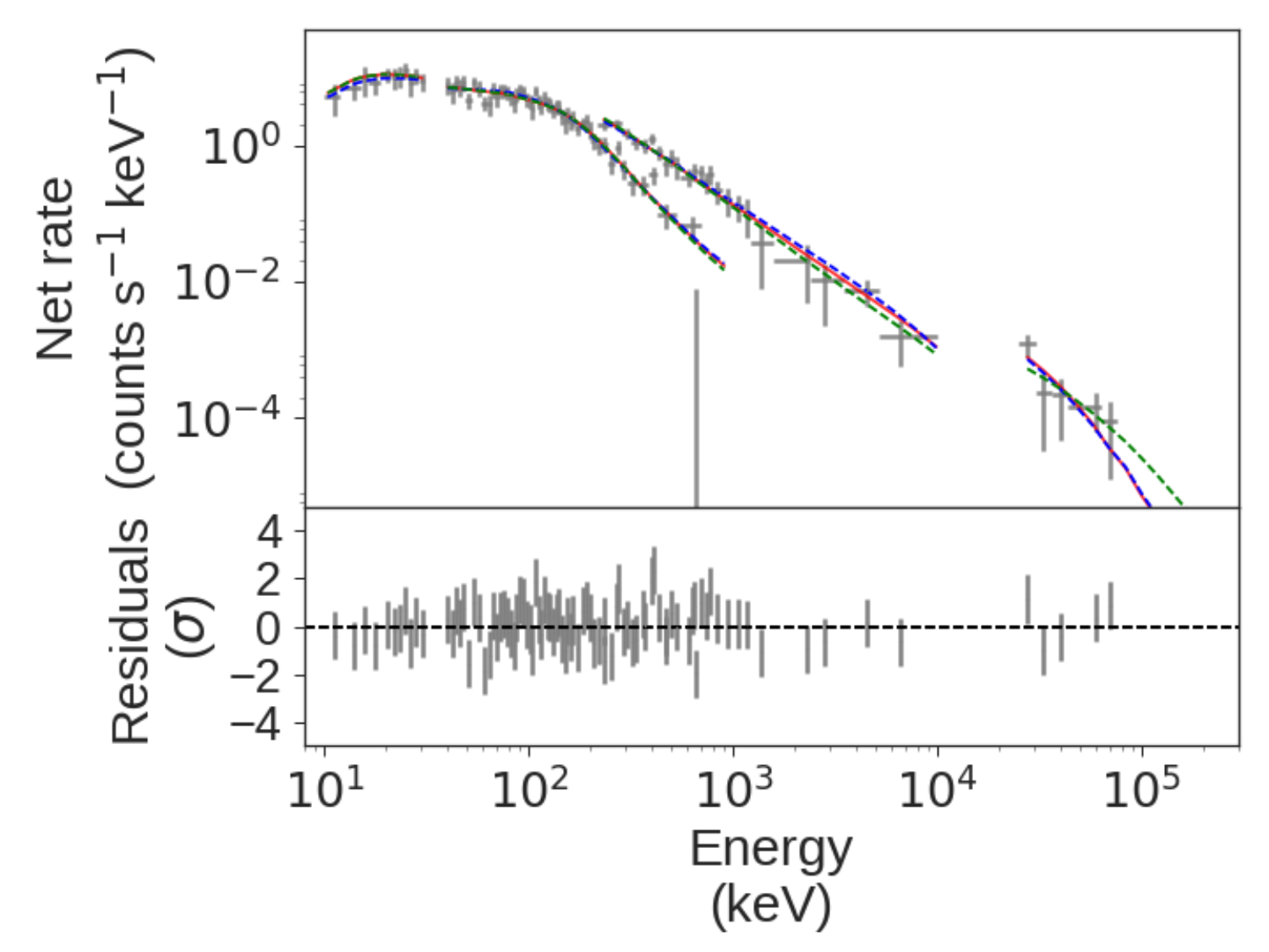}{0.333\textwidth}{(Interval 3)}
}
\gridline{
\fig{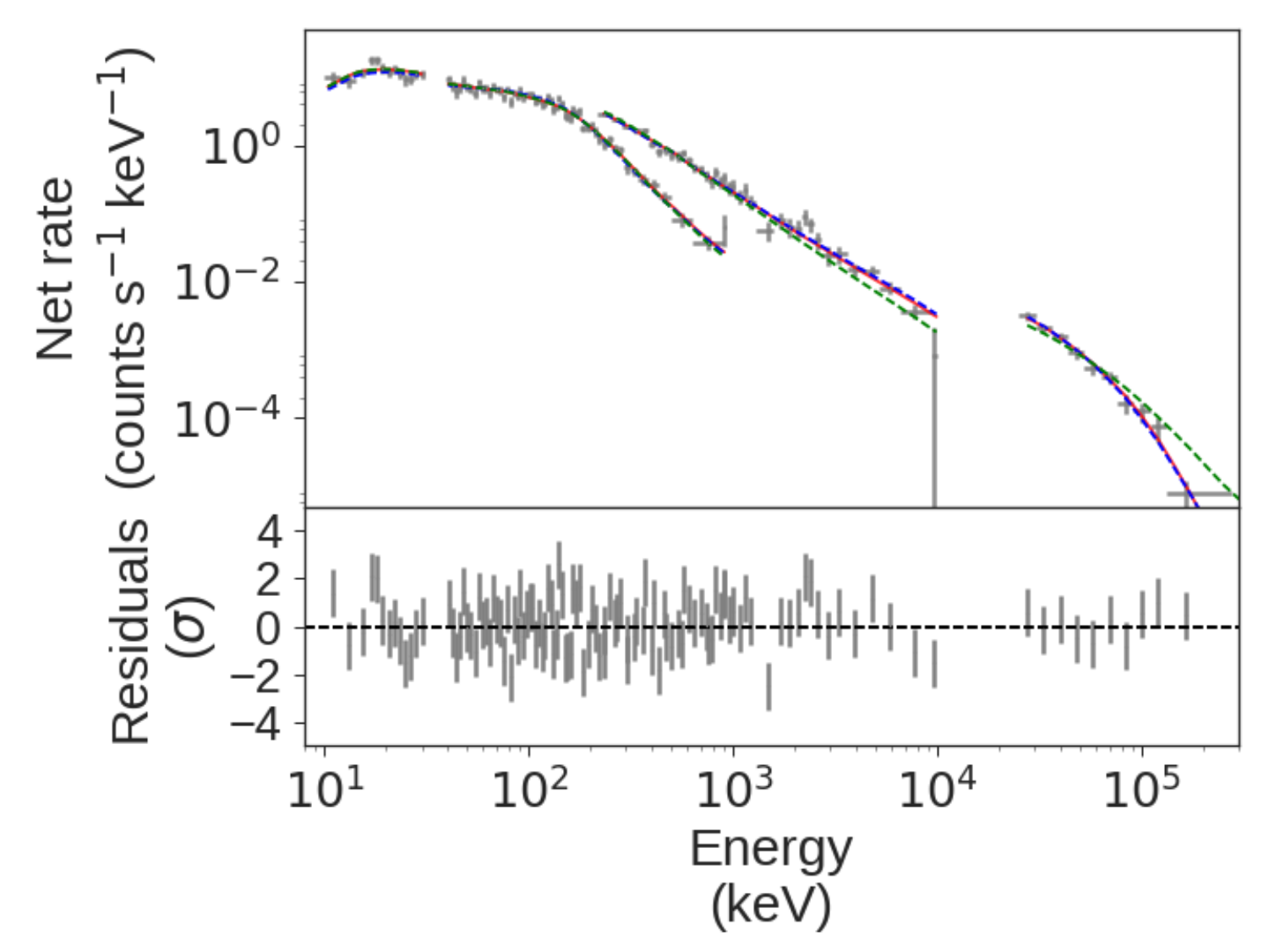}{0.333\textwidth}{(Interval 4)}
\fig{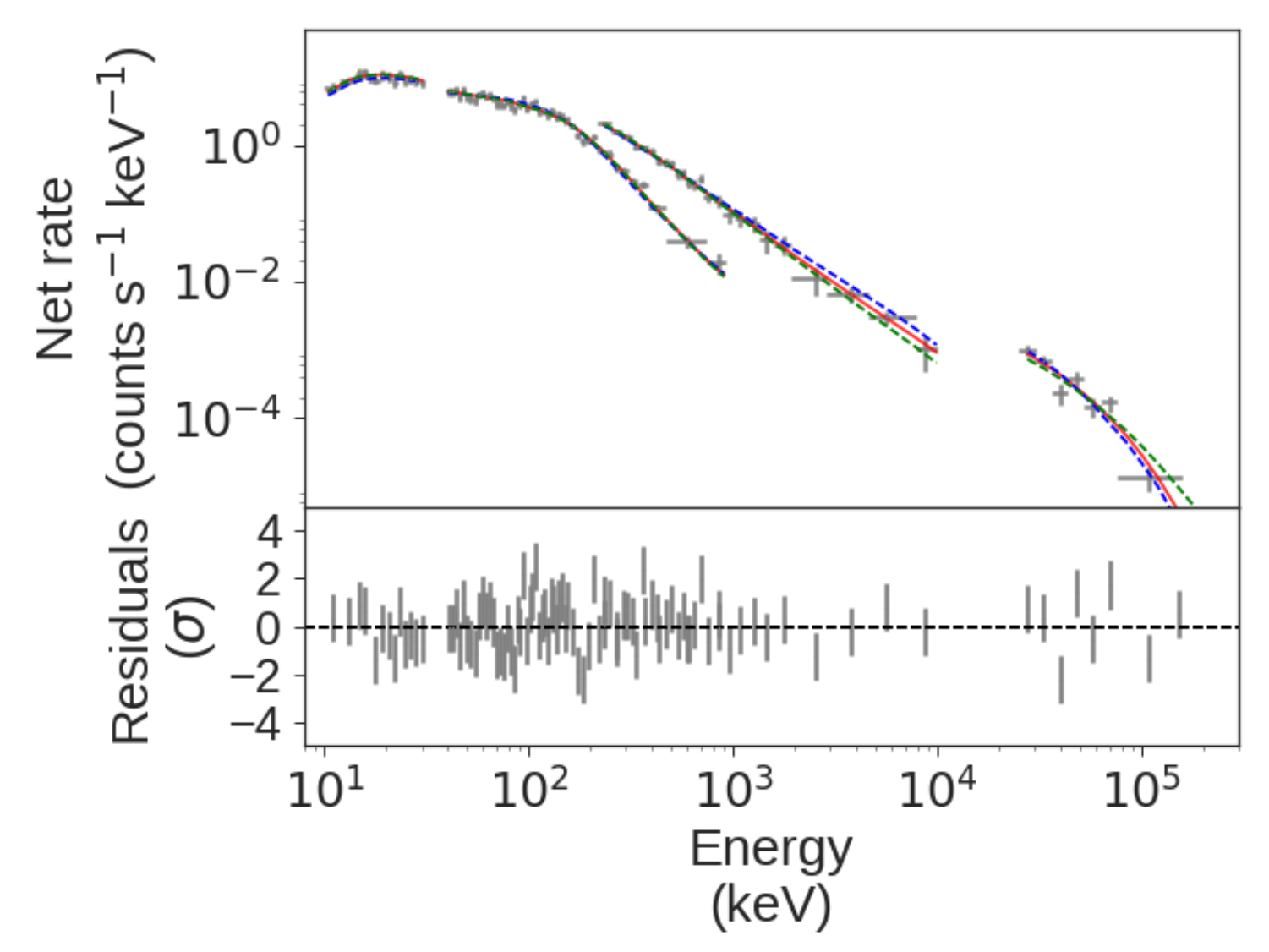}{0.333\textwidth}{(Interval 5)}
\fig{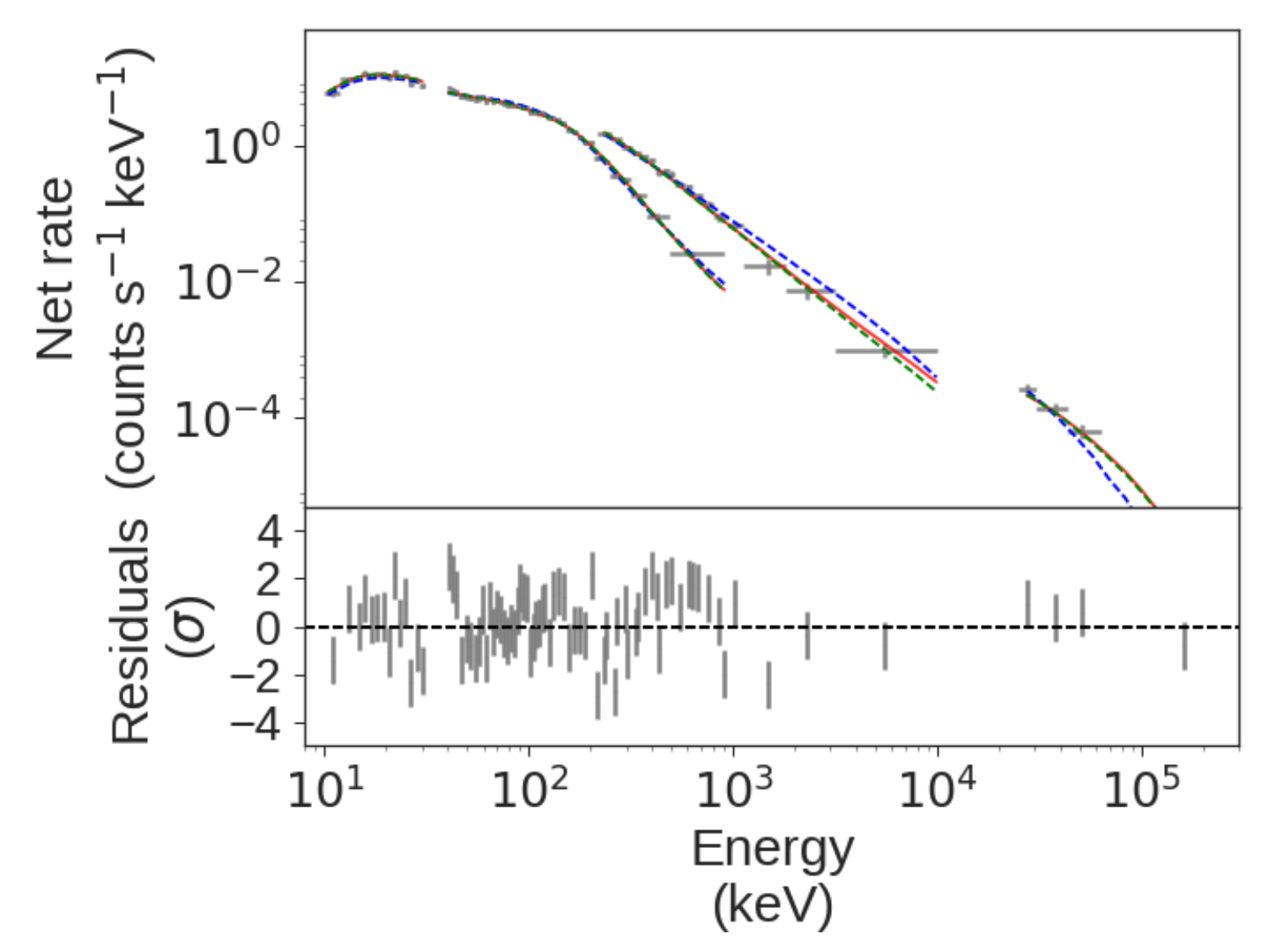}{0.333\textwidth}{(Interval 6)}
}
\gridline{
\fig{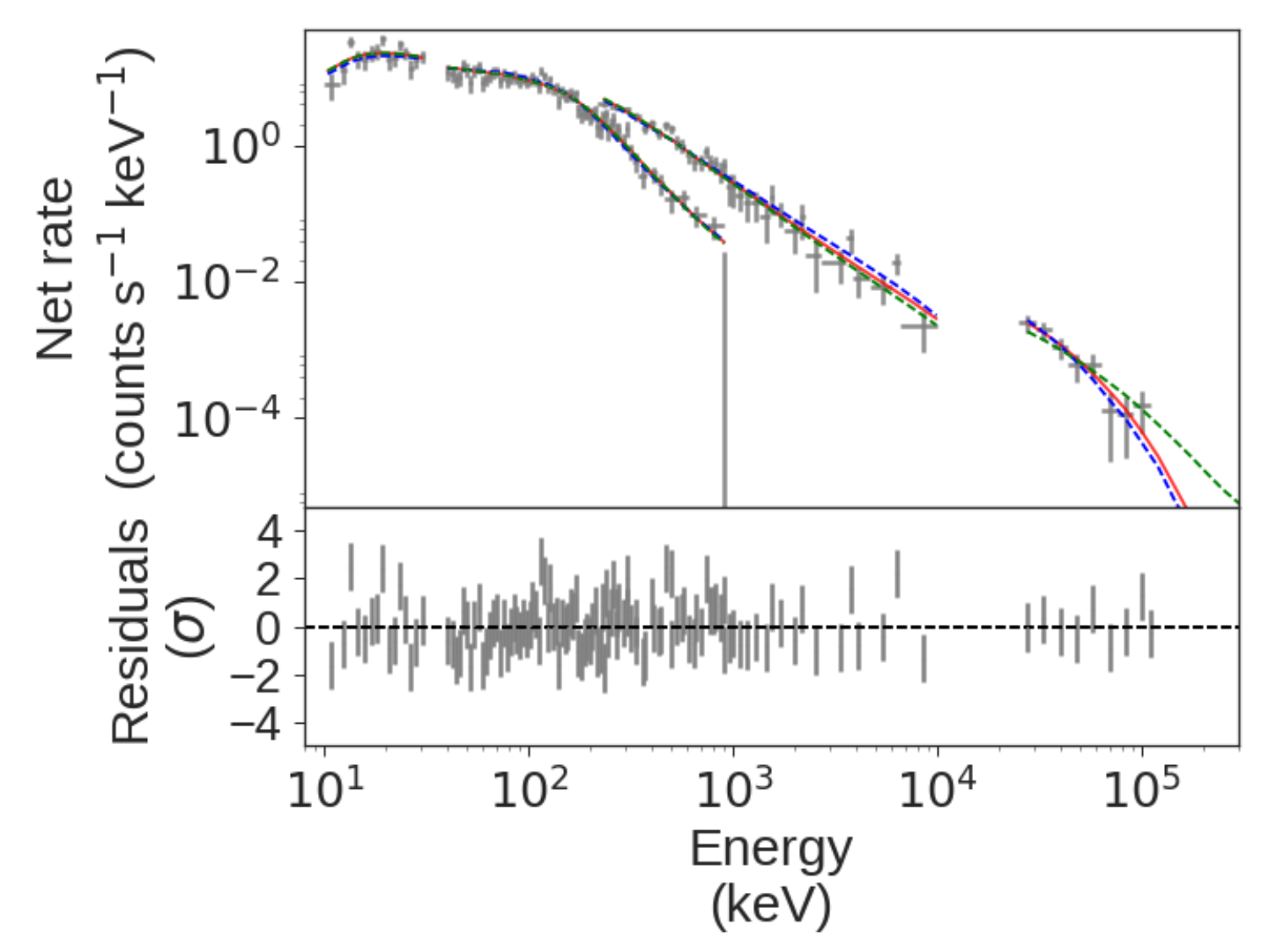}{0.333\textwidth}{(Interval 7)}   \fig{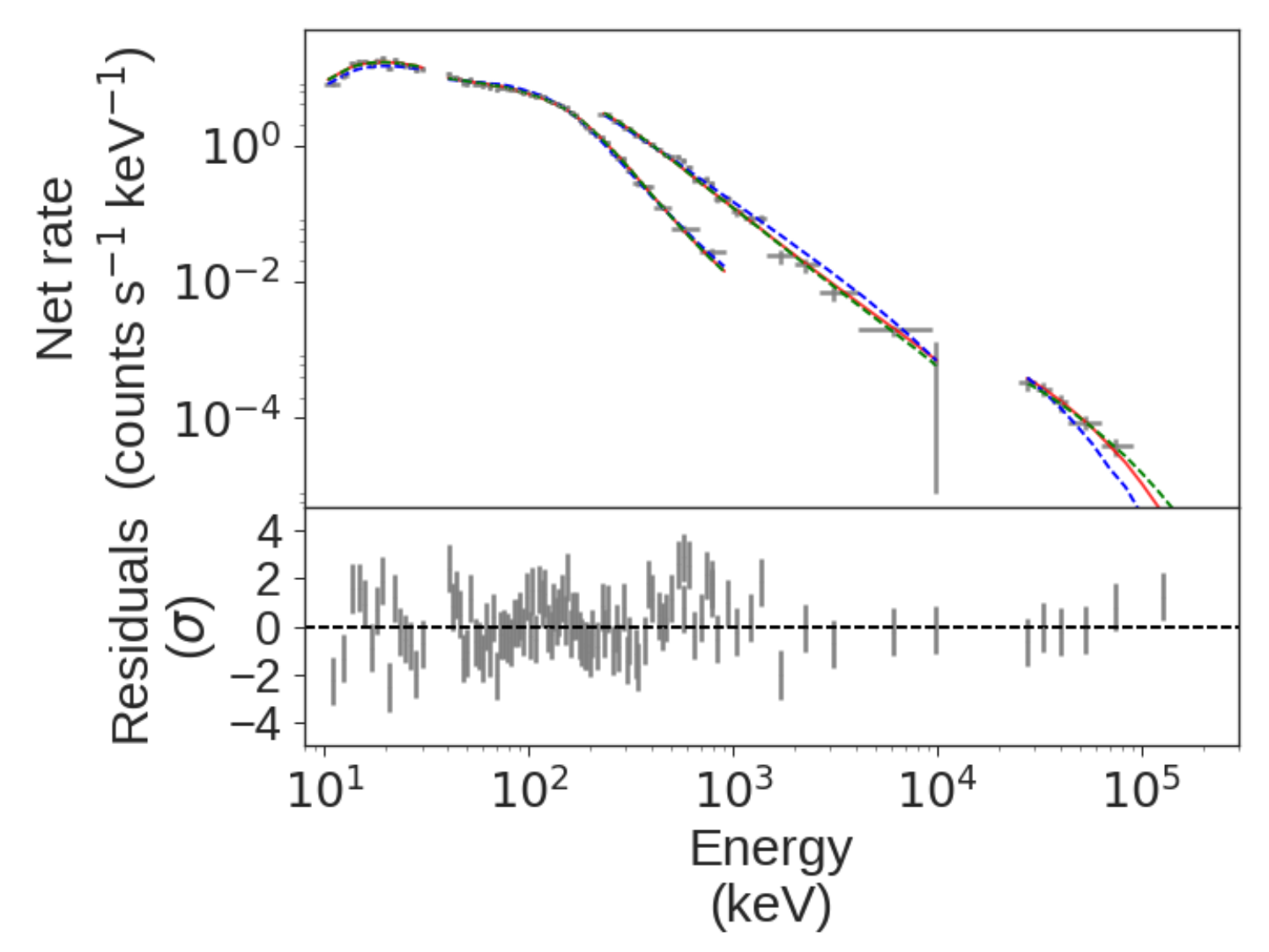}{0.333\textwidth}{(Interval 8)}
\fig{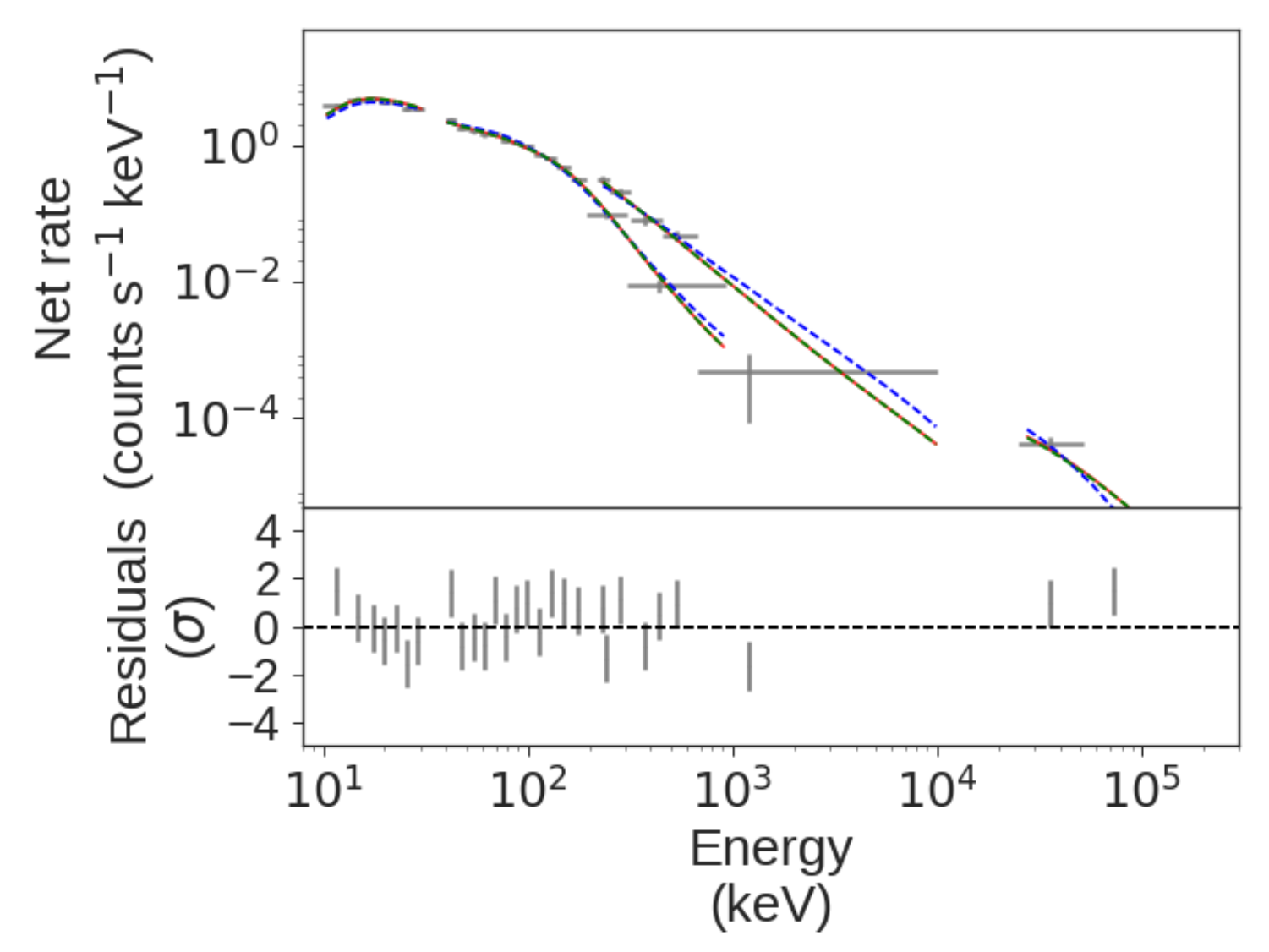}{0.333\textwidth}{(Interval 9)}
}
\caption{Counts spectra for GRB~100724B for the NaI, BGO and LAT detectors (gray points). The lines correspond to the models $f_{BHec}$ (continue line), $f_{GT}$ (blue dashed line), and $f_{BG}$ (green dashed line), convolved with the response of the instruments. The residuals are relative to the $f_{BHec}$ model.}
\label{fig:counts_spectra_100724029}
\end{figure*}

\begin{figure*}
\gridline{
\fig{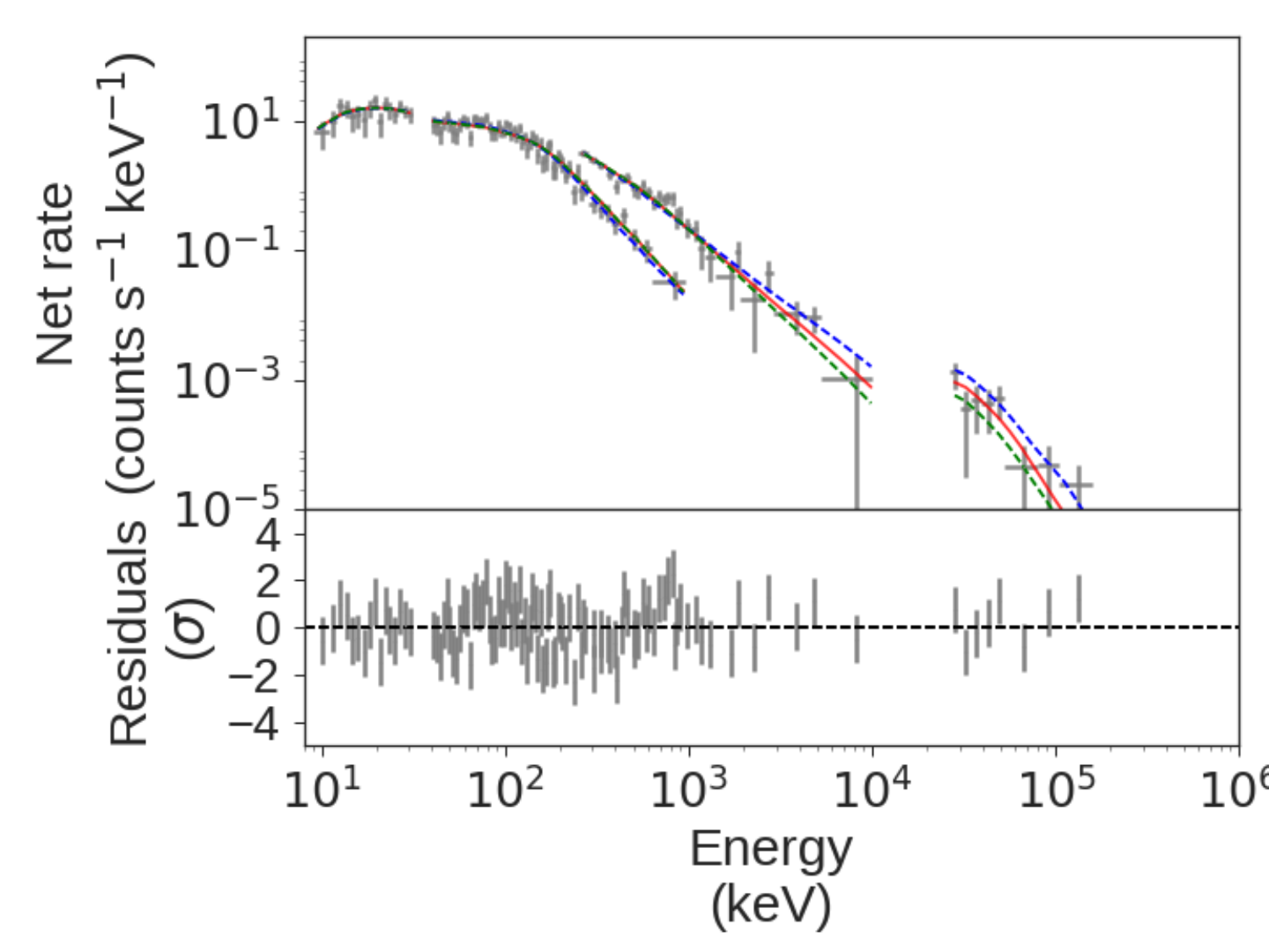}{0.333\textwidth}{(Interval 1)}
\fig{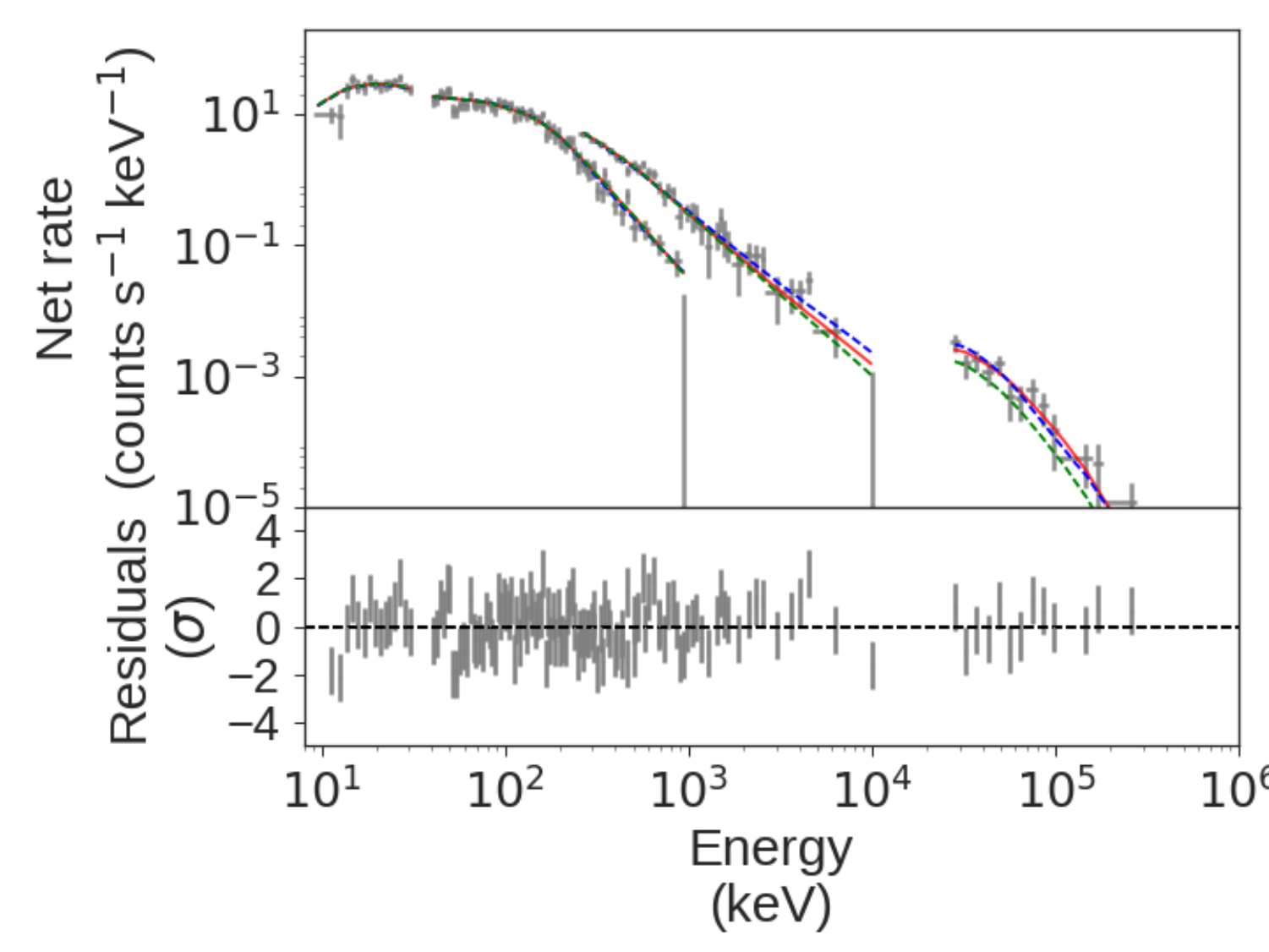}{0.333\textwidth}{(Interval 2)}
\fig{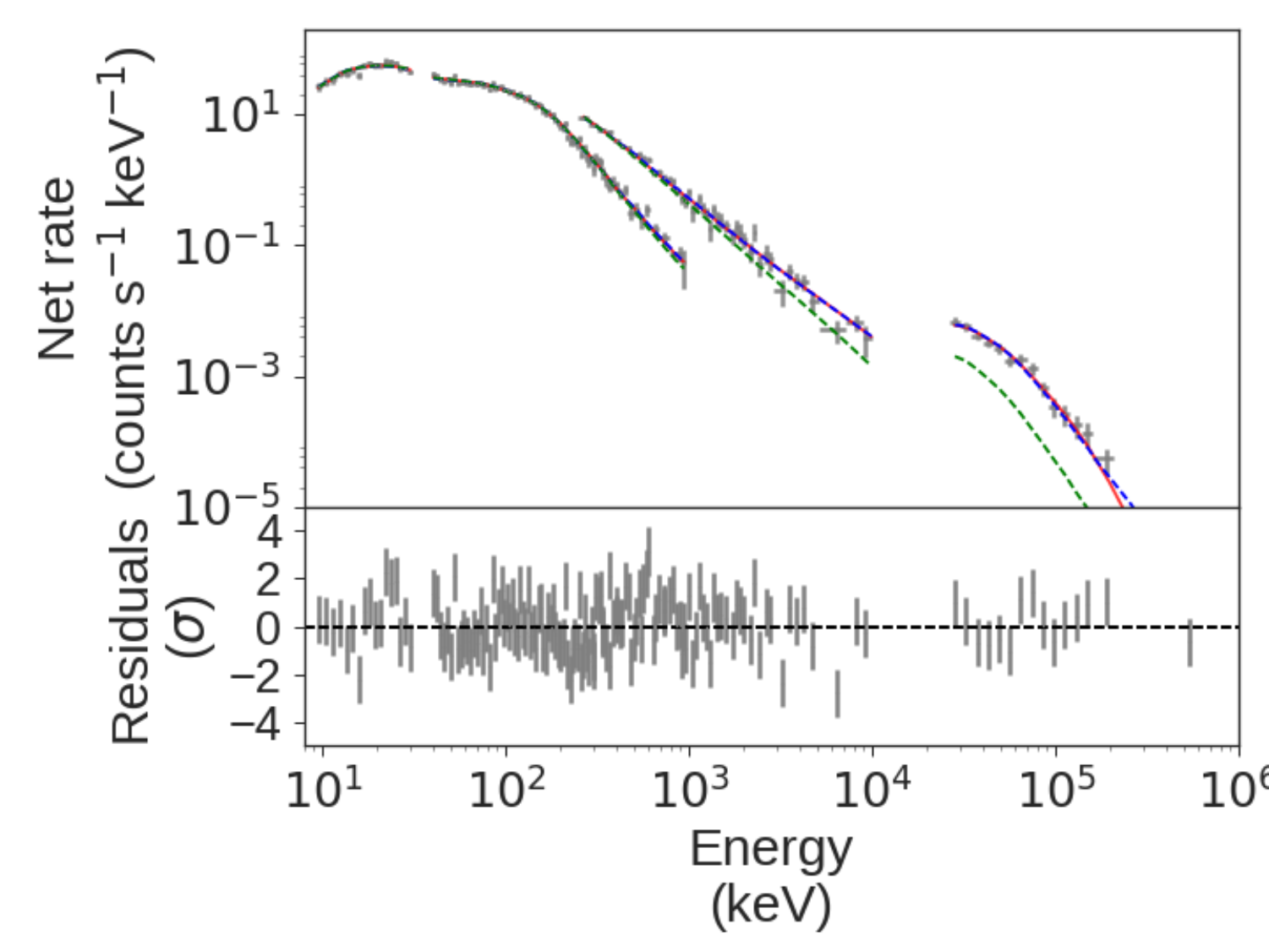}{0.333\textwidth}{(Interval 3)}
}
\gridline{ \fig{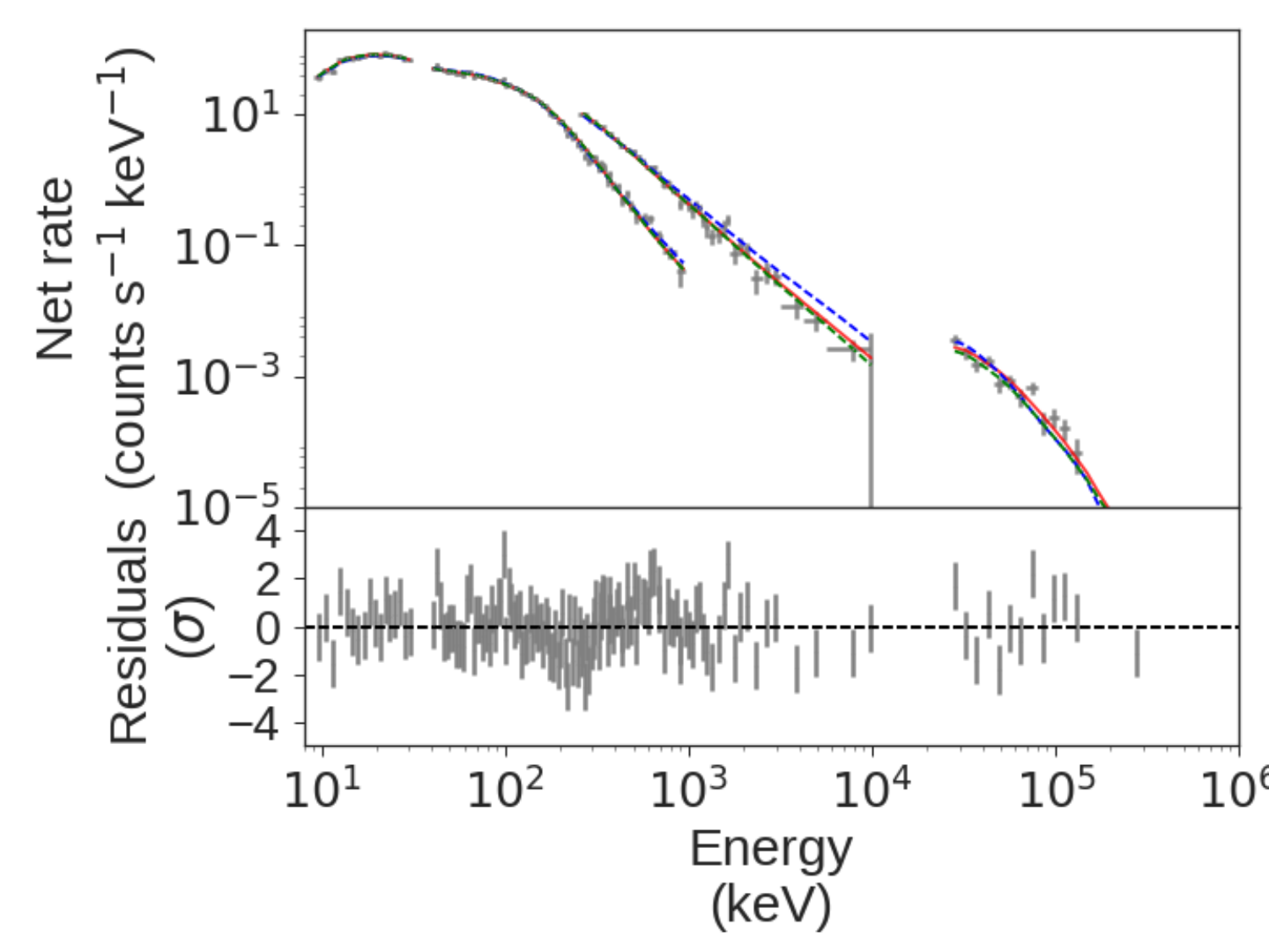}{0.333\textwidth}{(Interval 4)}
\fig{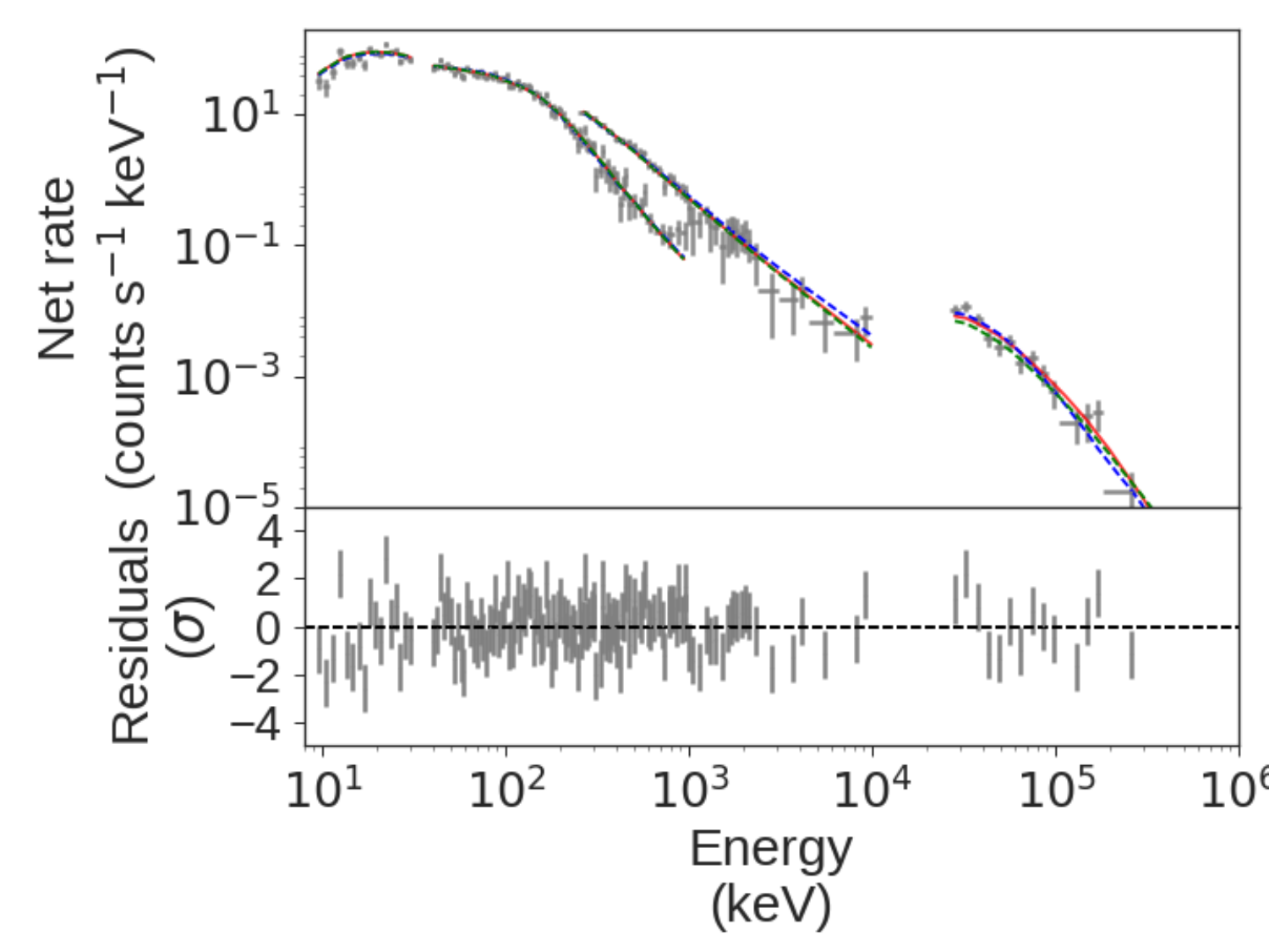}{0.333\textwidth}{(Interval 5)}
\fig{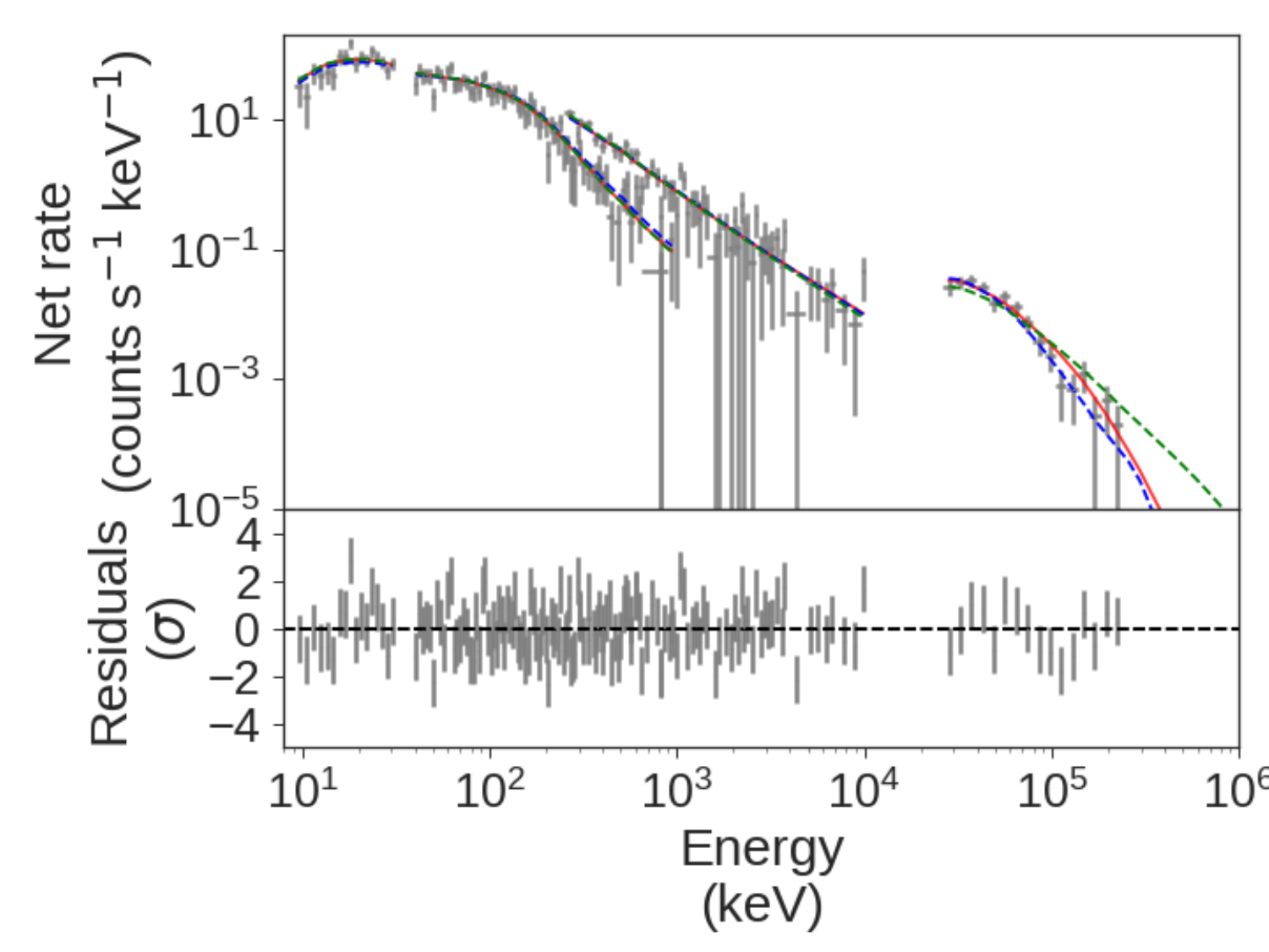}{0.333\textwidth}{(Interval 6)}
}
\gridline{
\fig{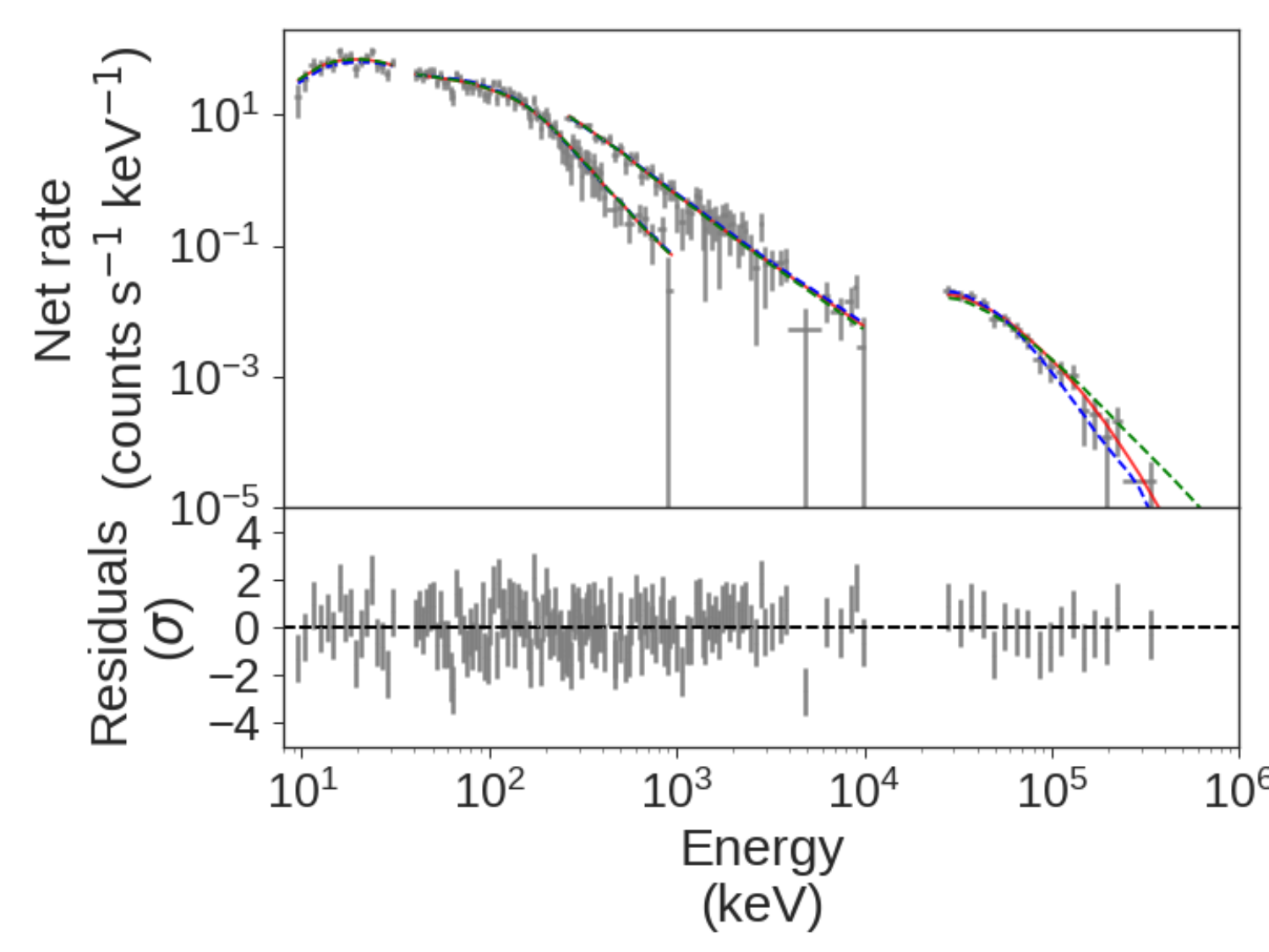}{0.333\textwidth}{(Interval 7)}
\fig{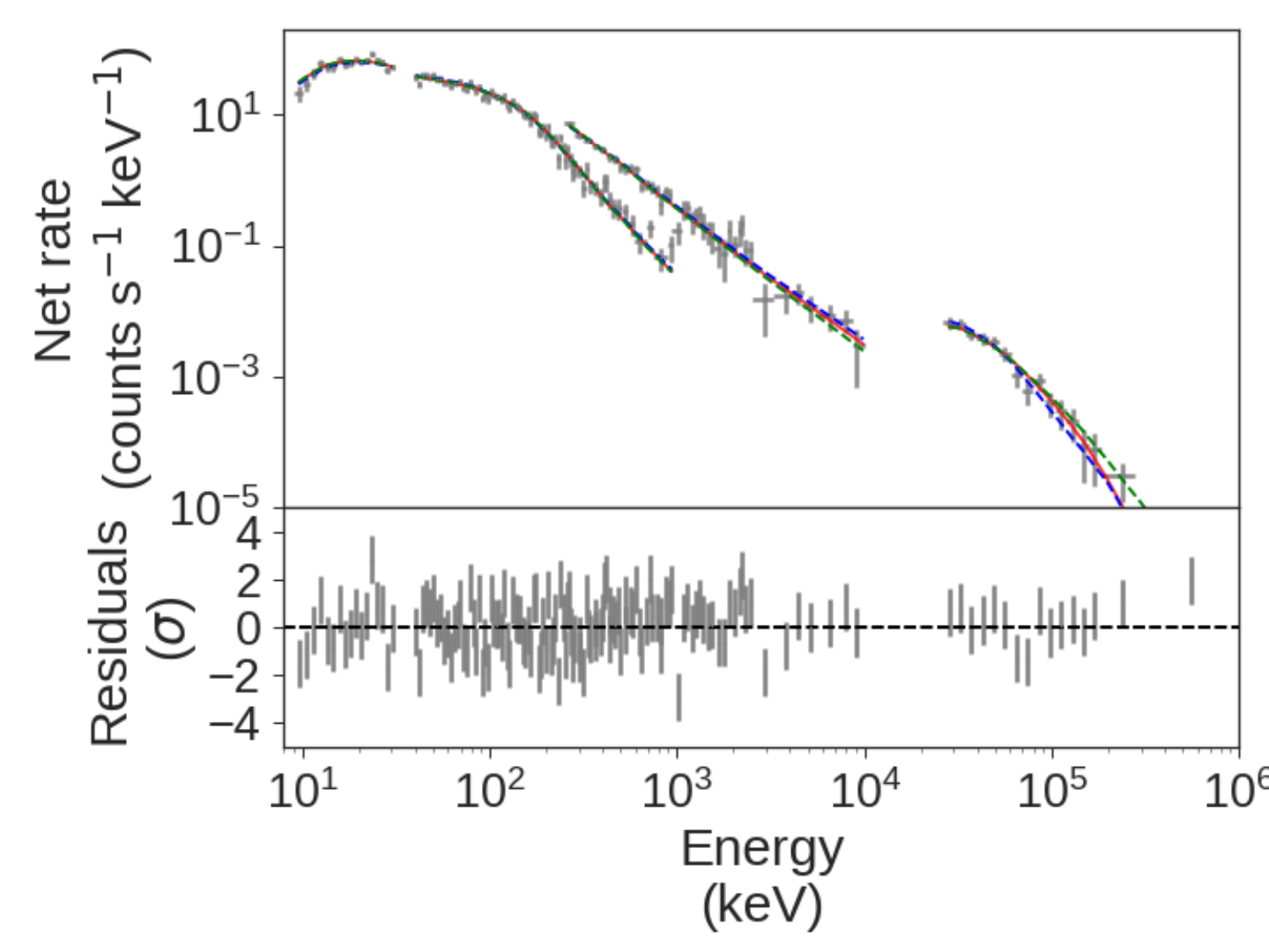}{0.333\textwidth}{(Interval 8)}
\fig{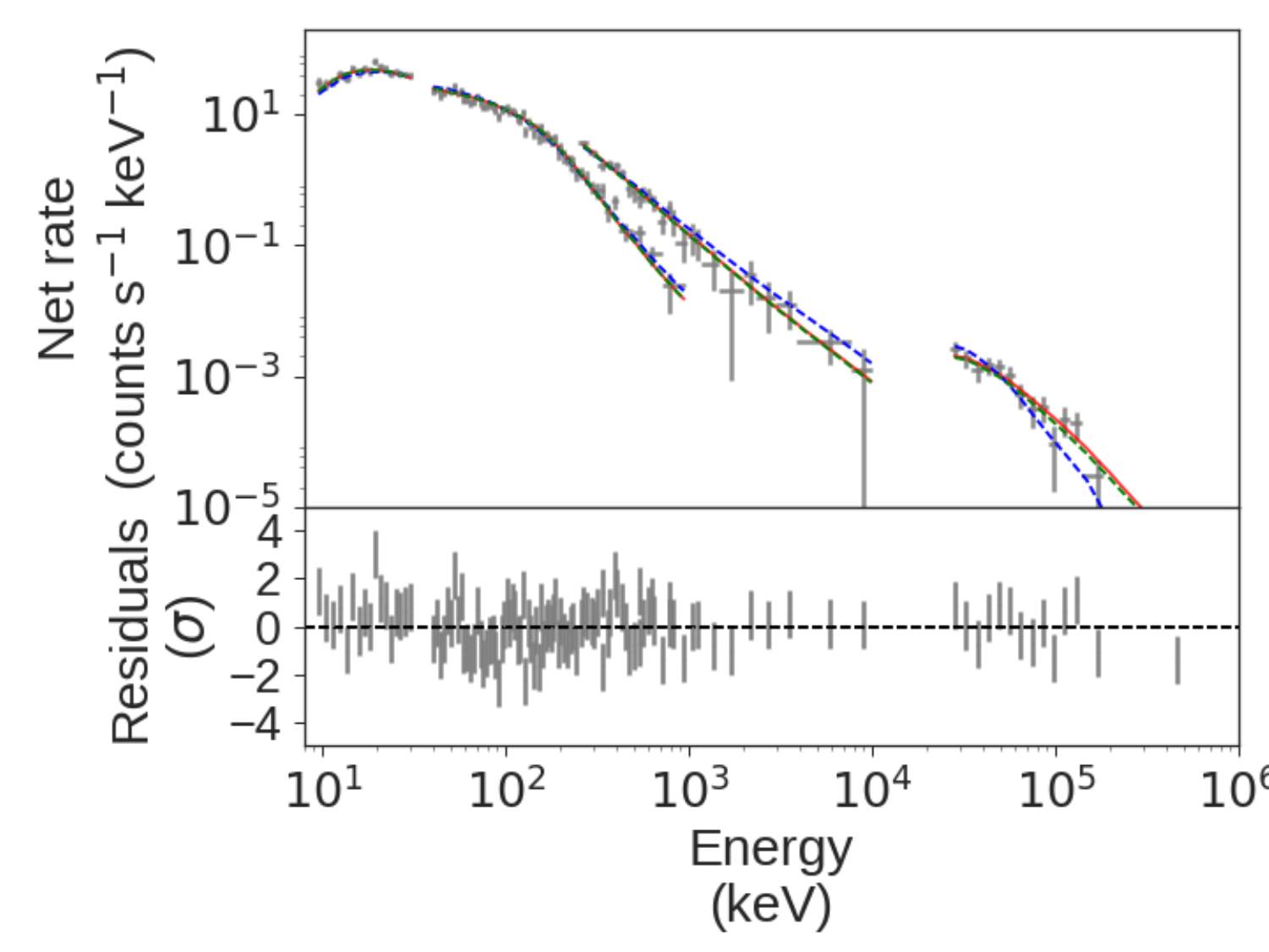}{0.333\textwidth}{(Interval 9)}
}
\gridline{ \fig{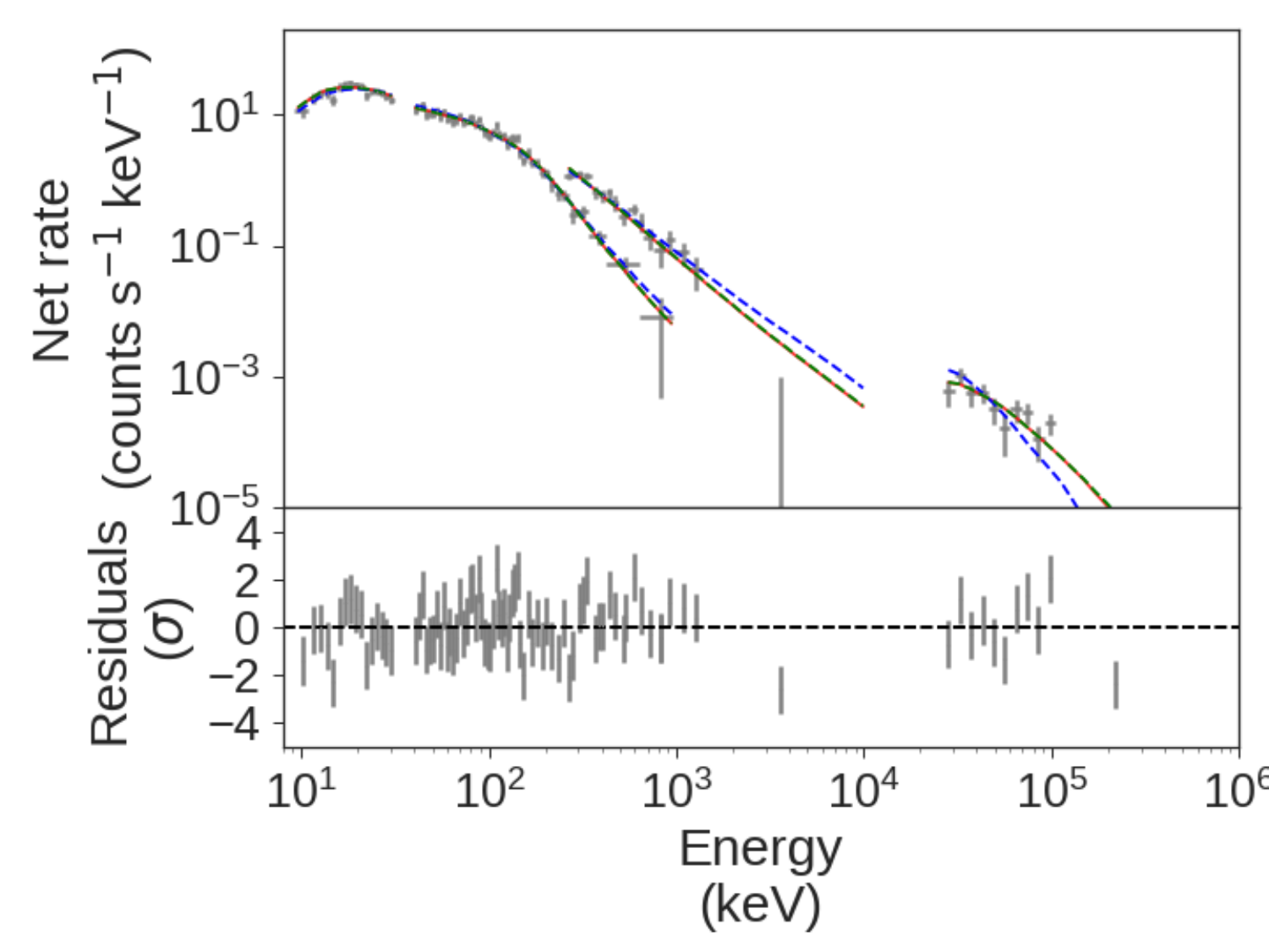}{0.333\textwidth}{(Interval 10)}
}
\caption{Counts spectra for GRB~160509A for the NaI, BGO and LAT detectors (gray points). The lines correspond to the models $f_{BHec}$ (continue line), $f_{GT}$ (blue dashed line), and $f_{BG}$ (green dashed line), convolved with the response of the instruments. The residuals are relative to the $f_{BHec}$ model.}
\label{fig:counts_spectra_160509374}
\end{figure*}

In Figure \ref{fig:nuFnu_phys_models} we show the best fit models for $f_{BHec}$, $f_{GT}$ and $f_{BG}$ for both GRBs. Even though the $f_{BG}$ model tends to predict a much higher flux at high photon energies, it is still fully compatible with the data due to the low statistics at high energies. This can be seen in the count spectra shown in Figures~\ref{fig:counts_spectra_100724029} and \ref{fig:counts_spectra_160509374}. 
On the other hand, the $f_{GT}$ model is very similar to $f_{BHec}$. The p-values for the goodness of fit test $p$ and $p_{LAT}$ for all three models, computed as described in section \ref{sec:spectral_analysis}, are $ \gtrsim 0.05$ for all intervals. We conclude that all 3 models appear to describe our data well. For high compactness the $f_{\rm GT}$ model features a fairly prominent pair annihilation line, visible for example in the best fit
models for GRB 160509A (middle right panel of Figure~\ref{fig:nuFnu_phys_models}). It is currently not detectable by \Fermi/LAT as it is smeared out by energy dispersion effects in the detector, and indeed it is not apparent once the model is folded with the response of the instrument (Figure~\ref{fig:counts_spectra_160509374}, blue dashed line).

\begin{figure*}
\centering
\includegraphics[width=0.9\textwidth]{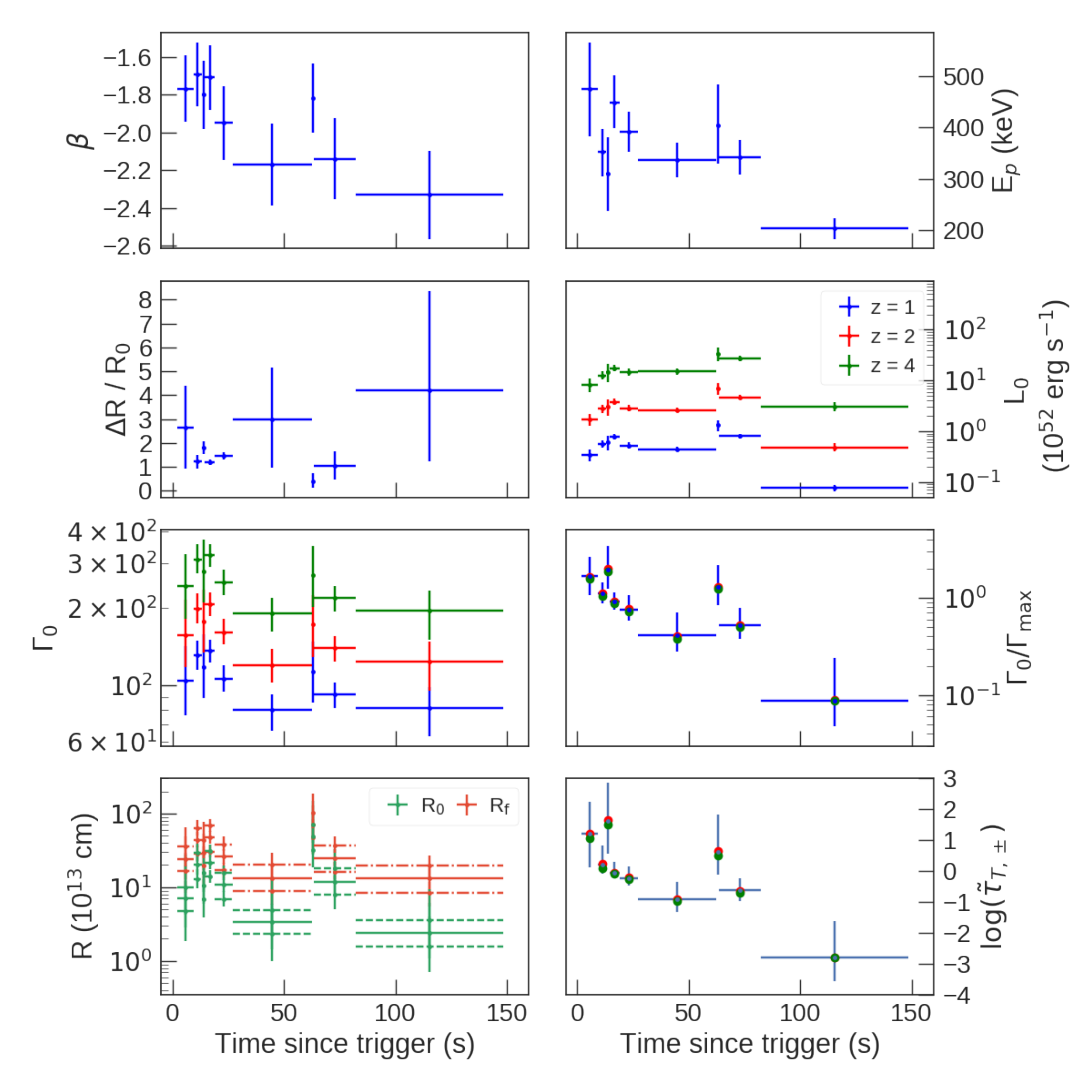}
\caption{Best fit parameters of the $f_{BG}$ model for the prompt $\gamma$-ray emission of GRB~100724B. 
Here $E_p$ and $\beta$ are for the Band spectrum peak photon energy and the photon indexes above it. The remaining parameters are for the  \citet{2008ApJ...677...92G} model:
$L_0 = 4\pi d_{L}^2(1+z)^{-\beta-2}F_{0}$ is the isotropic equivalent luminosity corresponding to the observed flux at $E_{\rm ph}=m_ec^2\Leftrightarrow\varepsilon=1$, $\Gamma_0$ is the bulk-LF at the emission onset radius $R_0$, and $\Delta R$ is the radial interval over which the emission takes place, ending at $R_f = R_0+\Delta R$. Also shown is the ratio of $\Gamma_0$ and $\Gamma_{\rm max}$ (identified with $\Gamma_{\rm\gamma\gamma,min}(E_c)$ and $\Gamma_{\rm\gamma\gamma,max}(E_c)$, respectively, in the text), and the implied Thomson optical depth in pairs (neglecting pair annihilation; $\tilde{\tau}_{T,\pm}$). Since the redshift of GRB~100724B is not known, quantities that depend on it ($L_0$, $\Gamma_0$, $R$, $\Gamma_0/\Gamma_{\rm max}$ and $\tilde{\tau}_{T,\pm}$) are shown for three representative values: $z=1$ (in {\it blue}), $z=2$ (in {\it red}), and $z=4$ (in {\it green}). For the emission radii $R$ (i.e. emission onset $R_0$ and turnoff $R_f$) a solid cross is used for $z=2$, and the modification of the central value for $z=1$ and $z=4$ is shown by horizontal dashed (for $R_0$) and dashed-dotted (for $R_f$) lines.
}
\label{fig:jg_results1}
\end{figure*}

\begin{figure*}
\centering
\includegraphics[width=0.9\textwidth]{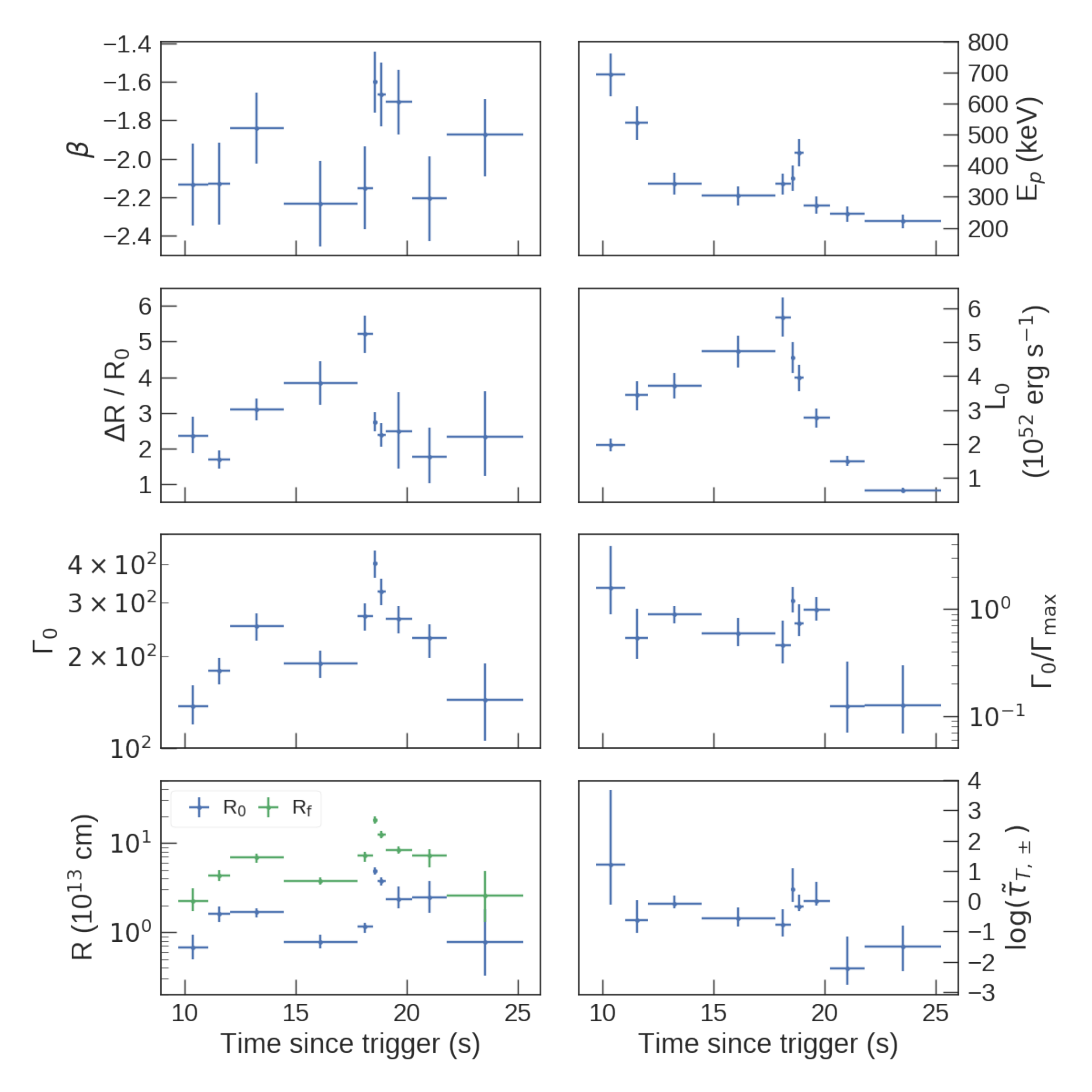}
\caption{Similar to Figure~\ref{fig:jg_results1}, but for GRB~160509A, 
which occurred at $z = 1.17$.}
\label{fig:jg_results2}
\end{figure*}

The best fit parameters of the $f_{BG}$ model for both GRBs are shown in Figures~\ref{fig:jg_results1} and \ref{fig:jg_results2}.  The values of  $\Delta R/R_0$ are of order unity (typically $\sim1-2$ and ranging from $\sim 0.5$ to $\sim 3.5$) for GRB~100724B, and somewhat larger, typically around $\sim 2-3$ (ranging between $\sim 1.5$ and $\sim 5.5$) for GRB~160509A. This is in reasonable agreement with the expectations of the internal shocks model (for which the physical setup of the \citet{2008ApJ...677...92G} model is particularly well suited), as are the typical inferred emission radii ($\sim 10^{13}-10^{15}\;$cm for GRB~100724B and $\sim 10^{13}-10^{14}\;$cm for GRB~160509A). The LF is relatively low when compared to what is inferred for other LAT-detected GRBs \citep{LATGRBcat}, ranging from $\sim 140$ to $\sim400$ for GRB~160509A, while for GRB~100724B  it depends on the unknown redshift but for typical redshifts it is broadly similar (ranging from $\sim70$ to $\sim310$ for $1\leq z\leq4$). This may account for the relatively low values of the cutoff energy $E_c$ (e.g. as inferred for the $f_{BHec}$ model and is shown in Figures~\ref{fig:spEvolution} and 
\ref{fig:spEvolution2}), of $\sim 15-50\;$MeV for GRB~100724B (except in the last time bin, where 
$E_c\sim200\;$MeV), and $\sim100\;$MeV (ranging between $\sim 20\;$MeV and $\sim400\;$MeV) for GRB~160509A.  In turn, this may demonstrate the fact that slower GRBs tend to be dimmer in the LAT energy range, thus producing a selection effect in favor of faster GRBs in the LAT GRB sample. This effect would be more 
pronounced when not accounting for LAT-LLE only detections (with no photons detected above $\sim100\;$MeV). 
Finally, the self consistency of the $f_{BG}$ (and \citealt{2008ApJ...677...92G}) model requires that $\tilde{\tau}_{T,\pm}<1$ and therefore $\Gamma_0/\Gamma_{\rm max} = \Gamma_{\rm\gamma\gamma,min}(E_c)/\Gamma_{\rm\gamma\gamma,max}(E_c) < 1$. This is satisfied, at least marginally, in all time bins (with the possible exception of time bin 3 in GRB~100724B). We conclude that $f_{BG}$ is a viable interpretation for both GRBs.\\

\begin{figure*}
\gridline{\fig{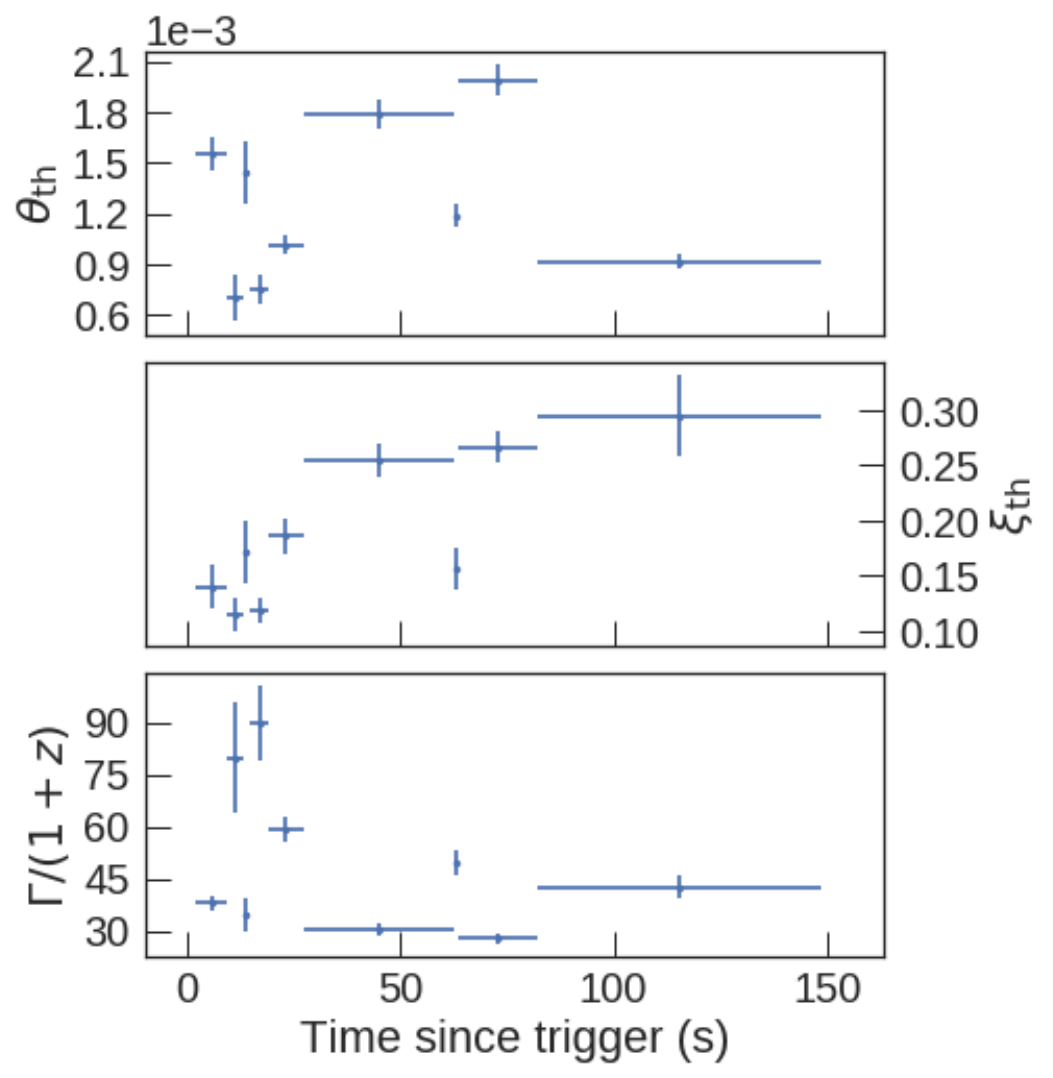}{0.5\textwidth}{}
          \fig{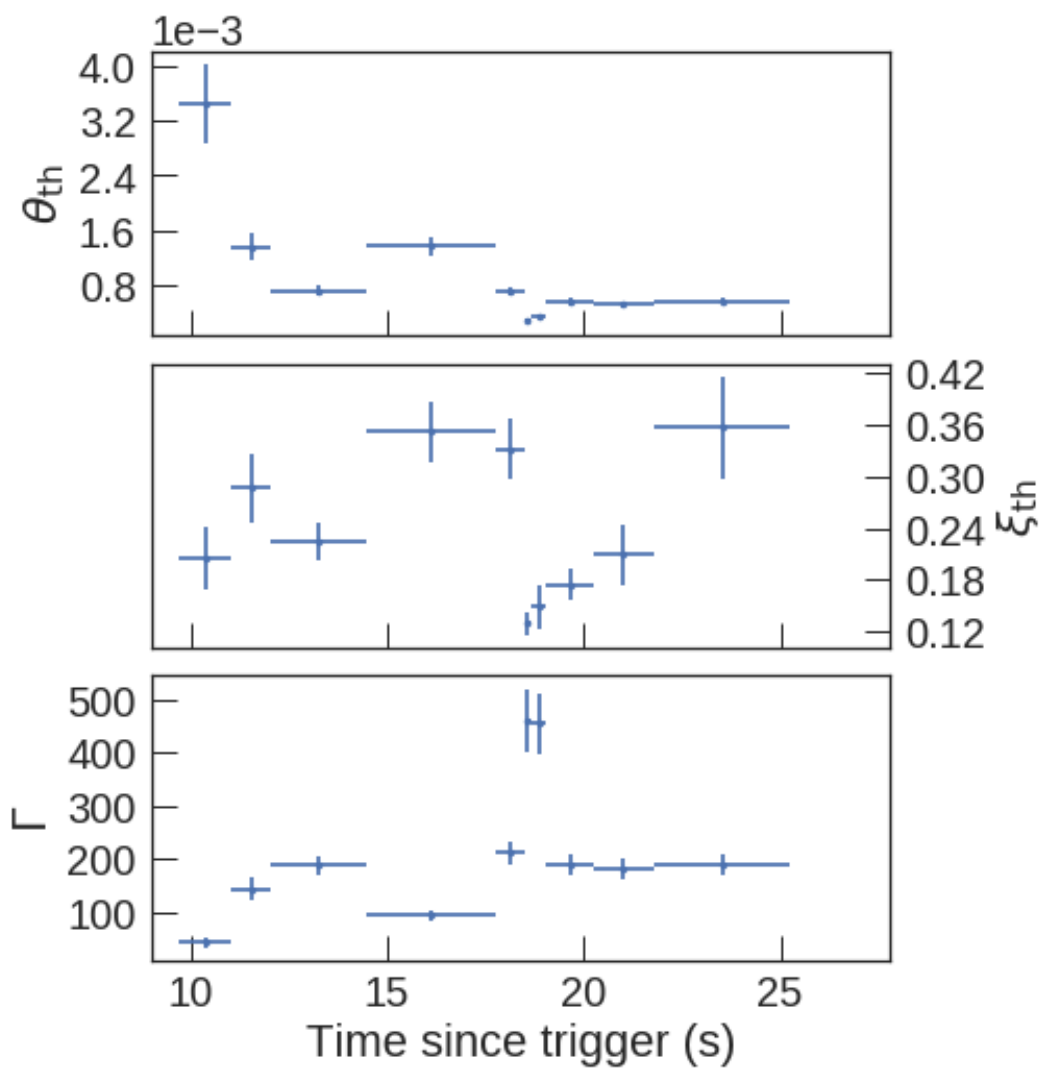}{0.5\textwidth}{}}
\caption{Best fit parameters of the $f_{GT}$ model for the main emission episodes of GRB~100724B (left) and GRB~160509A (right). Here $\theta_{\rm th}=k_BT'_{\rm th}/m_ec^2$ 
is the non-dimensional comoving temperature of the quasi-thermal radiation field and 
$\xi_{\rm th}=\ell'_{\rm th}/\ell'_{\rm heat}$ is the ratio of the quasi-thermal radiation 
field compactness to the heating compactness. Smaller $\xi_{\rm th}$ corresponds to 
larger heating compactness, and therefore, harder final spectra.}
\label{fig:ram_results}
\end{figure*}

The best fit parameters of the $f_{GT}$ model for both GRBs are shown in Figure~\ref{fig:ram_results}. 
The bulk-LF of the outflow in the $f_{GT}$ model was  determined by laterally shifting the comoving spectrum in energy by the factor $\Gamma/(1+z)$ to fit the observed spectrum. We show the temporal evolution of $\Gamma$ for both GRBs in Figure \ref{fig:ram_results}, where the fluctuations in $\Gamma$ are correlated with fluctuations in the observed flux, and therefore luminosity, for both bursts. In the case of GRB 160509A, this behavior clearly coincides with the two broad peaks observed in the BGO and LLE emission. Since there is no redshift available for GRB 100724B, $\Gamma$ could only be determined in the engine-frame. Assuming a typical redshift of $z\sim 2$ we obtain $\Gamma\sim 90 - 270$ throughout 
the entire prompt emission phase. For GRB 160509A, $\Gamma$ varies by a factor 
$\sim 5$ during the prompt phase and peaks at $\Gamma\lesssim 500$. In light of the fact that the underlying numerical model is one-zone, such a high value for 
$\Gamma$ is typically found from one-zone estimates \citep[e.g.][]{2001ApJ...555..540L} as compared to that obtained 
from models of \citet{\Granotpaper} and \citet{\Hascoetpaper}.

In the $f_{GT}$ model, as the outflow expands to larger radii, the comoving 
temperature of the quasi-thermal radiation field should drop due to adiabatic 
cooling. This behavior is clearly seen in the evolution of $\theta_{\rm th}$ 
in the case of GRB 160509A; the existence of a similar trend is less clear for 
GRB 100724B. 

The appearance of a quasi-thermal spectrum at smaller radii 
and, consequently, larger $\theta_{\rm th}=k_BT_{\rm th}/m_ec^2$ is quite 
naturally explained in the $f_{GT}$ model. Such an emission, with no 
high-energy component, is expected to escape from optically thin regions of 
the outflow before any dissipation has occurred. Since it originates at smaller 
radii, it should arrive at the observer earlier than the main burst. Thus, we 
associate this quasi-thermal component to the precursors observed in both GRBs. In particular, the precursor of GRB~100724B can be described well with the quasi-thermal spectrum predicted by the GT model $f_{\rm th}$ (in the observer-frame) in \eqt{\ref{eq:mg14_qt}}, yielding $\alpha = -0.79^{-0.17}_{+0.20}$ and $kT = 43^{-5}_{+7}\;$keV. The precursor of GRB~160509A can be similary fit, with best fit parameters $\alpha = -1.13 \pm 0.07$ and $kT = 87 \pm 21\;$keV.

\subsection{Comparison to other \Fermi/LAT GRBs}
\label{sec:comp-othetr-GRBs}

\begin{figure*}
\centering
\includegraphics[width=0.49\textwidth]{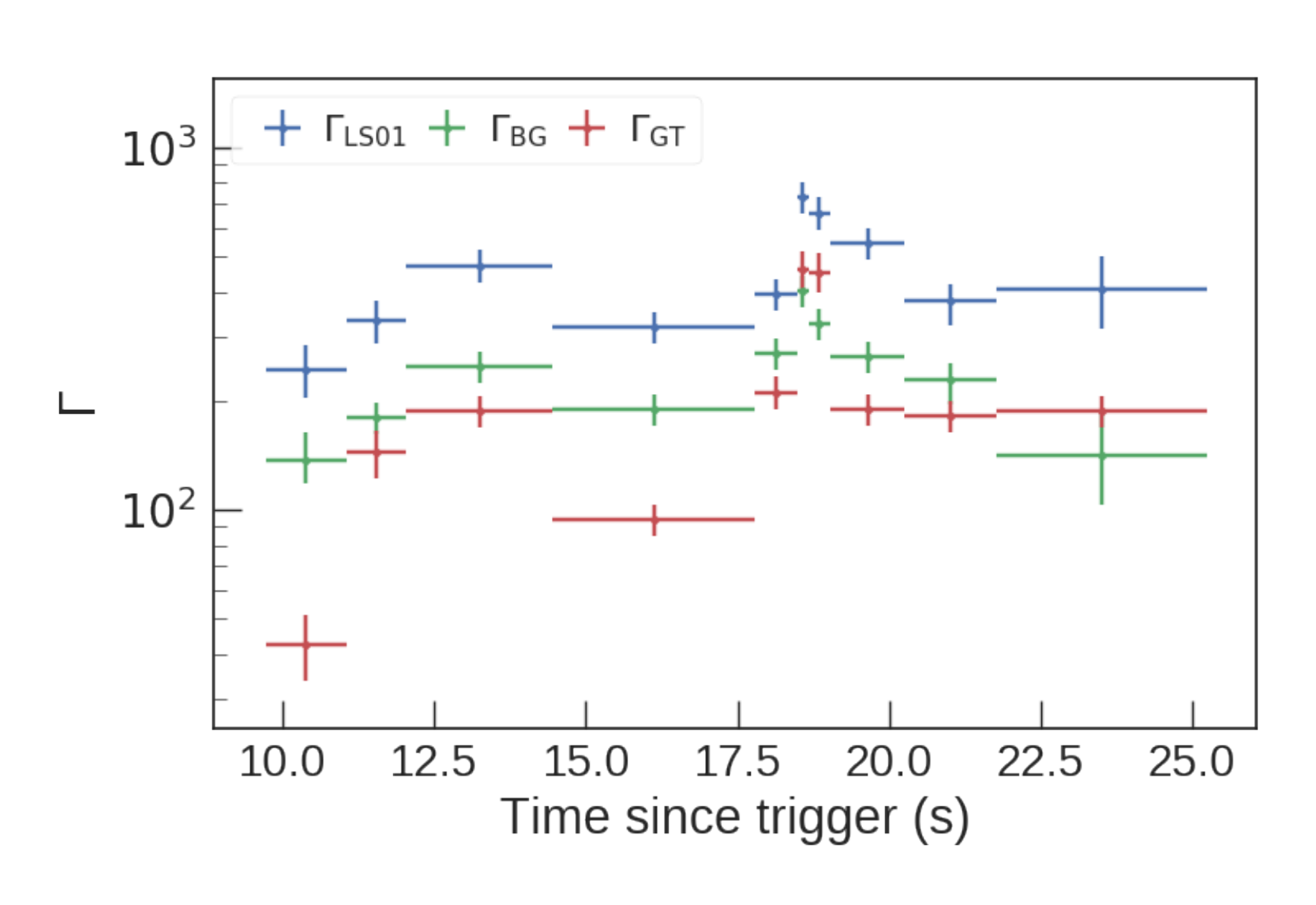}\quad
\includegraphics[width=0.49\textwidth]{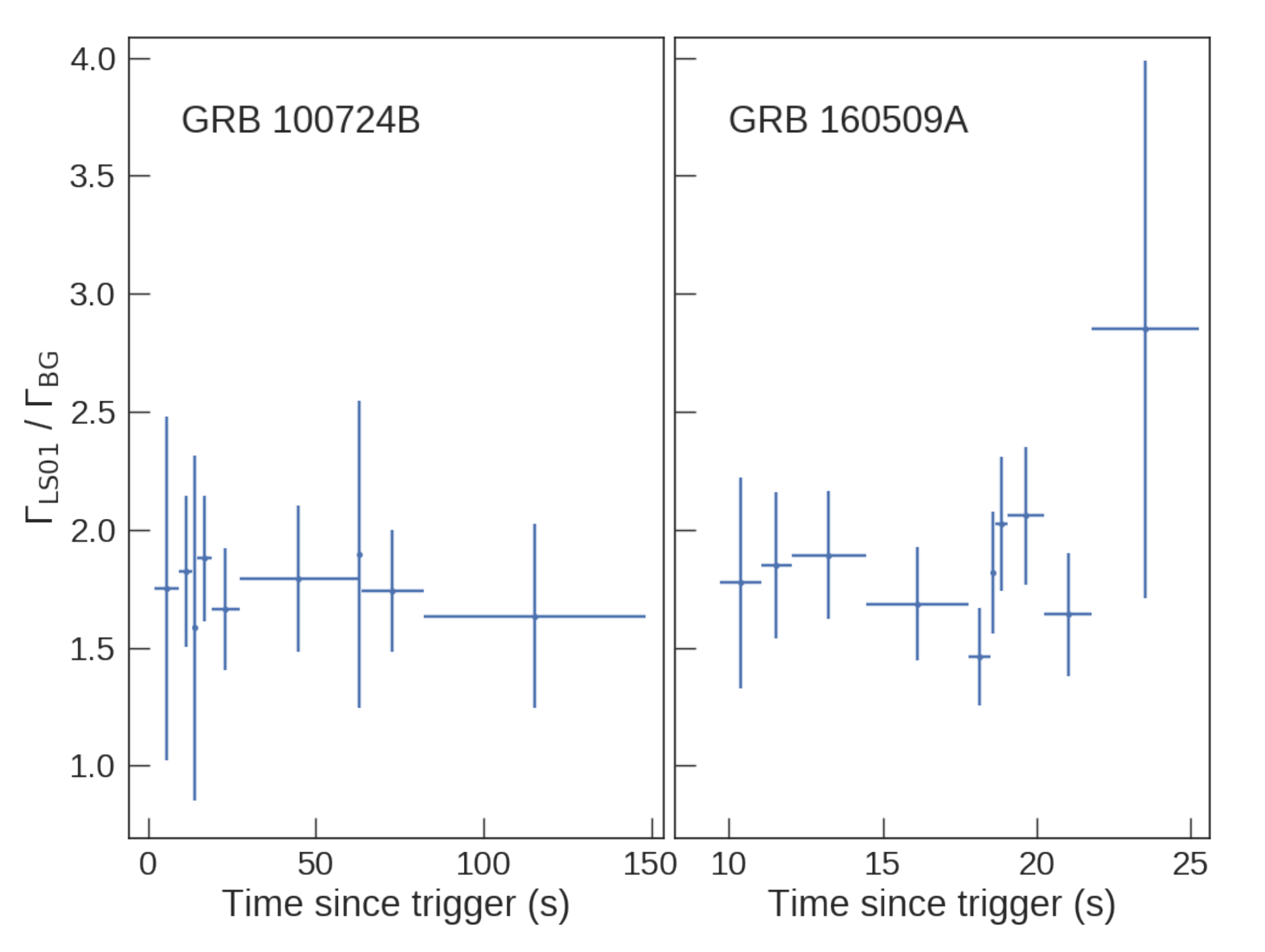}
\caption{(\textbf{Left}) Comparison of the bulk-LFs obtained from three different models for GRB 160509A. The three models are the ($f_{BG}$) semi-analytic model of \citet{\Granotpaper}, ($f_{GT}$) numerical model of \citet{GT14}, and (LS01) 
analytic model of Lithwick \& Sari (2001). For all time intervals, LS01 consistently yields the highest $\Gamma$. (\textbf{Right}) Ratio of $\Gamma$ obtained from the LS01 model to that obtained from the $f_{BG}$ model.}
\label{fig:Gamma-all-models}
\end{figure*}

The most striking property of GRBs 100724B and 160509A is the clear need for a high-energy spectral cutoff in their prompt emission with respect to the extrapolation of the low-energy component. For GRB 100724B the cut-off energy in its time-resolved spectrum typically lies in the range $E_c\sim 20 - 60\;$MeV with high statistical significance, and in the case of GRB 160509A the cutoff typically appears at energies $E_c\sim80 - 150\;$MeV.

In earlier LAT GRBs, for example GRB~080825C, there was marginal evidence for a cutoff at an energy of $E_c\sim 1.8\;$MeV \citep{2009ApJ...707..580A}, which if true does not have a good natural explanation. In GRB~090926A \citep{2011ApJ...729..114A} there was a high-energy spectral cutoff at $E_c\sim1.4\;$GeV in the time-integrated spectrum, and at $E_c\sim0.4\;$GeV in one time bin of the time-resolved spectrum, which has been nicely interpreted as arising due to intrinsic opacity to pair production in the source, in which case it implies a bulk-LF of $\Gamma\sim 300-700$ for the prompt emission region, depending on the exact model assumptions about the emission. 

The upper end of this range corresponds to a simple one-zone model in which the radiation in the outflow's frame is uniform, isotropic and time-independent. In this case, if the photon number spectrum is described by a power-law for photon energies $\varepsilon_{\rm pk} < (\varepsilon\equiv E/m_ec^2) < \varepsilon_c$, such that $f(\varepsilon) = f_{\rm pk}(\varepsilon/\varepsilon_{\rm pk})^{\beta}$, where $\varepsilon_{\rm pk}$ and $\varepsilon_c$ are, respectively, the dimensionless peak and cutoff energies, then an estimate of the bulk-$\Gamma$, which corresponds to the condition that $\tau_{\gamma\gamma}(\varepsilon>\varepsilon_c)>1$, is given by
\citet[][LS01 hereafter; \eqt{5} therein]{2001ApJ...555..540L},
\ba
\label{eq:LS-Gamma}
\Gamma &&= (1+z)^{\frac{-1-\beta}{1-\beta}}\left(\frac{11}{180}\frac{m_e\sigma_T d_L^2}{(-1-\beta)t_{\rm mv}}
f_{\rm pk}\varepsilon_{\rm pk}^{-\beta}\varepsilon_c^{-1-\beta}\right)^{\frac{1}{2-2\beta}} \\
&&= 323 \left(\frac{1+z}{3}\right)^{3/8}
\left(\frac{f_{\rm pk,7}}{t_{\rm mv,-1}}\right)^{5/32}d_{L,28.7}^{5/16}\varepsilon_{\rm pk,-0.3}^{11/32}\varepsilon_{c,2.3}^{3/16}\quad,\nonumber
\ea
where the numerical values are for $\beta = -2.2 = -11/5$, which is typical of the values measured for the prompt GRB. The above estimate also assumes a redshift of $z=2$, luminosity distance $d_L=4.8\times10^{28}d_{L,28.7}$~cm, variability time $t_{\rm mv}=10^{-1}t_{\rm mv,-1}$~s, peak photon number flux $f_{\rm pk}=10^7f_{\rm pk,7}~{\rm cm}^{-2}~{\rm s}^{-1}~{\rm erg}^{-1}$ at a peak energy $E_{\rm pk}=250\varepsilon_{\rm pk,-0.3}$~keV, and cutoff energy $E_c=100\varepsilon_{c,2.3}$~MeV. 

We use the analytic one-zone method of LS01 to calculate the bulk-$\Gamma$ for the case of GRB 160509A and compare it with $\Gamma$ obtained from the $f_{BG}$ and $f_{GT}$ model in Figure \ref{fig:Gamma-all-models}. The right panel of Figure~\ref{fig:Gamma-all-models} shows how the LS01 method of estimating bulk-$\Gamma$ generally yields a value that is a factor $1.5-2$ times higher than that given by a fully time-dependent model (which yields the lower limit on $\Gamma$ in the case of GRB 090926A) where the radiation field starts from zero at the emission onset and is calculated self-consistently as a function of time, space and direction \citep{2008ApJ...677...92G,2012MNRAS.421..525H,GG17}. In GRB~110731A there is a (slightly marginal) detection of a cutoff at $E_c\sim 0.4\;$GeV, which similarly implies $\Gamma\sim 300-600$ if interpreted as due to intrinsic pair production in the source \citep{2013ApJ...763...71A}. 

\begin{table}
\centering
\begin{tabular}{l|c|c|c|c}
\hline
& 080916C & 090926A & 090510 & 160509A \\
\hline
$z$ & 4.35 & 2.1062 & 0.903 & 1.17 \\
$E_c^*$~(GeV) & $3.0^*$ & 0.4 & $30.5^*$ & 0.08 \\
$\Gamma_{\rm min}^{**}$ & $887\pm21^a$ & $720\pm76^b$ & $1218\pm61^c$ & - \\
$\Gamma_{BG}$ & 451 & 319 & 628 & 363 \\
$\beta$ & $-2.21$ & $-1.71$ & $-1.85$ & $-1.60$ \\
$L_{0,52}~({\rm erg/s})$ & 55.78 & 3.42 & 1.24 & 6.18 \\
$t_{v,z}$~(ms) & 374 & 48 & 6.3 & 23 \\
\hline
$\psi_{(1+z)}$ & $-0.20$ & $-0.18$ & $0.02$ & \\
$\psi_\Gamma$ & $0.41$ & $-0.58$ & $0.81$ & \\
$\psi_\beta$ & $0.58$ & $0.41$ & $0.26$ & \\
$\psi_{L_{0,52}}$ & $-0.81$ & $0.50$ & $0.46$ & \\
$\psi_{t_{v,z}}$ & $1.03$ & $0.63$ & $-0.37$ & \\
\hline
\end{tabular}
\caption{Ratio of intrinsic parameters as described in \eqt{\ref{eq:ratios}} with 
GRB 160509A as the reference system. In systems where no spectral cutoff was observed 
(marked with ${}^*$), the maximum observed photon energy is quoted. ${}^{**}$ Minimum bulk-$\Gamma$ (or the actual inferred $\Gamma$ when a cutoff is observed) calculated using a one-zone analytical model employing a more elaborate radiation field spectrum as compared to a simple power-law used in LS01. In all three cases, the inferred $\Gamma_{\rm min}\sim2\Gamma_{BG}$:~${}^a$\citet{Abdo+09c}, ${}^b$\citet{2011ApJ...729..114A}, ${}^c$\citet{Ackermann+10}}
\label{tab:ratios}
\end{table}

\begin{figure}
\centering
\includegraphics[width=0.49\textwidth]{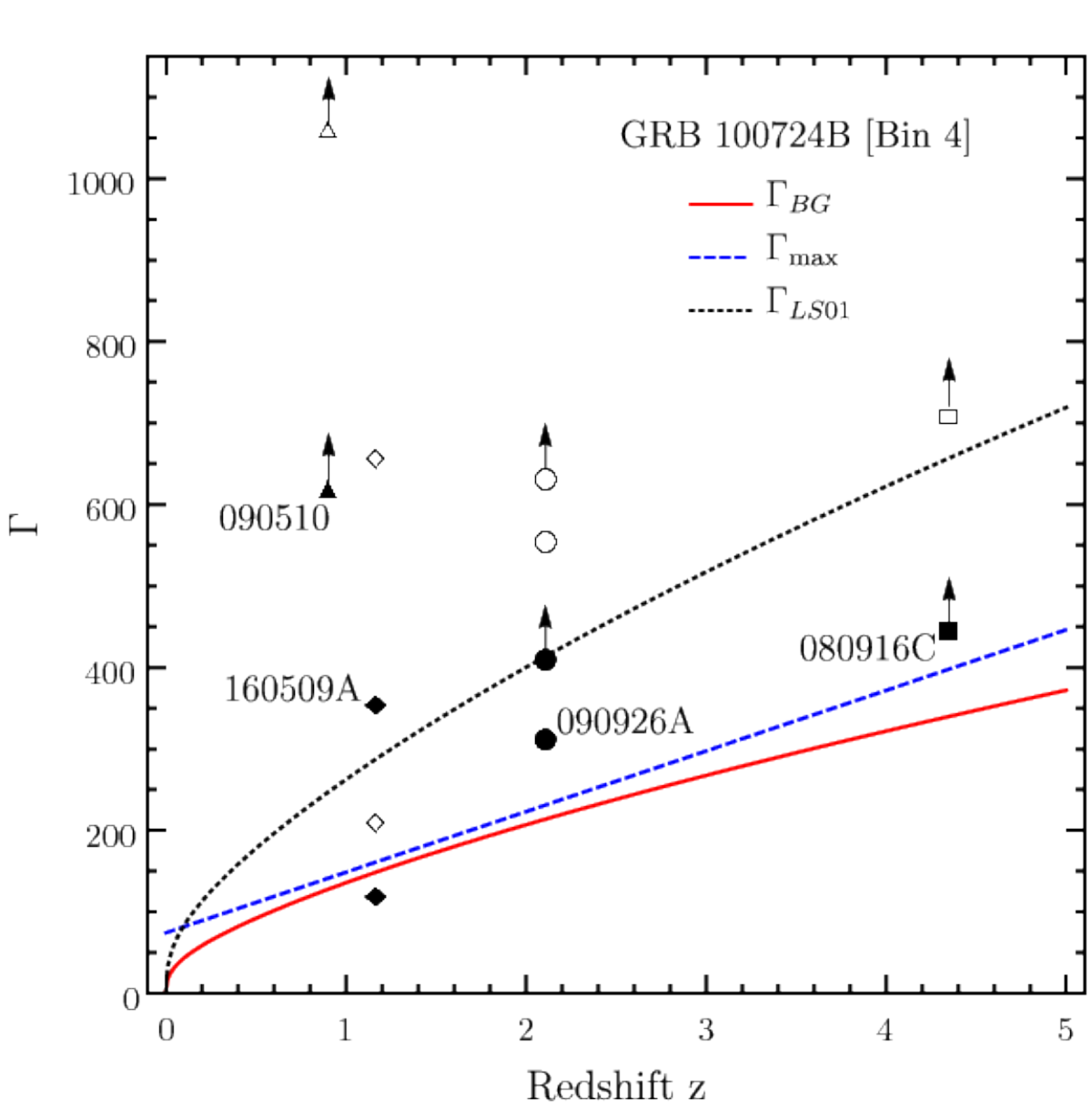}
\caption{Bulk-$\Gamma$ for GRB 100724B and other GRBs obtained from the $f_{BG}$
(filled symbols) and LS01 (open symbols) models. Lower limits 
($\Gamma = \Gamma_{\gamma\gamma,\rm min}$) are indicated by arrows for those GRBs 
(or for a given time-resolved spectrum) that did not show any spectral cutoff. 
For GRB 160509A, the smallest and largest $\Gamma$ estimates are shown. In the case 
of GRB 090926A, we show the $\Gamma$ estimate for time-bin (c) that showed a cutoff 
at $E_c = 0.4$~GeV. The solid red line shows the redshift evolution of the 
bulk-$\Gamma$ for GRB 100724B as obtained from the $f_{BG}$ model, while the estimate 
from LS01 for the same GRB is shown using the dotted black line. The dashed blue line 
shows $\Gamma_{\gamma\gamma,\rm max}(E_c)$.}
\label{fig:100724B-Gamma-z}
\end{figure}

In Figure \ref{fig:100724B-Gamma-z} we compare the bulk-$\Gamma$ estimates obtained from the $f_{BG}$ and LS01 models for several GRBs. Since GRB 100724B lacks redshift information, we show the evolution of $\Gamma$ with redshift. Other \Fermi/LAT detected GRBs that do not show any spectral cutoff, namely GRBs 080916C and 090510, are also shown. However, for these $\Gamma$ should be interpreted as a lower limit $\Gamma_{\gamma\gamma,\rm min}$. 

There are large differences in the observed cutoff energies $E_c$ between different GRBs (see, e.g., Table~\ref{tab:ratios}).
To understand which properties among the 
different GRBs are leading to different cutoff energies (or lack thereof, in which case 
the highest energy observed photon was used), we can express 
\eqt{\ref{eq:lorentz_factor_bg}} (which relies on Eq.~(126) of \citealt{\Granotpaper}; see footnote~\ref{foot:correction}) 
in terms of the intrinsic parameters, such that the cutoff energy in the central engine frame 
(quantities in this frame are expressed with a subscript $z$) is
\ba
E_{c,z} =&& (1+z)E_c 
\\ \nonumber
=& &5.11\left[\frac{C_2}{13.2}\left(\frac{-\beta}{2}\right)^{5/3}
\frac{t_{v,z}}{1\,{\rm s}}\,\frac{\Gamma_2^{2-2\beta}}{L_{0,52}}\right]^{-1/(1+\beta)}{\rm GeV}\ ,
\label{eq:Ecz}
\ea
where $t_{v,z}=t_v/(1+z)$. 
We compare each given GRB (subscript `$i$') with GRB~160509A (in particular 
the results of time-bin 6; subscript `$0$') where we quantify the effect of 
a change in the parameter $\xi = \{\Gamma,~\beta,~L_{0,52},~t_{v,z},~z\}$
as follows (see Table~\ref{tab:ratios}),
\be
\psi_{\xi,i} = \frac{\log(E_{c,0}(\xi_i)/E_{c,0})}{\log(E_{c,z,i}/E_{c,z,0})}\ ,
\label{eq:ratios}
\ee
where $E_{c,z,0}(\xi_i)$ is the value of $E_{c,z}$ for GRB~160509A obtained from \eqt{\ref{eq:Ecz}} by replacing the parameter $\xi$ by its value for GRB $i$ and keeping all other parameters fixed to their measured values for GRB~160509A. In GRBs where a cutoff was not observed, and instead only a lower limit on $E_c$ was derived, which was used to derive a lower limit on $\Gamma$ (GRBs 080916C and 090510), we use these lower limits for this comparison. 

When examining the origin of the large differences in the observed cutoff energies 
$E_c$ between different GRBs, we find that the effect due to differences in redshifts
between the different GRBs is sub-dominant, as can be seen in Table~\ref{tab:ratios}. This further implies that typically most of the differences between the observed break energies $E_c$ are intrinsic (i.e. the differences in $E_{c,z}$).

One might naively expect that the dominant intrinsic parameter that would account for the different $E_{c,z}$ and $E_c$ values would be the Lorentz factor, $\Gamma$, since in \eqt{\ref{eq:Ecz}} it appears with a power larger than that of $t_{v,z}/L_{0,52}$ by a factor of $2-2\beta\sim6$ (for typical values of $\beta\sim-2$). However, out of the three GRBs we considered for comparison here, which have a disparate set of intrinsic properties, only in one  of them, GRB~090510, does it appear to be dominant (by a factor of $\sim 2$) over the effects of the differences in $t_{v,z}$ and $L_{0,52}$, while in  GRB~090926A it has a comparable effect and in GRB 080916C it is sub-dominant.  This occurs since despite the larger power of $\Gamma$ in the expression for $E_c$, it varies by a smaller factor between different GRBs compared to $t_{v,z}$ and $L_{0,52}$. The dependence of $E_c$ and $E_{c,z}$ on $\beta$ is non-trivial (see \eqt{\ref{eq:Ecz}}), but the effect of its variation between different GRBs is typically comparable to that of the other physical parameters. The comparison of the intrinsic properties of different GRBs in 
Table~\ref{tab:ratios} ultimately shows that it is likely that differences in many of 
these properties jointly contribute to the appearance of high-energy spectral cutoffs 
(or lack thereof). This obviously constitutes a broad set of possibilities, and it is 
not unlikely to find GRBs with spectral cutoffs where only the difference in bulk-$\Gamma$ makes the dominant contribution.

Before GRB~100724B \citep{LATGRBcat} there was no clear direct evidence for a high-energy spectral cutoff at an energy $E_c\ll 1\;$GeV. The prompt GRB spectrum of some GRBs is consistent with a Comptonized spectrum featuring a power-law with an exponential cutoff, but with a typical peak energy $E_p\lesssim 1\;$MeV, so this cannot really be considered as a {\it high-energy} cutoff, and most likely has a different physical origin. On the other hand, there was indirect evidence for a high-energy cutoff at tens of MeV from the extrapolation of the GRB spectrum and LAT upper-limits \citep{2011MNRAS.416.3089B,2011A&A...525A..53G,2012ApJ...754..121F}.
Therefore, GRB~100724B shows the first clear-cut detection of a high-energy cutoff 
at well below a GeV. Other \Fermi-detected GRBs were shown to have similar sub-GeV cutoffs by \citet{2015ApJ...806..194T}.


\section{Discussion}\label{sec:dis}

\textbf{GRBs must be compact sources in order for their short variability time scales to be explained, and at the same time they must be optically thin above the pair production threshold in order to produce the observed high-energy $\gamma$-rays. This is the so-called ``compactness'' problem \citep{1999PhR...314..575P}, which in the fireball model is solved by assuming a high bulk Lorentz factor $\Gamma$ for the emitting shells. However, $\Gamma$ cannot be arbitrarily large; thus the system will become optically thick above a certain energy, producing a spectral cutoff. The observed high-energy spectral cutoffs in both GRBs (100724B and 160509A) can therefore be interpreted this way. Both theoretical models considered in this paper ($f_{BG}$ and $f_{GT}$) produce a high-energy spectral cutoffs due to pair production opacity, and they describe the data well. Still, it is prudent to ask if instead the spectral cutoffs can be explained due to some intrinsic limitation of the emission process in producing high-energy photons, which would introduce a cutoff at an energy different (and possibly lower) than the cutoff expected from the pair-production opacity.}

One such case, 
for example, would be if the underlying mechanism for the prompt emission 
was synchrotron. Then an exponential cutoff would be observed at fluid-frame energies 
$E' > E'_{\rm syn,max} = \gamma_{e,\rm max}^2(B'/B_Q)m_ec^2$, where 
$\gamma_{e,\rm max}m_ec^2$ is the maximum energy to which electrons are accelerated in 
the dissipation region, $B'$ is the local magnetic field in the fluid frame, and 
$B_Q = m_e^2c^3/e\hbar$ is the quantum critical field. In this case, $\gamma_{e,\rm max}$ 
will be limited if the synchrotron cooling time of electrons, 
$t'_{syn} = 6\pi m_ec/(\sigma_T B'^2\gamma_e)$, is shorter than their acceleration 
time. The shortest viable length scale over which electrons are accelerated is given by their Larmor radius, $r_L' = \gamma_e m_ec^2/(eB')$, which yields an acceleration time 
$t_{\rm acc}' = r_L'/c$. Comparison of the two timescales gives the factor 
$\gamma_{e,\rm max}^2B' = 6\pi e/\sigma_T$, where $\sigma_T = (8\pi/3)(e^2/m_ec^2)^2$ is 
the Thomson cross-section. From this we immediately find that the maximum energy of 
synchrotron photons is $E'_{\rm syn,max} = (9/4\alpha_F)m_ec^2$, where 
$\alpha_F = e^2/\hbar c = 1/137$ is the fine-structure constant 
\citep[e.g.][]{1983MNRAS.205..593G,1992ApJ...396..161D,2010ApJ...718L..63P,2013ApJ...774...76A}.
In the observer frame, this limiting energy translates to
\be
E_{\rm syn,max} = \frac{\Gamma}{(1+z)}\frac{9}{4\alpha_F}m_ec^2 = 0.159~\Gamma~(1+z)^{-1}~{\rm GeV}~.
\ee
It is much higher than the cutoffs observed in the two GRBs discussed in this work. In other GRBs photons approach or even exceed $E_{\rm syn,max}$, which suggests very efficient electron acceleration. Since the efficiency of electron acceleration is not expected to drastically change between different GRBs, this would support a different origin for the high-energy cutoffs in GRBs 100724B and 160509A, in agreement with our interpretation of an intrinsic opacity to pair production origin. 

In most GRBs there is no observed high-energy cutoff with respect to the extrapolation of the low-energy component, so the maximal observed photon energy $E_{\rm max}$ is used as a lower limit for any possible cutoff energy, which in turn sets a lower limit, $\Gamma_{\gamma\gamma,{\rm min}}$, on the LF of the emitting region, $\Gamma$, through the condition that $\tau_{\gamma\gamma}(E_{\rm max})<1$. In cases where a high-energy cutoff is actually observed at an energy $E_c$ and may be attributed to intrinsic pair production,
then $\Gamma_{\gamma\gamma,{\rm min}}$ can serve as an actual estimate of $\Gamma$ when $E_{\rm max}$ is replaced by $E_c$, as shown in \eqt{\ref{eq:LS-Gamma}}.  For the $f_{BG}$ model $\Gamma_{\gamma\gamma,{\rm min}}(E_c)$ is given by \eqt{\ref{eq:lorentz_factor_bg}}.

This is valid only as long as $\Gamma_{\gamma\gamma,{\rm min}}$ is lower than
\be
\label{eq:Gamma_max}
\Gamma_{\gamma\gamma,{\rm max}}(E_c) = (1+z)\frac{E_c}{m_e c^2} = 
176\left(\frac{1+z}{3}\right)\left(\frac{E_c}{30\;{\rm MeV}}\right)\ , 
\ee
for which $E_c$ corresponds to a comoving photon energy of $E'_c > m_e c^2$, so that photons near the cutoff can
pair-produce with other photons of comparable energy. When the effects of pair cascades on the spectral cutoff are 
ignored, then for a given $\Gamma$ the cutoff energy always satisfies
\be
E_c \geq E_{sa} = \frac{\Gamma}{1+z}m_e c^2 = 51.1\left(\frac{3}{1+z}\right)\left(\frac{\Gamma}{300}\right)\;{\rm MeV}\ ,
\ee
as long as the high-energy photon index is lower than $-1$ ($\beta < -1$ where $dN_{\rm ph}/dE\propto E^\beta$, as is almost always the case in GRB prompt spectra), so that $dN_{\rm ph}/d\log E = E(dN_{\rm ph}/dE)$ decreases with photon energy $E$. Here $E_{sa}$ is the minimal energy of photons that can ``self-annihilate", i.e. interact with other photons of the same energy.
This occurs because of the following reason. Let us denote by $E_1$ the photon energy above which the optical depth to pair production is large, $\tau_{\gamma\gamma}(E_1)=1$, and also denote by $\varepsilon=E/m_ec^2$ and 
$\varepsilon'=\varepsilon(1+z)/\Gamma = E/E_{sa}$ the observer and comoving frame photon energies in units of the electron rest energy.

If $E_1 >E_{sa} \Leftrightarrow \varepsilon'_1>1\Leftrightarrow \Gamma_{\gamma\gamma,{\rm min}}(E_c)<\Gamma_{\gamma\gamma,{\rm max}}(E_c)$ 
then the Thomson opacity of the $e^\pm$ pairs that are produced is small \citep{GG17},
\be
\tau_{T,\pm}\sim (\varepsilon'_1)^{2(1+\beta)}=
\left[\frac{\Gamma_{\gamma\gamma,{\rm min}}(E_c)}{\Gamma_{\gamma\gamma,{\rm max}}(E_c)}\right]^{2(-1-\beta)}\ll 1\ ,
\ee
and can therefore be ignored for our purposes. Therefore, we call this the ``thin'' regime. In this regime
 there will be a cutoff at $E_c=E_1>E_{sa}\Leftrightarrow\varepsilon'_c = \varepsilon'_1>1$. 

However, if $E_1<E_{sa}\Leftrightarrow\varepsilon'_1<1\Leftrightarrow\Gamma_{\gamma\gamma,{\rm min}}(E_c)>\Gamma_{\gamma\gamma,{\rm max}}(E_c)$ then the situation changes. In this regime photons in the energy range $E_1 < E < E_{sa}\Leftrightarrow\varepsilon'_1<\varepsilon'<1$ will on the one 
hand have a large initial optical depth to pair production, $\tau_{\gamma\gamma}(E)\sim (E/E_1)^{-1-\beta} = (\varepsilon'/\varepsilon'_1 )^{-1-   \beta}>1$.
However, on the other hand since they can only pair-produce or annihilate with photons of energy 
$\varepsilon'_{\rm an}\geq1/\varepsilon'>1>\varepsilon'$, and there are fewer such photons, they all quickly annihilate, 
so that the optical depth $\tau_{\gamma\gamma}(E)$ rapidly drops below its initial value to well below unity,  
and most of the photons in this energy range remain with no photons that they can pair-produce with. Therefore, only
photons of energy $E>E_{sa}\Leftrightarrow\varepsilon'>1$ can fully annihilate, and there is a cutoff only at 
$E_c=E_{sa}>E_1\Leftrightarrow\varepsilon'_c=1>\varepsilon'_1$. The latter implies that in this regime $\Gamma = \Gamma_{\gamma\gamma,{\rm max}}(E_c)$,
i.e. the bulk-LF of the emission region is given by \eqt{\ref{eq:Gamma_max}}. 
In this regime the Thomson optical depth 
of the pairs that are produced (neglecting their possible annihilation, hence the tilde in its notation) is large 
\citep{GG17},
\be
\tilde{\tau}_{T,\pm}\sim (\varepsilon'_1)^{1+\beta}=
\left[\frac{\Gamma_{\gamma\gamma,{\rm min}}(E_c)}{\Gamma_{\gamma\gamma,{\rm max}}(E_c)}\right]^{2(1-\beta)}\gg 1\ .
\ee
Therefore, we call this the ``thick''  regime.

Since typically $\beta \sim -2$, $\tilde\tau_{T,\pm}$ in the thick regime scales as a fairly high power ($\sim 6$) of the ratio
$\Gamma_{\gamma\gamma,{\rm min}}(E_c)/\Gamma_{\gamma\gamma,{\rm max}}(E_c)$, so even if 
$\Gamma_{\gamma\gamma,{\rm min}}(E_c)$ exceeds $\Gamma_{\gamma\gamma,{\rm max}}(E_c)$ only by a factor of a few 
one might already have $\tilde\tau_{T,\pm}\gtrsim 10^2$. In this case, the spectrum is modified due to Compton scattering by cold pairs, which also brings the cutoff energy below $E_{sa}$ \citep{GG17}. In addition, large 
$\tilde\tau_{T,\pm}$ decreases the radiative efficiency due to adiabatic losses and washes away much 
of the temporal variability, as photons must diffuse out of the emission region
on a diffusion time larger than the dynamical time.  On the other hand, in this regime pair annihilation can become important, and the expansion of the emitting region also dilutes its opacity 
and causes a non-isotropic photon distribution in the comoving frame, which further suppresses pair production.

Altogether the results for the LF and cutoff energy in the two regimes 
discussed above can be summarized by
\be
\Gamma \approx \min\left[\Gamma_{\gamma\gamma,{\rm min}}(E_c),\,\Gamma_{\gamma\gamma,{\rm max}}(E_c)\right]\ . 
\ee
\be
E_c \approx \max\left[E_{sa}(\Gamma),E_1(\Gamma)\right]\ . 
\ee



\subsection{$f_{GT}$ vs $f_{BG}$ Model}
Spectrally, the one 
important place where the $f_{GT}$ model differs from the $f_{BG}$ model is where the cutoff energy lies with respect to $m_ec^2$. If the comoving radiation field compactness is 
high the $f_{GT}$ model always yields the comoving frame cutoff energy $E_c' < m_ec^2$, such 
that $\Gamma_{\gamma\gamma,\rm min}(E_c) > \Gamma_{\gamma\gamma,\rm max}(E_c)$. As 
argued above, this consequently would yield a high pair Thomson depth and significantly alter the observed spectrum and temporal variability. However, since pair-production 
and pair-annihilation effects are self-consistently accounted for in the $f_{GT}$ model, 
the $e^\pm$ pair Thomson depth is always regulated to $\tau_{T,e^\pm}\sim 1 - 5$ in the 
dissipation region. What the $f_{GT}$ model does not account for is the pair opacity 
accumulated over the line of sight as the photon travels from its emission point to the 
observer while interacting with other photons en route. This is the essence of the $f_{BG}$ 
model. Still, this additional pair opacity effect will not significantly alter the spectrum obtained in 
the $f_{GT}$ model as the high-energy spectrum is already exponentially suppressed due 
to pair-production in the dissipation region. As a result, the cutoff energy cannot be made  appreciably
smaller due to additional pair-opacity effects.

When comparing the two spectral models, we find that the $f_{BG}$ model yields $\Gamma$ values
that are on average comparable to that obtained from the one-zone $f_{GT}$ model. 
Both models have additional parameters, other than the ones used for fitting in this 
work, that introduce some degeneracy in the final outcome.

A potential test for both models is that photons above 
the cutoff energy $E_c$ are expected to arrive preferentially near the beginning of 
pulses in the lightcurve, as compared to near their peak or during their tails. However, 
this requires good photon statistics within a single spike of the lightcurve, which was 
not available so far with \Fermi/LAT but may become possible in the future with the 
Cherenkov Telescope Array (CTA; e.g. \citealt{2013APh....43..252I}).

\subsection{Comparison with Other Work}
In a recent work, using \textit{Swift} X-ray data along with ground-based optical, infrared
and radio data, \citet{Laskar+16} analyzed the afterglow emission and 
determined $\Gamma(t_{\rm dec}) \approx 330$ for GRB~160509A at the deceleration time $t_{\rm dec}\approx460\;$s for a constant density circumburst environment ($k=0$). In comparison, 
we find a mean bulk-LF of $\Gamma_{\rm BG} \sim 220$ from the $f_{BG}$ model 
and $\Gamma_{GT}\sim 215$ from the $f_{GT}$ model over the entire duration of the prompt phase. The apparent discrepancy between our results and that of \citet{Laskar+16} critically
depends on the density profile of the circumburst medium, $\rho_{\rm ext}\propto R^{-k}$. They find a 
much lower $\Gamma(t_{\rm dec}) = 34$ (and $t_{\rm dec}\approx170\;$s) for the wind case ($k=2$), where the density of the surrounding 
medium is determined by stellar winds from the progenitor star. The actual value of 
$\Gamma$ may be somewhere in the middle depending on the value for $k$, which is likely also intermediate ($0<k<2$), as suggested by some afterglow modelings \citep[e.g.,][]{2013ApJ...779L...1K}, and is viable given the uncertain wind velocity and mass loss rate history at the massive star progenitor's last years (which determine the density profile around the deceleration radius corresponding to the afterglow onset). In that case, the
results of this work would be consistent with that obtained from the multi-wavelength 
afterglow analysis.

Moreover, the effective duration of the prompt emission in GRB~160509A is $\sim20-30\;$s (see Fig.~\ref{fig:bb_160509}), i.e. much shorter than its $T_{90}\approx 370\;$s, which is dominated by a weak and very soft emission episode around $T\sim 300-400\;$s. This may suggest either an earlier deceleration time, $t_{\rm dec}\sim30\;$s for a relativistic reverse shock (or ``thick shell''), or alternatively if $t_{\rm dec}\gg30\;$s the one would expect a Newtonian reverse shock (or ``thin shell"), in which case the correspondingly weaker reverse shock would tend to imply a lower value for $\Gamma(t_{\rm dec})$, which could be consistent with the values we derived for models $f_{BG}$ and $f_{GT}$. 

Finally, for strong internal shocks a good part of the outflow energy may reside in internal energy of the baryons just after the shells collide.  It eventually transforms back to kinetic energy above the internal shock emission radius leading to a larger asymptotic $\Gamma(t_{\rm dec})$ compared to $\Gamma$ of the emitting plasma during the internal shocks themselves. A similar effect may arise in a Poynting-flux-dominated outflow if the emission occurs during the acceleration phase.

In the work of \citet{2015ApJ...806..194T}, a total of eight GRBs (including GRB~100724B) that 
were detected by \Fermi were found to have spectral cutoffs between tens of 
MeV and several 100 MeV. They derived the bulk-$\Gamma$ for these GRBs using a simple one-zone 
analytical model, akin to the LS01 model, and found that for majority of the cases 
$\Gamma > \Gamma_{\gamma\gamma,\rm max}$. This led them to estimate the actual bulk-LF 
by its maximum value given by $\Gamma_{\gamma\gamma,\rm max}$. As shown by \citet{GG17}, estimating 
$\Gamma$ in this way can lead to underestimating its true value by as much as an order of magnitude, since 
in this case the spectral break energy is modified by pair cascades. It is clear from 
Figure \ref{fig:100724B-Gamma-z} that $\Gamma$ obtained from simple one-zone analytic 
models will exceed $\Gamma_{\gamma\gamma,\rm max}$ in the case of GRB 100724B 
(unless $z\ll1$ which is unlikely), whereas the much more detailed and self-consistent 
model of \citet{\Granotpaper} generally yields $\Gamma \lesssim \Gamma_{\gamma\gamma,\rm max}$ 
for all redshifts.

Recently, a sub-photospheric dissipation model of \citet{2006ApJ...642..995P} was used in the work of 
\citet{Ahlgren+15} to fit the time-resolved spectra of GRB 100724B using the code developed 
in \citet{PW05}. The underlying GRB model producing the prompt-phase spectrum has many 
similarities with the model of \citet{GT14}, in particular the continuous and slow-heating of 
electrons which then Compton up-scatter the soft thermal emission to produce the spectrum 
above the peak. The major difference between the two models is that the model of \citet{GT14}
assumes a Poynting-flux-dominated baryon-pure outflow whereas the model advanced in \citet{2006ApJ...642..995P} assumes a kinetic energy dominated baryonic jet. They also find a much larger mean $\Gamma \approx 443$ for GRB 100724B, while assuming a redshift $z=1$, in comparison to $\Gamma_{GT}\approx 180$ and $\Gamma_{BG}\approx10^2$ using the two models considered in this work. More importantly, the model fit in \citet{Ahlgren+15} lacks a spectral cutoff at high energies as sharp as the model considered in this work. Consequently, it yields a poorer fit in the 1 MeV to 1 GeV energy range (compare the upper right panel in Figure 2 in \citet{Ahlgren+15} to Fig. \ref{fig:counts_spectra_100724029}). A spectral cutoff is naturally and self-consistently produced in both the $f_{GT}$ and $f_{BG}$ models.

\section{Conclusions}\label{sec:conc}
We presented a detailed time-resolved analysis of two bright \Fermi/LAT GRBs, GRB~100724B and GRB~160509A, that provide the clearest examples of sub-GeV high-energy cutoffs during the prompt emission. We characterized phenomenologically the high-energy cutoffs, which we measure respectively in the range 20-60 MeV and 80-150 MeV. We have shown through the fitting of two physical models that the observed cutoff can be interpreted as the result of intrinsic opacity to pair production at the source, while it appears to be too low to be explained as originating from the limitation of the particle acceleration process.

In particular, a semi-phenomenological model of an impulsive relativistic outflow with detailed $\gamma\gamma$ opacity computation presented 
in \citet{2008ApJ...677...92G} can describe the data well and self-consistently, yielding estimates for the emission onset radius $R_0\sim 10^{13}-10^{14}\;$cm and for the fractional size of the emission zone $\Delta R/R_0 \sim 1-5$ that are consistent with the internal shocks model. The one-zone photospheric model of \citet{GT14} can also describe the data well. Moreover, it predicts a drop in the comoving temperature of the seed quasi-thermal radiation field which is clearly observed in GRB~160509A (but not in GRB~100724B as the details of the model depend on the redshift, 
which is lacking in this case). The estimate for the bulk Lorentz factors derived by 
using the model of \citet{\Granotpaper} are typically in the range $\Gamma_0 \sim 100-300$, and they are on average comparable to 
those obtained from the one-zone photospheric model of \citet{GT14}. These estimates are a factor of
a few to several smaller than the lower limits derived for bright LAT GRBs and a factor of $\sim2$ smaller than values inferred from high-energy cutoffs, which were generally obtained for LAT-detected GRBs from a one-zone analytical model \citep[see for example][]{2015ApJ...806..194T}. Indeed such a factor of $\sim 2$ difference exists also when deriving $\Gamma$ for the same GRB using different models (see, e.g., Figures~\ref{fig:Gamma-all-models} and \ref{fig:100724B-Gamma-z}). 

Because of opacity to intrinsic pair production, slower GRBs tend to be fainter in the LAT energy range and are therefore more difficult to detect. This may produce a selection bias against deriving lower Lorentz factors from the detection of high-energy cutoffs. We also note that our measurement for $\Gamma_{0}$ is in line with the upper limit estimated in \citet{2017MNRAS.465..811N} for GRBs observed but not detected by the LAT.

We find that the differences in observed cutoff energies $E_c$ between different GRBs are predominantly intrinsic, and arise not only from the different Lorentz factor $\Gamma$ of their emission regions, but also from differences in other intrinsic parameters, namely their variability times $t_{v,z}$, isotropic equivalent luminosities $L_{0,52}$, and high-energy photon index $\beta$.  

The two GRBs analyzed in this work have relatively low inferred Lorentz factors compared to other \Fermi/LAT GRBs. They were still detected by \Fermi/LAT despite their relatively low cutoff energies of $E_c\lesssim100\;$MeV, since they are extremely bright at $\lesssim\;$MeV energies.  This may introduce a bias in the \Fermi/LAT GRB sample against GRBs with low Lorentz factors $\Gamma$, as well as short variability times (corresponding to small emission radii),  as these would lead to low cutoff energies $E_c$, which would make them more difficult to detect with \Fermi/LAT. $E_c$ also decreases as the isotropic equivalent luminosity ($L_{0,52}$) increases, so that highly luminous GRBs would require a higher Lorentz factor in order to be detected 
by  \Fermi/LAT.  This may introduce an apparent positive correlation between 
the isotropic equivalent luminosity $L_{\rm iso}$ and $\Gamma$, such that 
$\Gamma\propto L_{\rm iso}^{1/(2-2\beta)}$ with all else being equal.
A positive correlation between $\Gamma$ and $L_{\rm iso}$ has indeed been claimed in the literature \citep[e.g.][]{2012ApJ...751...49L}.
The possible apparent correlation we point out is not expected to be very tight, and is not expected to appear in the time-resolved spectroscopy of a single GRB (in which such a correlation would most likely be of intrinsic origin). This correlation may be modified by the fact that more luminous GRBs may be detected for 
a slightly lower $E_c$ with possible correlations with $\beta$ or $t_v$.

\acknowledgements

The \Fermi/LAT Collaboration acknowledges generous ongoing support from a number of agencies and institutes that have supported both the development and the operation of the LAT as well as scientific data analysis. These include the National Aeronautics and Space Administration and the Department of Energy in the United States, the Commissariat \`a l'Energie Atomique and the Centre National de la Recherche Scientifique / Institut National de Physique Nucl\'eaire et de Physique des Particules in France, the Agenzia Spaziale Italiana and the Istituto Nazionale di Fisica Nucleare in Italy, the Ministry of Education, Culture, Sports, Science and Technology (MEXT), High Energy Accelerator Research Organization (KEK) and Japan Aerospace Exploration Agency (JAXA) in Japan, and the K.~A.~Wallenberg Foundation, the Swedish Research Council and the Swedish National Space Board in Sweden.
 
Additional support for science analysis during the operations phase is gratefully acknowledged from the Istituto Nazionale di Astrofisica in Italy and the Centre National d'\'Etudes Spatiales in France. This work performed in part under DOE Contract DE- AC02-76SF00515.

JG and RG acknowledge support from the Israeli Science Foundation under
Grant No. 719/14. RG is supported by an Open University of Israel Research Fund.

\bibliography{mnemonic,biblio}


\appendix

\section{High-energy detection and localization of GRB 100724B}
\label{sec:localization}
Shortly after the beginning of the prompt emission, due to the \Fermi satellite's orbital motion the location of GRB~100724B on the sky started to move closer and closer to the Earth Limb (EL), which is a very powerful source of gamma rays.
Given the altitude of \Fermi, the EL appears in LAT standard data as a curved band centered around $\eta = 113\,^{\circ}$, where $\eta$ is the angle with the zenith of the spacecraft (Zenith angle). Thus, at a given energy $E$ there is potentially significant contamination from the EL for all regions of the field of view at a zenith angle $\eta \gtrsim 113\,^{\circ} - PSF(\theta, E)$, where $PSF(\theta, E)$ is the size of the PSF at off-axis angle $\theta$ and at the energy $E$ as measured for example by the 90\% containment radius. Due to the orbital motion of the \Fermi spacecraft, the zenith angle $\eta(\vec{p})$ of a fixed point in the sky $\vec{p}$ is continuously changing. The procedure suggested by the LAT team to limit the contamination from the Earth Limb proceeds as follows. Let us fix a Region Of Interest (ROI) in sky coordinates, centered around the position $\vec{p}_{s}$ of the source, and with a radius $R$: there are time intervals in which such ROI is clean from EL contamination (\textit{good time intervals}, GTIs) and other intervals in which it is not (\textit{bad time intervals}, BTIs). Let us fix a minimum energy for our analysis $E_{\rm min}$.  The GTIs are all the time intervals in which:

\be
\eta(\vec{p}_{s}, t) + R < 113^{\circ} - PSF(\theta, E_{\rm min}),
\label{eq:zenithCut}
\ee

The computation of such GTIs is performed by the tool \textit{gtmktime}, part of the \Fermi Science Tools. The analysis is then performed only on the GTIs.

We first search for high-energy emission from GRB 100724B during the prompt emission in a circle of $15^{\circ}$ around the location provided by \Fermi/GBM. Given the high zenith angle of the source, the condition in \eqt{\ref{eq:zenithCut}} does not return any GTI unless we increase $E_{\rm min}$ up to 300$\;$MeV. There are only 2 photons above that energy, and the source is not detected. However, high-energy emission from GRBs has been proved to last much longer than the prompt emission \citep{LATGRBcat}. We therefore consider a longer time interval, covering up to $15\;$ks after the beginning of the prompt emission, and adopt $E_{\rm min} = 1\;$GeV to recover exposure. The condition in \eqt{\ref{eq:zenithCut}} returns now 10962$\;$s of GTIs in the first 
$15\;$ks after $t_{0}$. In this case 6 events survive the cuts, among which a $\sim 10\;$GeV photon at $t_{0} + 2239\;$s. Note that there are still no photons during the prompt emission with this selection. We detect and localize the source producing a Test Statistic (TS) map for this time interval, with a likelihood model containing the appropriate Galactic and isotropic templates provided by the LAT collaboration as well as all the point sources from the 3FGL catalog \citep{2015ApJS..218...23A}. We model GRB~100724B as a point source with a power-law spectrum, and we keep the parameters for all 3FGL sources fixed to their catalog values. No 3FGL source is detected in the small time window of our analysis, thus this choice is irrelevant for our final results. 

\begin{figure}[t!]
\centering
\includegraphics[width=0.5\textwidth]{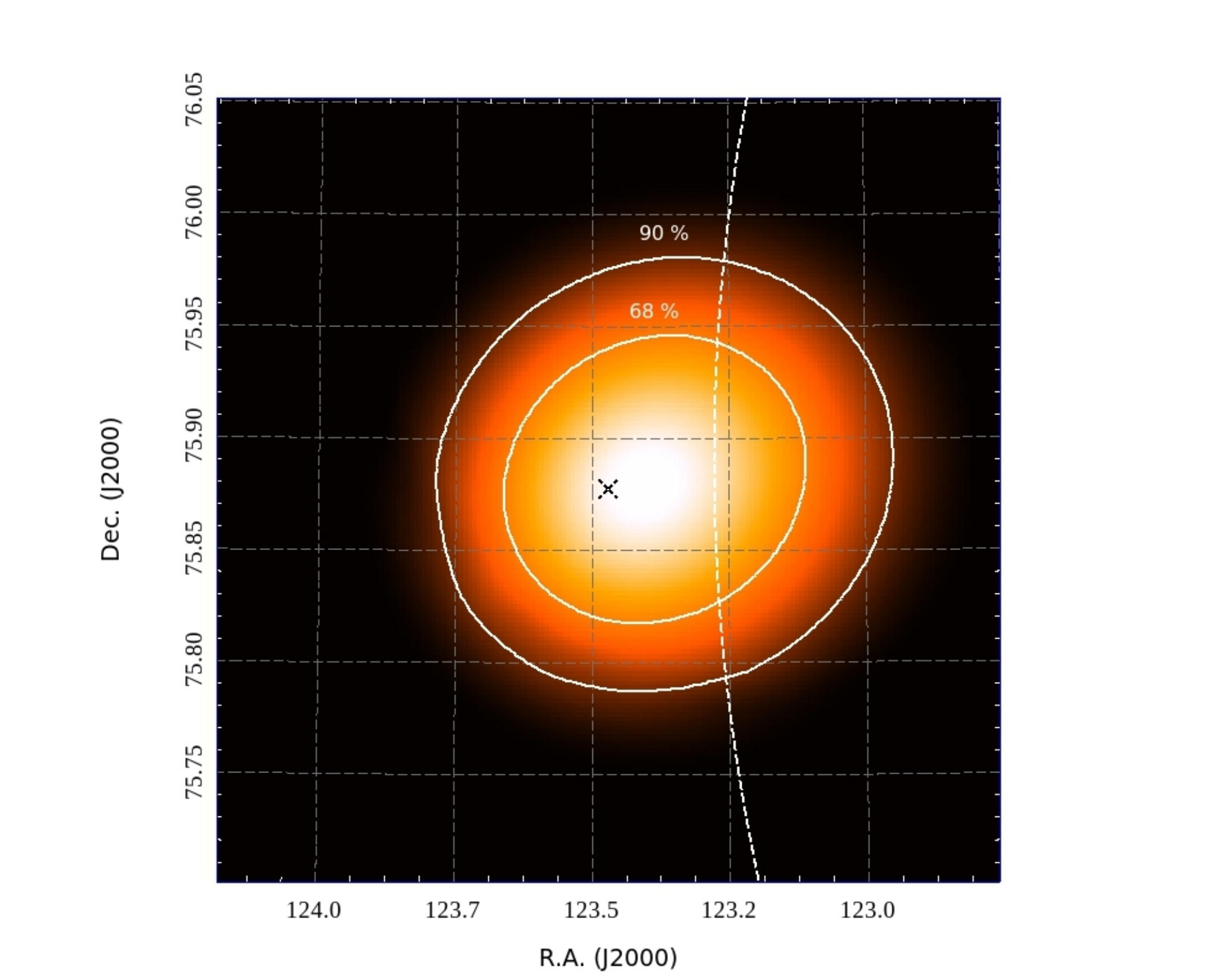}
\caption{Localization of GRB~100724B: the white contours are respectively the 68\% and 90\% containment regions, while the dotted line is the margin of the much larger 68\% containment region provided in \citet{LATGRBcat} based on the prompt interval. The black cross is the best fit localization. }
\label{fig:localization}
\end{figure}

This analysis results in a firm detection of the GRB with a significance of $\sim 7 \sigma$. The best fit power law has a photon index of  $\alpha = -1.8 \pm 0.4$, a value typical for the high-energy emission of GRBs \citep{LATGRBcat}, with an average flux of $8.26 \times 10^{-10}$ erg cm$^{-2}$ s$^{-1}$ (1-100 GeV). Our localization is shown in Figure~\ref{fig:localization}. The cross marks the best fit position, corresponding to $R.A. = 123.47\,^{\circ}$ and $Dec. = 75.88\,^{\circ}$ (J2000), while the white lines correspond to the 68\% and 90\% c.l. containment regions.  We used this position in the following analysis. The most accurate localization available in the literature before this work was reported in the \Fermi/LAT GRB Catalog \citep{LATGRBcat}, and corresponds to R.A. = $119.59\,^{\circ}$, Dec. = $75.86\,^{\circ}$ (J2000) with a 68\% containment radius of $0.88\,^{\circ}$ (dashed line in Figure~\ref{fig:localization}). This localization was based on the detection during the prompt phase, obtained by relaxing the zenith angle cut using a threshold of $110\,^{\circ}$, and it is compatible with the one found in this work, but it features a much larger containment radius. The excess we find is not related to any known source, and lies within the GBM localization region for GRB~100724B. The source is not detected in any other time interval. We therefore identify it as the high-energy counterpart of the burst.

\section{Likelihood Ratio Test}
\label{sec:LRT}

\begin{figure*}
\gridline{\fig{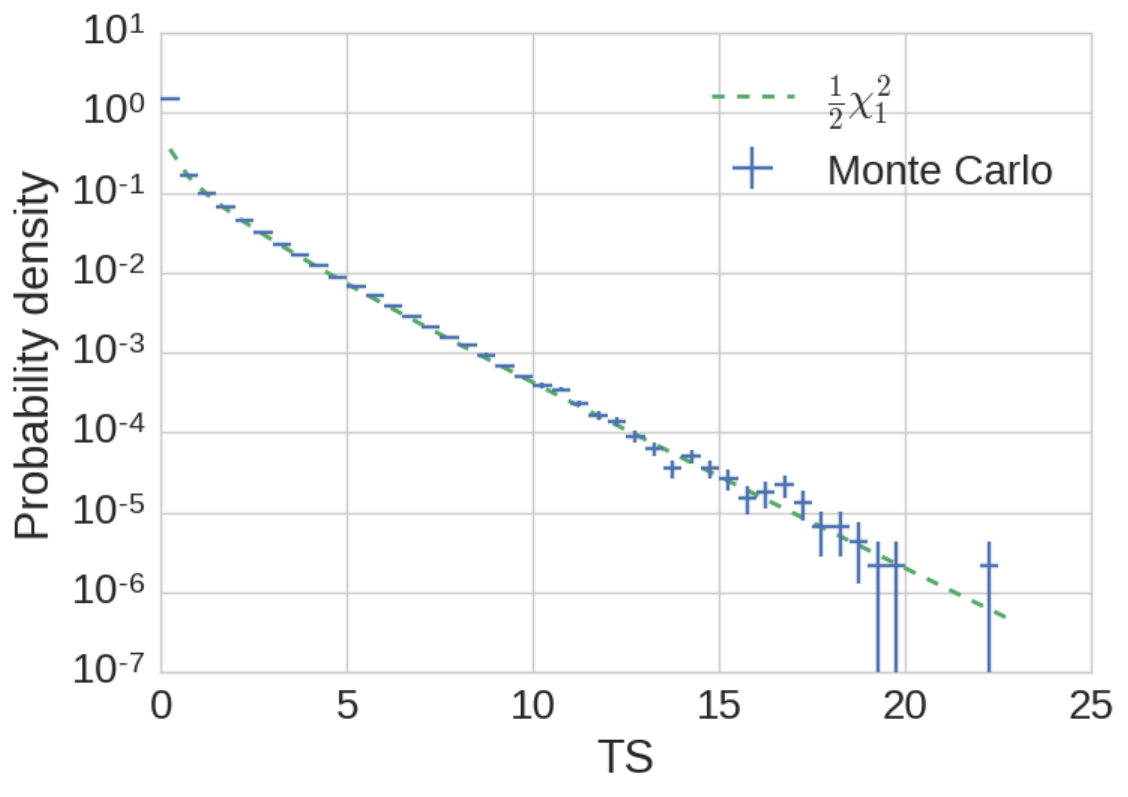}{0.45\textwidth}{(1)}
          \fig{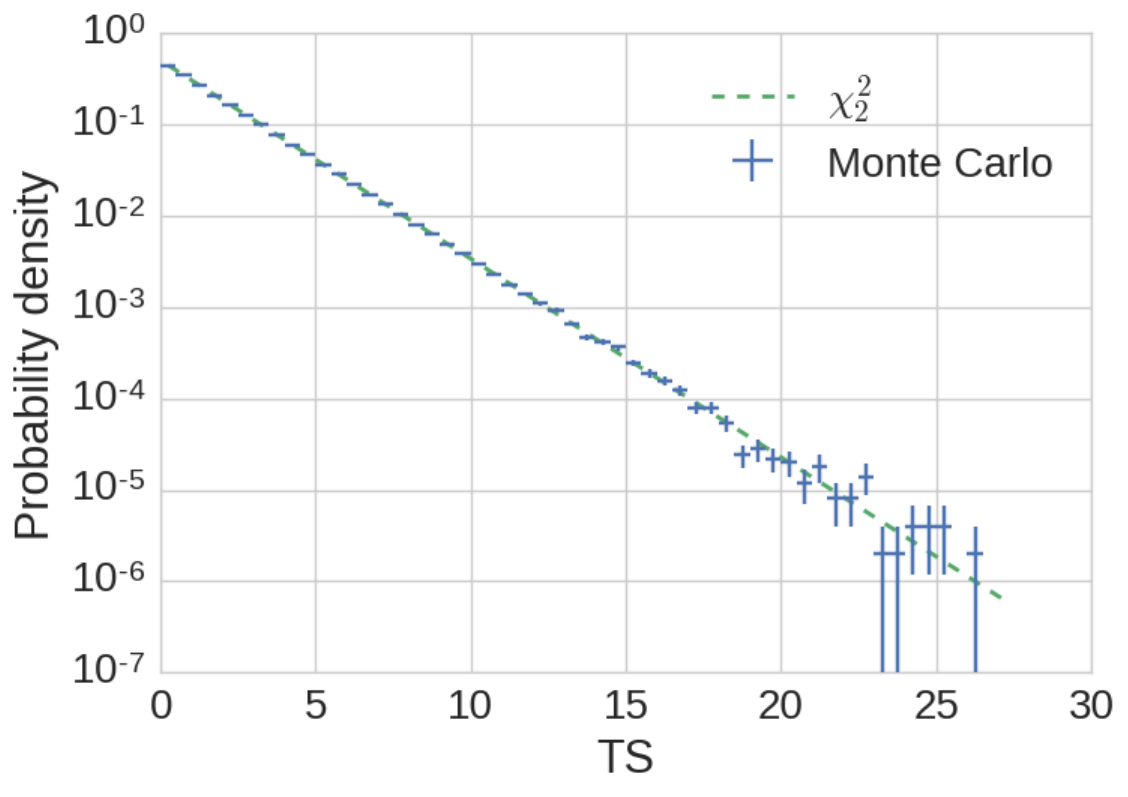}{0.45\textwidth}{(2)}}
\caption{Distribution of $TS$ from Monte Carlo simulations. The null hypothesis is $m_{0}=f_{Band}$ in both cases, while the alternative hypothesis $m_{1}$ is $f_{BHec}$ in panel 1 and $f_{BB}$ in panel 2. The distribution is well described respectively by $\frac{1}{2}~\chi^2_1$ and $\chi^2_2$ (dashed lines).}
\label{fig:ts_bhec}
\end{figure*}

The Likelihood Ratio Test (LRT) is a procedure for model selection. It is a statistical test that helps to select between two nested models $m_{0}$ and $m_{1}$ the one that best captures all significant features of the data. The model $m_{0}$ is the null-hypothesis, and represents the simpler model, while $m_{1}$ is the alternative hypothesis and is more complex than $m_{0}$. The Test Statistic (TS) is twice the difference in the log-likelihood $S$ between the two models. Wilks' theorem states that under certain assumptions $TS$ is asymptotically distributed as a $\chi^2$ with $n$ degrees of freedom, where $n$ is the difference in number of parameters between $m_{0}$ and $m_{1}$. In the cases of interest here such theorem is not guaranteed to hold, as our setup violates some of its hypotheses \citep[see][for details]{2002ApJ...571..545P}. Also, we might not be close enough to the asymptotic regime. We have then to rely on Monte Carlo simulations to calibrate the distribution of $TS$. In particular, we perform 1 million simulations of the null hypothesis $m_{0}$ using the responses of the instruments. Then, we fit each simulated dataset with both $m_{0}$ and $m_{1}$, recording the values of the log-likelihood (respectively $S_{0}$ and $S_{1}$). We can then study the distribution of $TS = 2~(S_{0} - S_{1})$. For the comparison between $m_{0} =f_{Band}$ and $m_{1}=f_{BHec}$ we find that the $TS$ is distributed as $ֿ\frac{1}{2}\chi^{2}_{1}$, as shown in panel 1 of Figure~\ref{fig:ts_bhec}. Instead, for the comparison between $m_{0}=f_{Band}$ and $m_{1} =f_{BB}$ we find that the TS is distributed as $\chi^2_{2}$, as expected from Wilks' theorem (panel 2). We also repeated this simulation exercise for each time interval used in our analysis, using 100,000 simulations, and verified that in each case we are in the same regime. We can then use these two distributions to determine the significance of the black body and the cutoff.

\section{Systematic uncertainties}
\label{sec:sys_effects}

The effective area correction we used during the fit, by introducing a multiplicative constant free to vary independently for each detector except one (see text), can help neutralizing effects due to systematic errors in the inter-calibration between the different detectors, introduced for example by underestimating or overestimating the effective area of a detector. It cannot, however, account for distortion in the spectrum introduced by errors in the energy-by-energy or channel-by-channel measurement of the instrument response model. In this section we study the impact of such uncertainties on our results.

We first briefly describe LLE data and the procedure we used to estimate the amount of systematic uncertainty in the relative response. We then proceed to study how such uncertainty, coupled with the uncertainty on the \Fermi/GBM response, can modify our results.

\subsection{LLE class}
\label{sec:vela_sys}

The LAT Low energy (LLE) technique is an analysis method designed to study bright transient phenomena, such as GRBs and solar flares, in the 30 MeV--1 GeV energy range. The LAT collaboration developed this analysis using a different approach than the one used in the standard photon analysis, which is based on sophisticated classification procedures \citep[a detailed description of the standard analysis can be found in][]{2009ApJ...697.1071A,2012ApJS..203....4A}. The idea behind LLE is to maximize the effective area below $\sim$ 1 GeV by relaxing the standard analysis requirement on background rejection. The basic LLE selection is based on a few simple requirements on the event topology in the three sub-detectors of the LAT namely: a tracker/converter (TKR) composed of 18 x--y silicon strip detector planes interleaved with tungsten foils; an 8.6 radiation length imaging calorimeter (CAL) made with CsI(Tl) scintillation crystals; and an Anti-coincidence Detector (ACD) composed of 89 plastic scintillator tiles that surrounds the TKR and serves to reject the cosmic-ray background.

First of all, an event passing the LLE selection must have at least one reconstructed track in the TKR and therefore an estimate of the direction of the incoming photon. Secondly, we require that the reconstructed energy of the event be nonzero. The trigger and data acquisition system of the LAT is programmed to select the most likely gamma-ray candidate events to telemeter to the ground. The on-board trigger collects information from all three subsystems and, if certain conditions are satisfied, the entire LAT is read out and the event is sent to the ground. We use the information provided by the on-board trigger in LLE to efficiently select events which are gamma-ray like. In order to reduce the number of photons originating from the Earth Limb in our LLE sample we also include a cut on the reconstructed event zenith angle (i.e. angle $<$100$^{\circ}$). Finally we explicitly include in the selection a cut on the region of interest, i.e. the position in the sky of the transient source we are observing. In other words, the localization of the source is embedded in the event selection and therefore for a given analysis the LLE data are tailored to a particular location in the sky.

The response of the detector for the LLE class is encoded in a response matrix, which is generated using a dedicated Monte Carlo simulation for each GRB, and is saved in the standard \texttt{HEASARC RMF} File Format\footnote{Described here: \url{http://heasarc.gsfc.nasa.gov/docs/heasarc/caldb/docs/memos/cal_gen_92_002/cal_gen_92_002.html\#Sec:RMF-format}.}. LLE data and the relative response are made available for any transient signal (GRB or Solar Flare) detected with a significance above 4$\sigma$ through the {\tt HEASARC} web site\footnote{FERMILLE, at \url{http://heasarc.gsfc.nasa.gov/W3Browse/fermi/fermille.html}}.

\subsubsection{Validation and systematic uncertainties}

Discrepancies between the actual response of the LAT and the response matrix derived from 
simulations can cause systematic errors in spectral fitting. We investigated the systematic uncertainties tied to the 
LLE selection by following the procedure described in \citet{LATperform}. In particular, we compared Monte Carlo with 
flight data, using the Vela pulsar (PSR J0835--4510) as a calibration source. The pulsed nature of the gamma-ray 
emission from this source \citep{LATVela2} gives us an independent control on the residual charged particle 
background. In fact, off-pulse gamma-ray emission is almost entirely absent, and a sample of ``pure photons'' can be 
simply extracted from the on-pulse region, after the off-pulse background is subtracted. Considering all time intervals during which the Vela pulsar was observed at an incidence angle $\theta < 80^\circ$, 
we estimate the discrepancy between the efficiency of the LLE selection criteria in the LAT data and in Monte Carlo to be 
$\sim$17\% below 100~MeV, decreasing to $\sim$8\% at higher energies, with an average value $\sim$9\% (note that this
average is weighted by the Vela spectrum). 

Additionally, we performed a spectral analysis of the Vela pulsar, comparing LLE results with the standard \Fermi likelihood analysis. The $>$100 MeV flux obtained from the LLE analysis is 16\% less than the flux reported by \citet{LATVela2}. 
This discrepancy can be attributed to the fact that the selection criteria between LLE and the standard LAT likelihood analysis are rather different with the former being much looser. 

Finally, we also studied the energy resolution using large samples of simulated events with the \Fermi/LAT full simulator. No significant bias was found, and the energy resolution for LLE is estimated to be $\sim$ 40\% at 30 MeV, $\sim$30\% at 100 MeV and $<$ 15\% for energies greater than 100 MeV.

\subsection{\Fermi/GBM detectors}

The systematic uncertainties on the responses of the \Fermi/GBM detectors have been studied before launch, giving a calibration uncertainty of $\sim$ 10\% for NaI detectors and $\sim$ 20\% for BGO detectors \citep{2009ExA....24...47B}. In-flight calibration efforts have been limited so far, but preliminary results show a systematic uncertainty of around 15\% in overall flux measurements, obtained comparing \Fermi/GBM with other instruments \citep{2009AIPC.1133..446V}.

\subsection{Effect of systematic errors on the measurement of the cutoff}

In this section we describe a method we have used to estimate the impact that systematic uncertainties in the response of the instruments have on the detection of the cutoff.

For all instruments used in this work, the response is encoded in a response matrix. A response matrix\footnote{As encoded in OGIP RSP files} is a bidimensional histogram having true energy on the x-axis and detected energy (or channels) on the y-axis, and a value proportional to the probability for a given photon of true energy $E_{true}$ to be detected at energy $E_{obs}$. Such responses are generated using simulations of the detectors, which must account for the geometry of the observation, the physical characteristics of the detector, and the physics of all the processes involved. All of these have uncertainties that are very difficult to study and to model, contributing to errors in the final response matrix. However, we have presented in the previous sections calibration studies that have been performed to estimate the overall systematic uncertainties of the effective area at different energies. 

The approach we decided to use is to deliberately distort the response matrix in many different ways while keeping the difference in effective area between the original and the distorted matrix within the systematic uncertainties. In particular, this is our procedure: i) we consider the original response, with M true-energy channels and N observed-energy channels, and we generate a M x N matrix of uncorrelated noise ii) we smooth the noise matrix with a ``diagonal kernel'' that introduces correlation among the elements in the matrix along the diagonal direction iii) we smooth again the noise matrix with a Gaussian kernel, which removes unphysical jumps between neighboring elements iv) we re-normalize the noise matrix to have only elements between $1 - f_{sys}$ and $1 + f_{sys}$, where $f_{sys}$ is the fractional systematic uncertainty for the detector under examination. We used respectively 0.1 for NaI detectors, 0.2 for the BGO detector and 0.15 for LLE. We then multiply it by the original matrix to give a distorted matrix with at most $f_{sys}$ fractional variations v) we renormalize the distorted matrix to have the same total effective area as the original one, since the bias in the total area is already encapsulated by the effective area correction factor used in the fit.

\begin{figure*}[t!]
\centering
\includegraphics[width=0.5\textwidth]{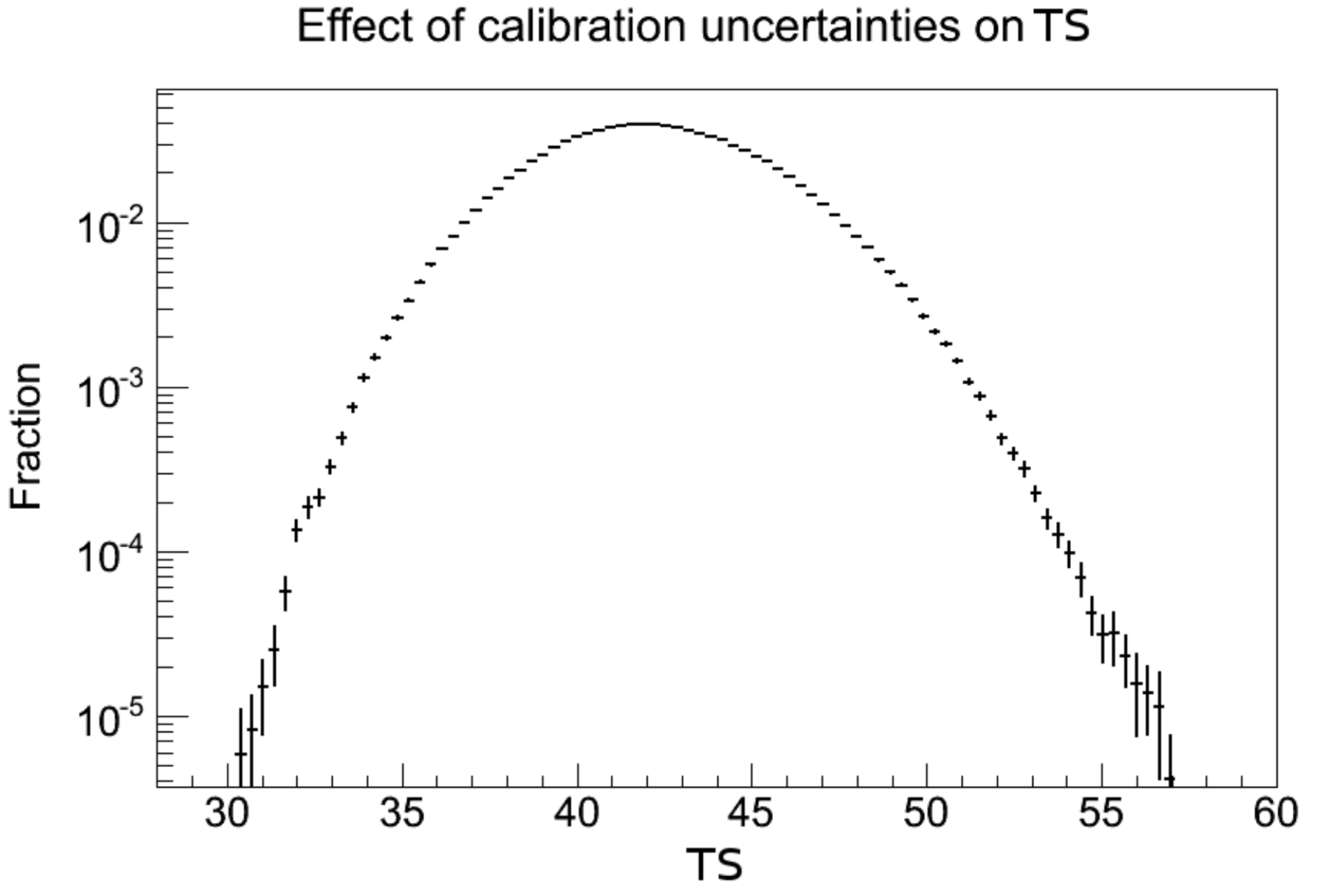}
\caption{Distribution of TS values for the cutoff obtained in a set of simulations where the response matrices of the instruments have been distorted to study the effect of systematic uncertainties.}
\label{fig:systematics}
\end{figure*}

Using this procedure we generated 500,000 distorted matrices for the data. For each realization we performed a fit, but using the distorted matrices in place of the original ones. We fit both $f_{Band}$ and $f_{BHec}$, and record the value of the best fit parameters as well as the value for TS. 
Among all intervals for both GRBs, interval 3 for GRB 100724B turned out to be the interval most affected by changes in the response matrices, and it consequently has the widest TS distribution, which is shown in Figure \ref{fig:systematics}. We show the distribution with a logarithmic y-axis, which emphasizes the extremes of the distribution, but it must be noted that the vast majority of realizations change the TS value by only a few units. Among all intervals for both bursts, there are very rare instances where TS changes by up to 15 units, corresponding to very specific cases where changes in effective area of all the instruments conspire to change more drastically the significance of the cutoff. We can then assume a reduction of 15 units in TS for all time intervals as the most pessimistic case. It still would translate in 4 intervals above $5~\sigma$ for GRB~100724B, and 5 intervals for GRB~150509A. However, our simulations show that this is extremely unlikely: out of 500,000 simulations per interval, only a handful resulted in such a big change. Much more likely, systematic uncertainties change the value of TS by just few units, which does not affect almost at all our results. We can conclude that the detection of the cutoff is not likely affected by systematic uncertainties in the response of our instruments.

\section{Comparison with AGILE observations of GRB 100724B}
\label{appendix:agile}
\begin{figure}[t!]
\centering
\includegraphics[width=0.45\textwidth]{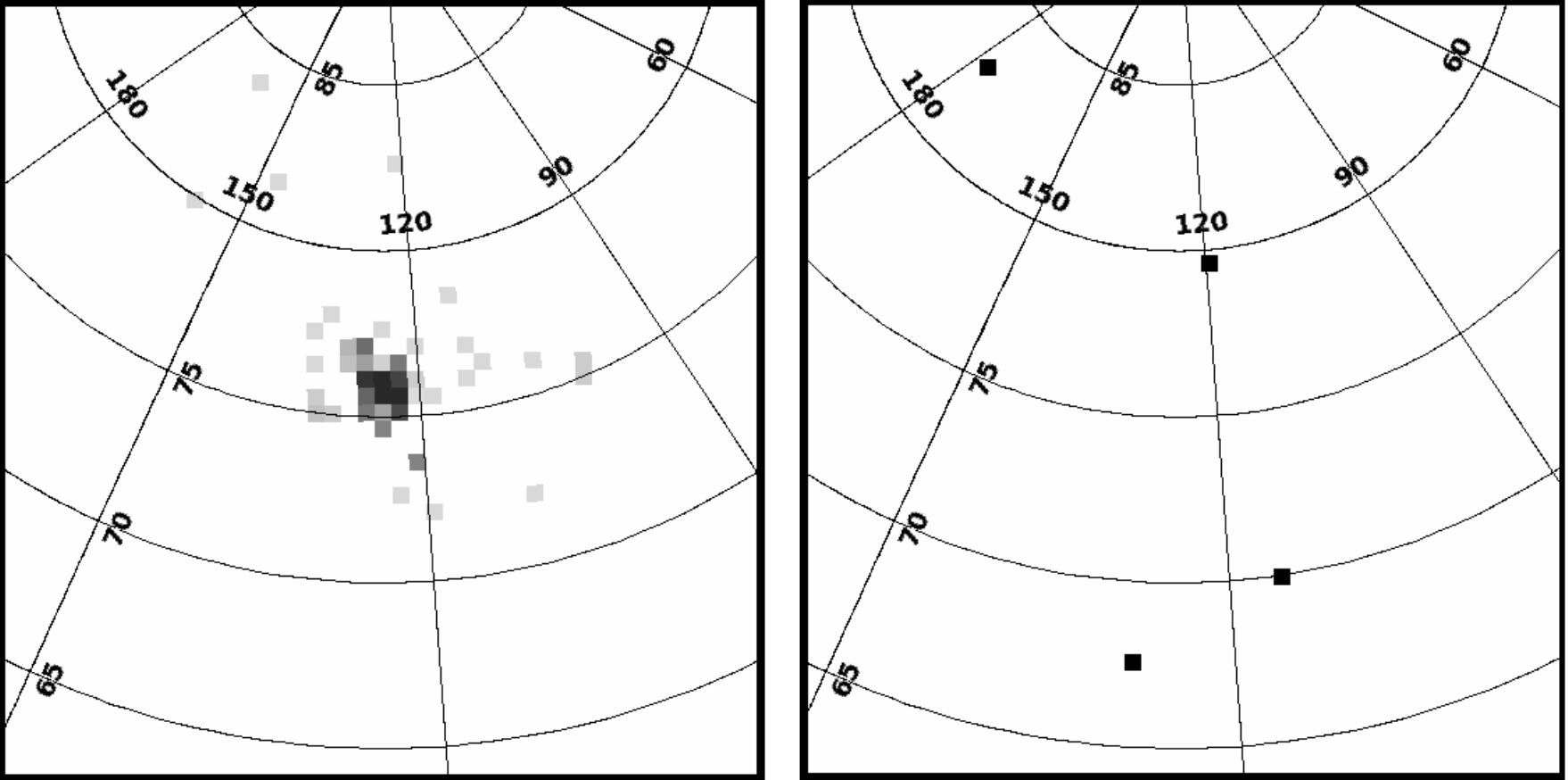}
\caption{Comparison between a simulated LAT observations of a source with the characteristics measured by {\it AGILE} for GRB~100724B  \citep{2011A&A...535A.120D} (left panel) and what we actually observed (right panel).}
\label{fig:agile}
\end{figure}

GRB~100724B was detected and studied by {\it AGILE} as well \citep{2011A&A...535A.120D}. However, their results are different from what we observed with the LAT. For example, they observe a signal between the trigger time and 90 s with photons up to 3.5 GeV. They report a photon fluence in the 22 MeV - 3.5 GeV energy band of $0.25 \pm 0.05$ ph.~cm$^{-2}$~s and an energy fluence in the same band of $(4.7 \pm 0.9 \times 10^{-5})$ erg~cm$^{-2}$. They also measure a photon index of $-2.04^{+0.31}_{-0.14}$. These values correspond to a source much brighter and harder than what we see. To demonstrate this, we have performed a simulation of a source with the flux measured by {\it AGILE}. We used the tool \textit{gtobssim}, part of the public \Fermi Science Tools, which takes into account all the aspects of the LAT as well as the real pointing history of the satellite. We then compared it with what we observe. In Figure~\ref{fig:agile} we show a counts map of the simulation (left panel) and what we see (right panel), both using events above 300 MeV where the Earth Limb contamination in the data is small. For the sake of this comparison we did not introduce any cut on the zenith angle. It is apparent that what we observe is incompatible with what has been measured by {\it AGILE}. According to the simulation we should have detected $N_{pred} = 60$ photons from the source above 300 MeV. We instead observe $N_{obs} = 4$ photons. Even assuming that they all come from the source, the Poisson probability of observing $N_{obs}$ when we expect $N_{pred}$ is $5 \times 10^{-21}$, i.e., essentially zero.

Given the extremely soft spectrum that we measure, we note that a possible culprit for such discrepancy could be the energy dispersion. For example, let us consider the photon that {\it AGILE} observed during the prompt emission with a reconstructed energy at 3.5 GeV. The LAT has an effective area above 1 GeV which is several times the one of {\it AGILE}/GRID, thus it should have detected several photons above 1 GeV in the same time interval, while we detect none. The energy resolution in {\it AGILE}/GRID is such that photons with energies below 100 MeV have a probability of being reconstructed well above 1 GeV of few percent \citep[see upper left panel in Figure 3 in][]{2013A&A...558A..37C}. Given the brightness and softness of this GRB, there are hundreds of photons below 100 MeV in the LAT, as clear from the LLE light curve in Figure \ref{fig:bblc}. {\it AGILE}/GRID 
observed 57 events. Thus it is entirely possible that one photon with a true energy below 100 MeV has a reconstructed energy above 1 GeV in {\it AGILE}/GRID. In \citet{2011A&A...535A.120D} it is not clear if and how the energy dispersion has been accounted for. However, ignoring the energy dispersion altogether or inaccuracies in its treatment can have a huge impact on the analysis of this burst. We note that in our analysis of the prompt emission, the energy dispersion in all instruments is accounted for.

\section{Spectral models}
\label{sec:spmodels}
In this appendix we report the expressions for the spectral models used in this paper. All formulae are for the differential photon flux in photons cm$^{-2}$ s$^{-1}$ keV$^{-1}$.

\subsection{Power law with exponential cutoff}

\be
f_{PHec}(E) = K~\frac{E}{E_{piv}}^{-\alpha}~e^{-\frac{E}{E_{c}}},
\label{eq:PHec}
\ee
where $\alpha$ is the photon index, $E_{c}$ is the cutoff energy, $K$ is the differential flux at $E_{piv}$, and $E_{piv}$ is the pivot energy, which we keep fixed to 1.

\subsection{Band model}
This is the model from \citet{1993ApJ...413..281B}, which consists of a low-energy power law and a high-energy power law joined by an exponential function:
\be
f_{Band}(E) = K~\begin{cases}
E^{\alpha} \exp{\left(\frac{-E}{E_{0}}\right)} & E < (\alpha-\beta)E_{0} \\ 
\left[ (\alpha-\beta)E_{0}\right]^{\alpha-\beta} \exp{(\beta-\alpha)} E^{\beta} &  E \ge (\alpha-\beta)E_{0}\\
\end{cases},
\label{eq:band}
\ee
where $\alpha$ is the low-energy photon index, $\beta$ is the high-energy photon index, $E_{0}$ is the break energy and $K$ the normalization constant. It is easy to show that the peak energy in the Spectral Energy Distribution $E^{2}f_{Band}(E)$ is $E_{p} = (2+\alpha)E_{0}$.

\subsection{Band model with exponential cutoff}
A Band model multiplied by an exponential function:
\be
f_{BHec}(E) = f_{Band}(E) \times \exp{\left(-~\frac{E}{E_{c}}\right)},
\label{eq:bhec}
\ee
where $E_{c}$ is the cutoff energy.

\subsection{Band model plus black body}
A Band model plus a Planck function:
\be
f_{BB}(E) = f_{Band}(E) + A~\frac{E^{2}}{\exp{\left(\frac{E}{kT}\right)}-1}
\label{eq:bb}
\ee


\subsection{Band model with power law break}
A Band model where the second branch is a broken power law instead of a power law:
\be
f_{Bbkpo}(E) = \begin{cases}
K E^{\alpha} \exp{\left(\frac{-E}{E_{0}}\right)} & E < (\alpha-\beta)E_{0} \\ 
K \left[ (\alpha-\beta)E_{0}\right]^{\alpha-\beta} \exp{(\beta-\alpha)} E^{\beta} & (\alpha-\beta)E_{0} \le E < E_{c}\\
K \left[ (\alpha-\beta)E_{0}\right]^{\alpha-\beta} \exp{(\beta-\alpha)} E_{c}^{\beta-\beta_{2}} E^{\beta_{2}} & E \ge E_{c}
\end{cases}
\label{eq:band_bknpo}
\ee

\subsection{Band model with smooth transition to a power law}
A Band model where the high-energy power law changes index smoothly:
\be
f_{Bgr}(E) = f_{Band}(E) \times \left[ 1 + \frac{E}{E_{c}}^{n\Delta\beta}\right]^{-1/n},
\label{eq:Bgr}
\ee
with $\Delta\beta$ fixed to:
\be
\Delta\beta = \frac{(\beta+1)(2-\beta)}{\beta-1},
\ee
as expected from theoretical considerations \citep{2008ApJ...677...92G}.

\end{document}